\documentclass[
 preprint,
 superscriptaddress,
 amsmath,amssymb,
 aps,
 pra,
]{revtex4-2}

\usepackage{graphicx}
\usepackage{dcolumn}
\usepackage{bm}
\usepackage[
left=0.7in,right=0.7in,top=0.8in,bottom=0.9in
]{geometry}
\usepackage{subfigure}
\usepackage{latexsym}
\usepackage{epsfig}
\usepackage{amsbsy}
\usepackage{array}
\usepackage{setspace}
\usepackage{lineno}
\usepackage{xcolor}
\usepackage[colorlinks,linkcolor=blue,anchorcolor=blue,urlcolor  = blue,citecolor=blue]{hyperref}
\usepackage{ulem}
\usepackage{txfonts}
\newcommand*\funit{ph/\ensuremath{\mu}m\ensuremath{^2}}

\newcommand*\fluence[2]{#1\ensuremath{\times}10\ensuremath{^{#2}}~\funit}
\newcommand*\iunit{W/cm\ensuremath{^2}}

\newcommand*\intensity[2]{#1\ensuremath{\times}10\ensuremath{^{#2}}~\iunit}
\newlength{\figurewidth}
\begin{document}

\title{Transient ionization potential depression in nonthermal dense plasmas at high x-ray intensity}

\author{Rui Jin}\email{rui.jin@cfel.de}
\affiliation {Center for Free-Electron Laser Science, DESY, Notkestrasse 85, 22607 Hamburg, Germany}
\affiliation {Department of Physics and Astronomy, Shanghai Jiao Tong University, 200240 Shanghai, China}

\author{Malik Muhammad Abdullah}
\affiliation {Deutsches Elektronen-Synchrotron DESY, Notkestrasse 85, 22607 Hamburg, Germany}

\author{Zoltan Jurek}
\affiliation {Center for Free-Electron Laser Science, DESY, Notkestrasse 85, 22607 Hamburg, Germany}
\affiliation {The Hamburg Centre for Ultrafast Imaging, Luruper Chaussee 149, 22761 Hamburg, Germany}

\author{Robin Santra}
\affiliation {Center for Free-Electron Laser Science, DESY, Notkestrasse 85, 22607 Hamburg, Germany}
\affiliation {The Hamburg Centre for Ultrafast Imaging, Luruper Chaussee 149, 22761 Hamburg, Germany}
\affiliation {Department of Physics, Universit\"at Hamburg, Jungiusstrasse 9, 20355 Hamburg, Germany}

\author{Sang-Kil Son}\email{sangkil.son@cfel.de}
\affiliation {Center for Free-Electron Laser Science, DESY, Notkestrasse 85, 22607 Hamburg, Germany}
\affiliation {The Hamburg Centre for Ultrafast Imaging, Luruper Chaussee 149, 22761 Hamburg, Germany}

\normalem
\date{\today}

\begin{abstract}
The advent of x-ray free-electron lasers (XFELs), which provide intense ultrashort x-ray pulses, has brought a new way of creating and analyzing hot and warm dense plasmas in the laboratory.
Because of the ultrashort pulse duration, the XFEL-produced plasma will be out of equilibrium at the beginning and even the electronic subsystem may not reach thermal equilibrium while interacting with a femtosecond time-scale pulse.
In the dense plasma, the ionization potential depression (IPD) induced by the plasma environment plays a crucial role for understanding and modeling microscopic dynamical processes.
However, all theoretical approaches for IPD have been based on local thermal equilibrium (LTE) and it has been controversial to use LTE IPD models for the nonthermal situation.
In this work, we propose a non-LTE (NLTE) approach to calculate the IPD effect by combining a quantum-mechanical electronic-structure calculation and a classical molecular dynamics simulation.
This hybrid approach enables us to investigate the time evolution of ionization potentials and IPDs during and after the interaction with XFEL pulses, without the limitation of the LTE assumption. 
In our NLTE approach, the transient IPD values are presented as distributions evolving with time, which cannot be captured by conventional LTE-based models.
The time-integrated ionization potential values are in good agreement with benchmark experimental data on solid-density aluminum plasma and other theoretical predictions based on LTE.
The present work is promising to provide critical insights into nonequilibrium dynamics of dense plasma formation and thermalization induced by XFEL pulses.

\end{abstract}
                              
                              
\maketitle

\section{\label{sec:intro}Introduction}

High energy density matter exists extensively in the Universe, from the hot dense plasmas such as those in supernovae and stellar interiors~\cite{Taylor1994,Rogers1994} to warm dense plasmas like those in planetary interiors~\cite{Chabrier2009,Helled2010,Guillot1999,Militzer2016}.
Understanding matter at extreme conditions is of crucial importance for not only astrophysical observation, but also emerging experiments in the laboratory such as inertial confinement fusion (ICF)~\cite{Lindl1995,Glenzer2010} and x-ray free-electron laser (XFEL)~\cite{Seddon2017,Glenzer2016} facilities. 
The typical energy density is above $10^{11}$~J/m$^3$~\cite{Drake2006}. 
In the laboratory, high energy density is achieved by imposing energy sources such as high intensity lasers or high pressure onto fluids or solids within ultrashort time. 
In particular, highly brilliant, spatially coherent XFEL pulses can uniformly heat bulk matter and transform it to a warm dense plasma on a  femtosecond time scale, which makes XFELs an ideal tool for creating and probing solid-density plasmas~\cite{Vinko2012}.

When a solid-density material is exposed to an intense x-ray pulse generated by XFELs, many electrons are ionized via x-ray photoionization and the system evolves through microscopic dynamical processes such as collisional ionization, recombination, and electron-ion relaxation processes. 
With such a high energy density and an ultrashort time scale, the plasma created by XFELs lies far from local thermodynamic equilibrium (LTE)~\cite{Bauche2015} during the interaction with the x-ray pulses. 
The electron-ion relaxation time scale is a few picoseconds~\cite{Agassi84,Silvestrelli06}, which is much longer than the typical duration of XFEL pulses (1--100 femtoseconds), so electrons and ions are not in thermal equilibrium within the time scale of the pulse duration. 
The electron-electron relaxation time scale is tens~\cite{Ostrikov16} or hundreds~\cite{Medvedev11} of femtoseconds, which is comparable to the XFEL pulse duration, and thus it is questionable whether the electronic subsystem is fully thermalized, especially at the early stage. 
Therefore, XFEL-created plasmas are theoretically treated by the LTE approach with the two-temperature method~\cite{Anisimov74,Rethfeld17}, where electrons are assumed to be instantaneously thermalized and ions remain cold, or by the non-LTE (NLTE) approach~\cite{Chung09,Hansen20,Scott:2001aa}, where detailed microscopic dynamics are taken into account in the kinetic simulation.

In dense plasmas, ionized electrons roam around ions, which influences the electronic structure of ions. 
Due to the screening effect of the plasma electrons and the dense environment, the atomic energy levels are shifted and the ionization potential is lowered in comparison with that for the isolated atom case. 
This phenomenon is called ionization potential depression (IPD) and it plays a crucial role for understanding and modeling atomic processes in dense plasmas. 
To describe this IPD effect, two distinct models have been available, Ecker--Kr\"oll (EK)~\cite{E-K1963} and Stewart--Pyatt (SP)~\cite{S-P1966}, both of which are based on thermal equilibrium. 
Recent experiments using an XFEL~\cite{Vinko2012,Ciricosta2012} and a high-power optical laser~\cite{Hoarty13,Hoarty2013} have triggered a controversial debate on the validity of these models, which has been followed by extensive studies on the IPD measurements~\cite{Ziaja13,Preston2013,Fletcher14,Ciricosta16a,Kraus19,Kasim:2018aa} and the theoretical treatments of the IPD effect~\cite{Crowley2014,Iglesias14,Calisti2015,Calisti15a,Stransky2016,Lin2017,Roepke2019,Rosmej18,Pain2019,Li:2019aa}, including \textit{ab initio} electronic structure calculations~\cite{xatom2014AA,Vinko2014,Bekx2020,Zeng2020,Driver2018,Hu2017}.
It is worthwhile to note that all of the employed methods are based on the LTE condition for the electronic subsystem. 
In other words, all above-mentioned methods incorporate the assumption of thermal equilibrium at a given electron temperature. 
Therefore, the IPD calculated may not be suitable for describing femtosecond dynamics of solid-density plasma formation by XFELs, where an equilibrium electron temperature often cannot be defined~\cite{Hau-Riege13}. 
In particular, the way that NLTE dynamics simulations are fed by LTE IPD input such as EK and SP might be questionable for calculation of XFEL-created dense plasmas.
Furthermore, nonthermal femtosecond phase transitions induced by an intense hard-x-ray pulse have been reported~\cite{Makita19,Hartley19}, thus necessitating an IPD treatment that is unrestricted, for both electrons and ions, by any thermal equilibrium condition.

In this work, we propose an NLTE approach to calculate the IPD effect, based on the real-space charge distribution obtained directly from real-space molecular-dynamics (MD) simulation trajectories.
In order to incorporate the NLTE plasma status, we employ a Monte Carlo-molecular dynamics (MC-MD) simulation tool, XMDYN~\cite{Murphy2014,Zoltan2016}. 
The IPD values are calculated by subtracting the ionization potential for an isolated atom and that for an atom embedded in the plasma with the same electron configuration. 
For both cases, the ionization potential is calculated quantum-mechanically, by using an atomic toolkit, XATOM~\cite{xatom2011,Zoltan2016}. 
To calculate the atomic electronic structure including the plasma effect, we have developed a dedicated tool, XPOT, interfacing between XMDYN and XATOM. 
This paper is organized as follows. 
In Sec.~\ref{sec:theo}, we provide the theoretical description of how to construct the classical environmental micro field from MD simulations and how to evaluate IPD values. 
In Sec.~\ref{sec:results}, we present an ensemble of calculated IPD values during an NLTE simulation of Al solid-density plasma created by an intense x-ray pulse. 
This result is compared with experimental measurement and other theoretical results. 
We conclude with a short summary and an outlook in Sec.~\ref{sec:con}.

\section{\label{sec:theo}Methodology}
 
\subsection{\label{sec:md}XMDYN: Monte Carlo-molecular dynamics scheme}
We briefly review the theoretical background of XMDYN~\cite{Murphy2014,Zoltan2016}, which is used in this study to simulate the creation of a nonequilibrium dense plasma. 
XMDYN is a versatile code based on the classical molecular dynamics (MD) and Monte Carlo (MC) approaches. 
In this method, ionized electrons and atomic ions are treated as classical particles roaming in 3D real space, and their dynamics are described by the MD technique. 
The dynamics of the electronic configuration of each atomic ion are computed in a stochastic framework using an MC algorithm. 
The electronic structure (orbitals and orbital energies) of each atomic ion is solved quantum-mechanically on the fly (i.e., when needed in a given numerical time step) by using the atomic toolkit XATOM~\cite{xatom2011,Zoltan2016}, based on the Hartree--Fock--Slater (HFS) method. 
Cross sections and rates of photoionization, Auger decay, and fluorescence are calculated based on the orbitals and orbital energies within XATOM. 
Further, the treatment of collisional ionization and recombination in XMDYN also relies on computed orbital energies.
The complex electron thermalization and energy transfer processes are realized  in the real-time MD evolution by the elementary classical many-body collisions (including electron--electron scattering and electron--ion scattering) as well as collisional ionization and recombination. 
Note that a trajectory tracing algorithm~\cite{Zoltan2016} is employed to treat the recombination process. 
Essentially, a classical electron is recombined with an atomic ion if it is found to orbit around the same trapping ion for $n_\text{rec}$ full periods, where the cycle-count threshold for recombination $n_\text{rec}$ is a small integer. 

This hybrid quantum--classical treatment is capable to treat some quantum effects, but admittedly not all, and there are no significant limitations in its ability to capture many-body interactions, which is hard to achieve within, for example, time-dependent density-functional theory (TDDFT)~\cite{Baczewski16}.
For the solid-density Al experiment considered here, the created plasma is not deep in the regime of quantum degeneracy, once the system has been heated.
Note that the standard functionals used in DFT and TDDFT do not treat exchange effects, which are connected to quantum degeneracy, in a rigorous manner.
At the moment, only a mixed quantum-classical framework, as adopted within XMDYN, is available to tackle nonequilibrium dynamics in the regime of fairly strong coupling.

This XMDYN--XATOM approach has been applied to explain many XFEL experiments, for example, explosion and fragmentation dynamics of C$_{60}$~\cite{Murphy2014, Berrah2019} and nanoplasma formation and disintegration dynamics of rare gas clusters~\cite{Tachibana2015,Kumagai2018,Kumagai2020}. 
It has been extended to simulate a bulk system within the supercell approach in combination with periodic boundary conditions~\cite{Abdullah2016,Abdullah2017,Abdullah2018}, and it has been an indispensable tool for the start-to-end simulation of single-particle imaging at the European XFEL~\cite{Yoon2016,Fortmann-Grote2017}. 
However, before the present work, there was the caveat that the quantum treatment of the atomic electronic structure did not take into account the plasma environment. 
The most obvious shortcoming of this was the lack of ionization potential depression emerging in the ionization dynamics of dense plasmas. In the following, we describe how we have addressed this challenge. 

\subsection{\label{sec:IPD}Electronic structure calculation}
The electronic structure of an isolated atom is obtained by solving the effective single-electron Schr\"odinger equation (atomic units are used unless specified otherwise),
\begin{align}
    \label{HFS}
	\left[ -\frac{1}{2} \nabla^2 + V_A^\text{iso}(\mathbf{r}) \right] \psi(\mathbf{r}) = \varepsilon \psi(\mathbf{r}).
\end{align}
Here, $V_A^\text{iso}(\mathbf{r})$ is the mean field potential for an isolated atom,
\begin{align}
	V_A^\text{iso}(\mathbf{r}) = 
	- \frac{Z}{r} + \int d^3 r' \; \frac{ \rho(\mathbf{r}') }{ | \mathbf{r} - \mathbf{r}' | } + V_\text{x}\left[ \rho(\mathbf{r}) \right],
	\label{V_iso}
\end{align}
where $Z$ is the nuclear charge, $V_\text{x}\left[ \rho(\mathbf{r}) \right] = -(3/2) \left[ ( 3 / \pi ) \rho(\mathbf{r}) \right]^{1/3}$ is the Slater exchange potential~\cite{Slater1951}, and $\rho(\mathbf{r}) = \sum_p^{N_{\text{elec},A}} \psi^\dagger_p(\mathbf{r}) \psi_p(\mathbf{r})$ is the electron density. 
Here, $N_{\text{elec},A}$ is the number of electrons assigned to atom $A$ and $p$ indicates the one-particle state index, i.e., $p=(n, l, m_l, m_s)$, where $n$, $l$, $m_l$, and $m_s$ are the principal quantum number, the orbital angular momentum quantum number, the associated projection quantum number, and the spin quantum number, respectively. 
In addition, the Latter tail correction~\cite{Latter1955} is employed to obtain the proper long-range potential. 
Assuming that the electron density is spherically symmetric, $V_A^\text{iso}(\mathbf{r})$ is also spherically symmetric and Eq.~(\ref{HFS}) can be reduced to a radial Schr\"odinger equation.
This equation is solved using a numerical grid technique with a sufficiently large radius~\cite{xatom2014AA}.

An atom embedded in a plasma environment experiences an additional potential $V_A^\text{env}(\mathbf{r})$ from the environment. 
The electrons assigned to each atom are treated quantum mechanically via direct Coulomb interaction and exchange interaction with electron density, whereas the electrons and atomic ions in the environment are considered as classical particles. 
For each atom, we employ a microcanonical ensemble at every realization at every time step, implying a fixed electron configuration.
Then, the potential for an atomic electron in a plasma calculation, $V_A^\text{pla}$, is given by
\begin{align}
V_A^\text{pla}(\mathbf{r}) = 
\begin{cases}
{\displaystyle V_A^\text{atom}(\mathbf{r}) + V_A^\text{env}(\mathbf{r})} &  \text{for }r < r_c, 
\\
{\displaystyle V_0} &  \text{for }r \geq r_c.
\end{cases}
\end{align}
Here we introduce a flat potential tail $V_0$ for $r \geq r_c$, using the muffin-tin approximation~\cite{Slater1937}. 
The energy levels below $V_0$ are considered as bound states, whereas above $V_0$ are continuum states.
Then, the excitation of an electron from a bound state to the continuum threshold located at $V_0$ defines the {\it inner} ionization~\cite{xatom2014AA}, and we use the associated excitation energy for defining the ionization potential.
The determination of $V_0$ and $r_c$ will be explained later. 
$V_A^\text{atom}(\mathbf{r})$ has the same form as shown in Eq.~(\ref{V_iso}), except $\rho(\mathbf{r})$ is determined self-consistently in the presence of the additional environmental potential. 
Note that no Fermi-Dirac occupation factor is involved in the density calculation, because we use a fixed electron configuration and the concept of temperature is not defined in the NLTE condition.
The Latter correction is not applied because of the muffin-tin approximation. 

In a given MC-MD simulation step, the environmental potential for atom $A$ is simply given by the sum of the static electric potentials of all charged particles in a supercell, excluding atom $A$ itself. 
In the MD simulations, the electric potential is approximated with a soft-core potential~\cite{Zoltan2016} to avoid the Coulombic singularity. 
Thus, $V_A^\text{env}(\mathbf{r})$ is evaluated as
\begin{equation}
\label{eqn:soft-pot}
V_A^\text{env}(\mathbf{r}) = - \sum_{i \neq A} \frac{q_i}{ \sqrt{\left| \mathbf{r} - \mathbf{r}_i \right|^2 + a^2} },
\end{equation}
where $q_i$ is the current charge state of the particle: $q_i = -1$ for electrons and $q_i = Z_i - N_{\text{elec},i}$ for atomic ions. 
Here, $a$ is the soft-core potential radius, which is chosen according to the numerical criteria suggested in Ref.~\cite{Zoltan2016}. 
We examined convergence in XMDYN varying $a$ as well as the time step $\Delta t$ and the cycle-count threshold for recombination $n_\text{rec}$, and found that $a$~=~0.6~a.u., $\Delta t$~=~0.5~attoseconds, and $n_\text{rec}$~=~3 were satisfactory for dense Al plasma simulations. 
Hence, in this study, we set $a = 0.6$~a.u.\ in Eq.~(\ref{eqn:soft-pot}). 
Note that in order to construct the environmental potential more efficiently, we sample the MC-MD simulation results using a time step of 0.25~fs, which is larger than the $\Delta t$ used for propagation in XMDYN, because the environmental potential, especially after applying the averaging schemes to be introduced below, varies smoothly on the femtosecond time scale.
We use the same time step for free electrons and ions for simplicity.

In order to perform an atomic calculation, we need to assume a spherically-symmetric potential. 
We impose it for the environmental potential by performing spherical averaging of $V_A^\text{env}(\mathbf{r})$, 
\begin{align}
V_A^\text{env}(r) &= \frac{ \int d\Omega_A \; V_A^\text{env}(\mathbf{r}) }{ \int d\Omega_A } 
= - \frac{1}{4 \pi} \sum_{i \neq A} \int d\Omega_A \; \frac{q_i}{ \sqrt{\left| \mathbf{r} - \mathbf{r}_i \right|^2 + a^2}}\notag 
\\
&=-\sum\limits_{i \neq A}q_{i} \frac{\sqrt{(r+r_{i})^2+a^2}-\sqrt{(r-r_{i})^2+a^2} }{2rr_{i}},
\label{eqn:avrg-pot}
\end{align}
where $\Omega_A$ is the solid angle around atom $A$. 
Note that the minimal image convention~\cite{Abdullah2017} is applied to evaluate the potential for crystalline structures. 
The minimal image convention basically chooses a translational equivalent image of the original supercell, so that the atom of interest sits exactly at the center.   

Next, we introduce another averaging scheme. 
The evaluation of Eq.~(\ref{eqn:avrg-pot}) is relatively simple, but $V_A^\text{env}(r)$ has to be calculated and stored for every single $A$ at every MC-MD step.  
In our calculation, one supercell typically contains a few hundred atoms. 
Each atomic case represents one realization of a stochastic process, and we can make an ensemble average with individual atomic realizations. 
One extreme is to average all individual atomic potentials in a supercell (global averaging scheme), 
\begin{equation}
\label{eqn:global-pot}
V^\text{env}_\text{global}(r) = \frac{1}{N_A} \sum^{\text{all}}_{A}V_A^\text{env}(r).
\end{equation}
Alternatively, one can group the atomic potentials according to individual charge states and average them, assuming that the short-range shape of the environmental potential is dominated by the ionic charge state (charge-selective averaging scheme),
\begin{equation}
\label{eqn:charge-pot}
V_{q}^\text{env}(r) = \frac{1}{N_q}\sum_{A}^{q_A = q}V_A^\text{env}(r), 
\end{equation}
where $q_A = Z_A - N_{\text{elec},A}$ and $N_q$ is the number of atoms corresponding to $q_A = q$ in a given time step.
This charge-selective averaging scheme preserves locality of the environmental potential. 
A comparison among the different averaging schemes with respect to calculated ionization potentials will be made at the end of Sec.~\ref{subsec:ipd}.

The global potential experienced by a quantum electron is obtained by connecting all atomic potentials $V^{\mathrm{pla}}_A$ such that the resulting potential is continuous in the interstitial regions via $V_0$. 
Moreover, we assume that the atomic potentials are spherically symmetric.
Thus, the connecting potential $V_0$ must be the same for all atoms. 
This approximation has been widely used in solid-state calculations and its validity has been tested by precise band structure studies~\cite{Korringa1947,Kohn1954,Anderson1989,Skriver1984}. 
Here, we use a similar procedure for determining a muffin-tin potential as described in Ref.~\cite{Bekx2020}. 
Strictly speaking, determination of $V_0$ requires information on all atomic potentials within a supercell. 
The total atomic potential for each atomic site, however, is to be determined self-consistently including $V_0$, which means that a self-consistent-field (SCF) calculation for the whole supercell is necessary. 
To avoid such a complication, exclusively for determining $V_0$ we approximate the total atomic potential as $V_A^\text{approx}(r) = -(q_A+1)/r + V_A^\text{env}(r)$, where $(q_A+1)$ accounts for the Latter correction~\cite{Latter1955}. 
We match this potential with the $B$th neighboring atom at a distance $r_{AB}$. 
The touching potential of $A$ and $B$ is given by $V_{AB} = V_A^\text{approx}(r_T) = V_B^\text{approx}(r_{AB} - r_T)$, where $r_T$ is the touching sphere radius with respect to atom $A$. 
The lowest value of this touching potential $V_{AB}$ for all atom combinations is chosen as the global muffin-tin potential $V_0$ (~=~$\min \lbrace V_{AB} \rbrace$). 

\begin{figure}[b]
	\includegraphics[width=\figurewidth]{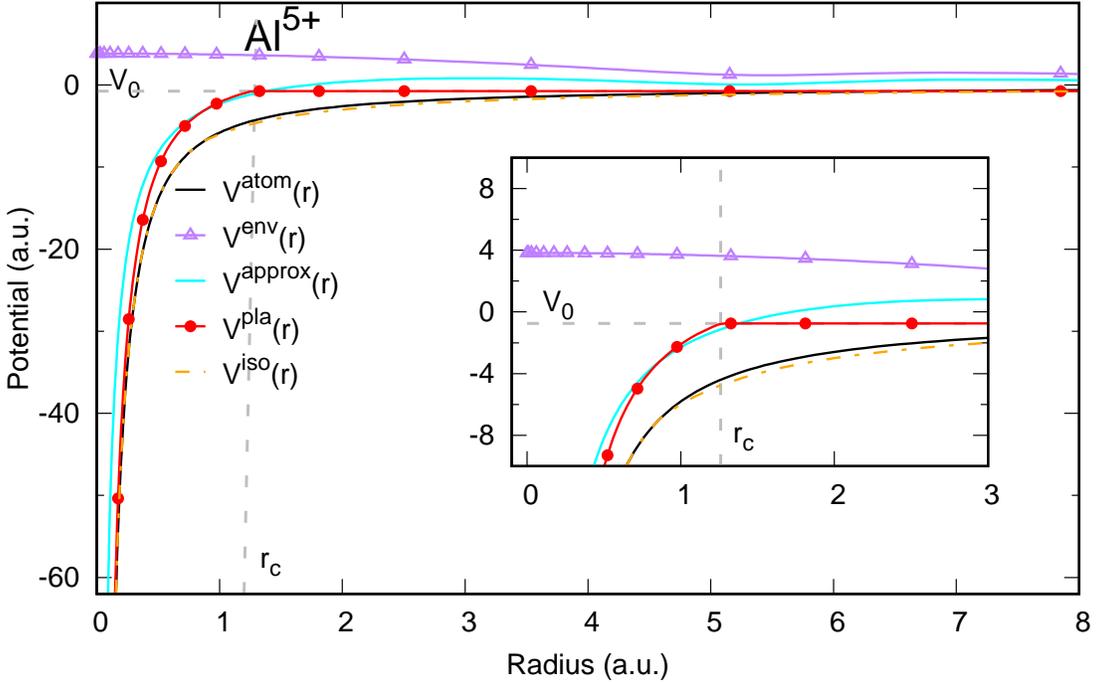}
    \caption{\label{fig:potential}Potentials for Al$^{5+}$ in the charge-selective averaging scheme. $V^\text{pla}$ is the total potential used for the electronic-structure calculation on an atom in a plasma environment. It is the sum of the atomic potential $V^{\mathrm{atom}}$ and the environmental potential $V^{\mathrm{env}}$. The approximation potential $V^\text{approx}$ is very close to $V^\text{pla}$ in the vicinity of $r=r_c$. The muffin-tin energy tail $V_0$ is fixed during the SCF iterations.}
\end{figure} 

For a given MC-MD step, $V_A^\text{env}(r)$ is constructed for all individual atomic sites and $V_0$ is determined with the above procedure, before entering individual atomic SCF calculations. 
The final expression for the plasma potential (with the charge-selective averaging scheme) is
\begin{align}
	V_A^\text{pla}(r) = 
	\begin{cases}
		{\displaystyle V_A^\text{atom}(r) + V_{q=q_A}^\text{env}(r)} & \text{for }r < r_c, 
		\\
		{\displaystyle V_0} & \text{for }r \geq r_c.
	\end{cases}
    \label{eqn:pla-pot}
\end{align}
Here $r_c$ is determined by matching the potential $V^{\mathrm{atom}}_A(r)+V^{\mathrm{env}}_{q=q_A}(r)$ with the fixed muffin-tin energy tail $V_0$ in each SCF iteration. 
Figure~\ref{fig:potential} shows all potentials involved in the plasma calculation of Al$^{5+}$ (electron configuration: $1s^2 2s^2 2p^4$): $V_A^\text{pla}(r)$, $V_A^\text{atom}(r)$, $V_{q=q_A}^\text{env}(r)$, and $V_A^\text{approx}(r)$. The approximation potential $V_A^\text{approx}$ approaches $V_A^\text{atom}(r) + V_{q=q_A}^\text{env}(r)$ as $r$ approaches $r_c$, indicating the approximation is valid.
For comparison, $V_A^\text{iso}(r)$ in the isolated-atom calculation is also shown with the dashed line. The subscripts are dropped in Fig.~\ref{fig:potential} as no confusion will occur for a single atomic species.
As shown in Fig.~\ref{fig:potential}, the curve of $V^\text{pla}(r)$ shows a kink at $r_c$ because of the way the plasma potential is constructed.
In the numerical grid employed here, the maximum radius is much larger than $r_c$ (typically $r_\text{max}$=50~a.u.) and smoothing around this kink position provides almost no change in calculated orbital energies and orbitals.
 
Note that $V_A^\text{pla}(r)$ is only used for the bound electrons in atoms and atomic ions to account for the plasma environment effect.
The classical particles (free electrons and atomic ions) move on the potential surface given by the sum of Coulomb interactions with all other classical particles in the system.

\subsection{\label{subsec:ipd}Ionization potential depression}
The ionization potential depression (IPD) for an atom is defined by the difference between the ionization potential of the atom in isolation and the ionization potential of the atom in a plasma environment. 
For the $j$th orbital in the electronic configuration $I$, the IPD is calculated as
\begin{equation}
\Delta E_{I,j} = \text{IP}^\text{iso}_{I,j} - \text{IP}^\text{pla}_{I,j},
\label{eqn:ipd}
\end{equation}
where the ionization potentials (IPs) for an isolated atom and for an atom in a plasma are defined by
\begin{align}
\text{IP}_{I,j}^\text{iso} &= -\varepsilon_{I,j}^\text{iso},
\\
\text{IP}_{I,j}^\text{pla} &= V_0 - \varepsilon_{I,j}^\text{pla},
\end{align}
which are obtained from the isolated-atom calculation using Eq.~(\ref{V_iso}) and from the plasma calculation using Eq.~(\ref{eqn:pla-pot}), respectively.
For the plasma case, the continuum states start at the muffin-tin flat potential $V_0$, so the ionization potential is defined by the difference between $V_0$ and the orbital energy calculated with the plasma environment.

In the following, we apply this procedure to XFEL-irradiated Al solid~\cite{Vinko2012, Ciricosta2012, Cho2012} and investigate IP and IPD values for given Al atomic charges. 
It is known that, for Al solid at room temperature, $3s$ and $3p$ electrons are not bound to the atom, but are delocalized and form conduction bands. 
In our previous average-atom-based study~\cite{xatom2014AA}, states were considered bound if their energy was below the muffin-tin flat potential. 
For example, for Al solid at $T=0$~eV, the $M$-shell states ($3s$ and $3p$ orbitals) are not bound to the atom, whereas at $T=80$~eV they become bound in the average-atom calculation. 
However, the recent all-electron quantum-mechanical crystalline calculation of Al plasmas~\cite{Bekx2020} indicates that the states above $2p$ are not fully localized even at high temperature ($T \leq 100$~eV), leading to finite valence energy bands. 
Following Ref.~\cite{Bekx2020}, for the assignment of the atomic charge we count only electrons in $1s$, $2s$, and $2p$ orbitals. 
Thus, for the electronic configuration $I_A = (n_{1s}^A, n_{2s}^A, n_{2p}^A)$, regardless of $M$-shell occupation number, the atomic charge is given by
\begin{equation}\label{eqn:Q_A}
Q_A = Z_A - \sum_j^{1s,2s,2p} n_j^A,
\end{equation}
where $n_j^A$ is the occupation number in the $j$th orbital of atom $A$. 
This assignment is consistent with the spectroscopic notation used in Refs.~\cite{Vinko2012, Cho2012}, where the occupation numbers of only the $K$- and $L$-shells are counted and the energy resolution in the experiment was insufficient to resolve the $M$-shell occupation number.

Note that $Q_A$ in Eq.~(\ref{eqn:Q_A}) used to assign an atomic charge for the IP and IPD calculation is different from $q_A$ in Eq.~(\ref{eqn:charge-pot}) used for charge-selective averaging. 
This originates from a subtlety of how $3s$ and $3p$ electrons are treated. 
In our quantum-mechanical calculations, the Al electronic configuration contains $1s$, $2s$, $2p$, $3s$, and $3p$ for both isolated-atom and plasma cases. 
More precisely, $V_A^\text{atom}$ in Eq.~(\ref{eqn:pla-pot}) contains the electron density including $3s$ and $3p$ electrons. 
In this way, $3s$ and $3p$ states are explicitly treated as if they are bound to the atom. 
However, this is an \emph{ad hoc} treatment because they may not be bound states as discussed above. 
In the current implementation, $3s$ and $3p$ states are included in our atomic quantum-mechanical treatment, but they are excluded when defining atomic charge. 
Thus, quantum effects in free electrons are partially incorporated, but at the same time they could be overestimated because $3s$ and $3p$ electrons cannot be delocalized in the current approach, until they are ionized and become classical particles.
This limitation could be overcome by introducing a quantum charge transfer treatment among individual atoms~\cite{Kumagai2018}. 
During the interaction with an intense x-ray pulse, many electrons are ionized, turning $3s$ and $3p$ electrons into classical particles within XMDYN, so that the difference between $Q_A$ and $q_A$ vanishes when no electron remains in the $M$ shells. 
Alternatively, electrons in $3s$ and $3p$ states could be treated as classical particles from the beginning and only the $K$ and $L$ shells could be treated quantum mechanically.
In this extreme case, quantum effects in free electrons would be completely neglected.
To overcome this drawback, one could apply the electron force field~\cite{Su:2007aa,Kim:2011aa} to free electrons.

\begin{figure}[b]
	\includegraphics[width=\figurewidth]{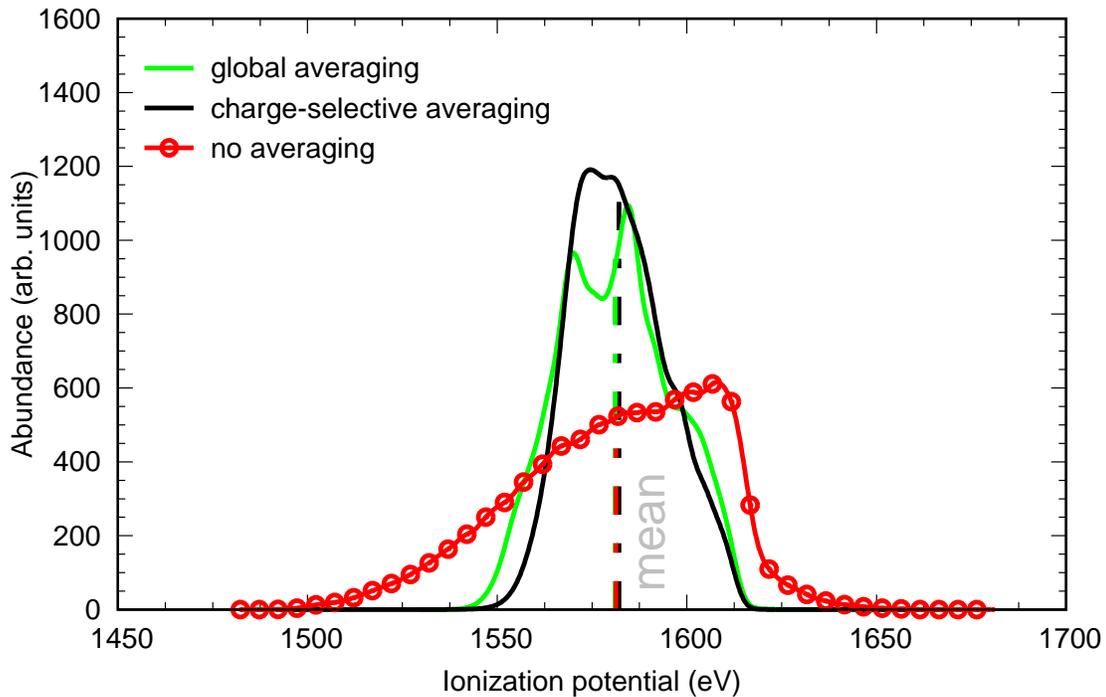}
	\caption{\label{fig:avrg-scheme}Time-integrated ionization potential distributions for Al$^{5+}$ in solid-density Al plasma using different averaging strategies.}
\end{figure} 

We perform a calculation of IP and IPD at every single MC-MD step, which provides the time evolution of the IP and IPD values.
We will examine not only time-resolved IP but also time-integrated IP values to be compared with time-integrated measurements.
In Fig.~\ref{fig:avrg-scheme}, time-integrated IP values for Al$^{5+}$ based on the averaging schemes of Eqs.~(\ref{eqn:global-pot}) and (\ref{eqn:charge-pot}) are compared with that obtained with no averaging.
Computational details regarding the time-integrated IP and its distribution will be discussed later in Sec.~\ref{sec:results}. 
As shown in Fig.~\ref{fig:avrg-scheme} the time-integrated IP distributions (manually convolved with a Gaussian of 4.7~eV full width at half maximum, FWHM) for Al$^{5+}$ ions in the global and charge-selective averaging schemes look similar. 
The no-averaging scheme leads to a wider and blue-shifted distribution, indicating that environmental fluctuations matter. 
However, the mean values of the IP distributions calculated with the three different schemes are very similar to each other within 1~eV.
The no-averaging scheme is computationally less attractive than the other two schemes, because snapshots of classically treated electron density without averaging often cause difficulties in SCF convergence and every atomic ion in a supercell has to be treated separately.
On the other hand, the global averaging scheme completely ignores different charge environments within the supercell.
Given these reasons, we decide to use the charge-selective averaging scheme for further calculations.

\section{\label{sec:results}Results and discussion}
In order to test our NLTE IP and IPD calculation procedure, we apply it to the simulation of solid density aluminum plasma and compare with a recent experiment~\cite{Vinko2012,Ciricosta2012}. 
In this experiment, a solid aluminum target was irradiated with intense XFEL pulses with a pulse duration of 80~fs (FWHM) and a peak intensity of \intensity{1.1}{17}, corresponding to a peak fluence of \fluence{1.0}{11}, providing time-integrated K-shell IP values for different charge states.

\begin{figure}
	\includegraphics[width=\figurewidth]{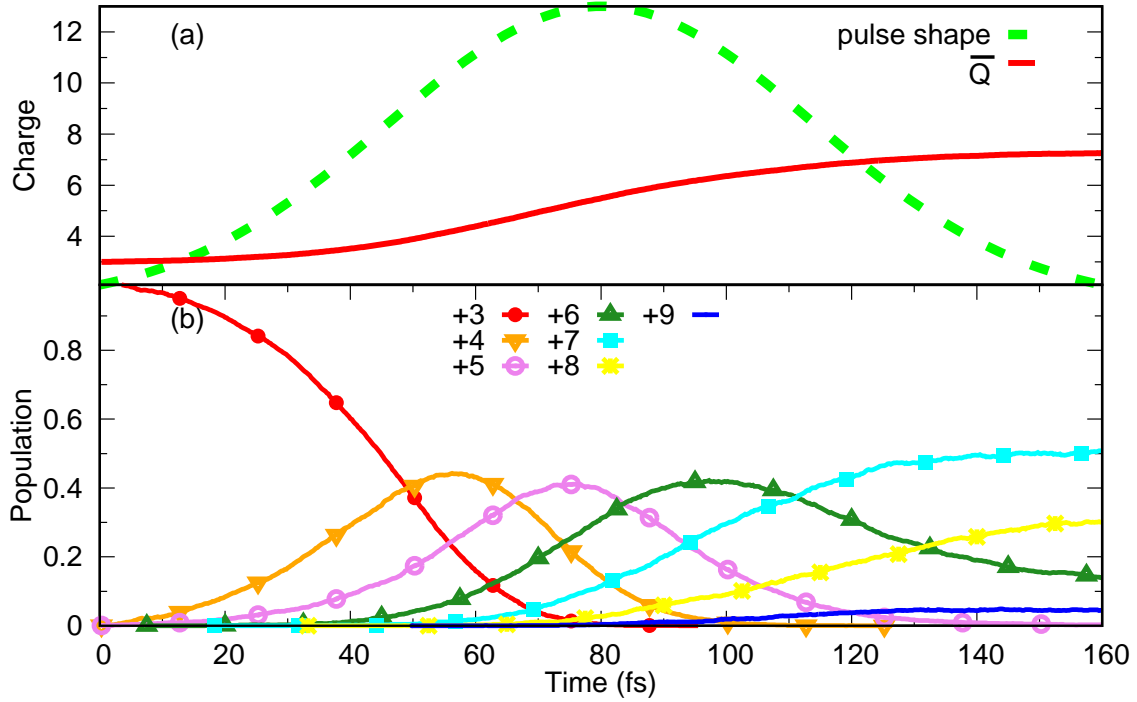}
	\caption{\label{fig:MD-CSD 1850}XMDYN simulation for a solid aluminum target irradiated by a short laser pulse with a photon energy of 1850~eV and a fluence of \fluence{1.0}{11}. (a) Average charge state ($\bar{Q}$) and Gaussian pulse profile. (b) Time evolution of charge populations.}
\end{figure}

\begin{figure}
	\includegraphics[width=\figurewidth]{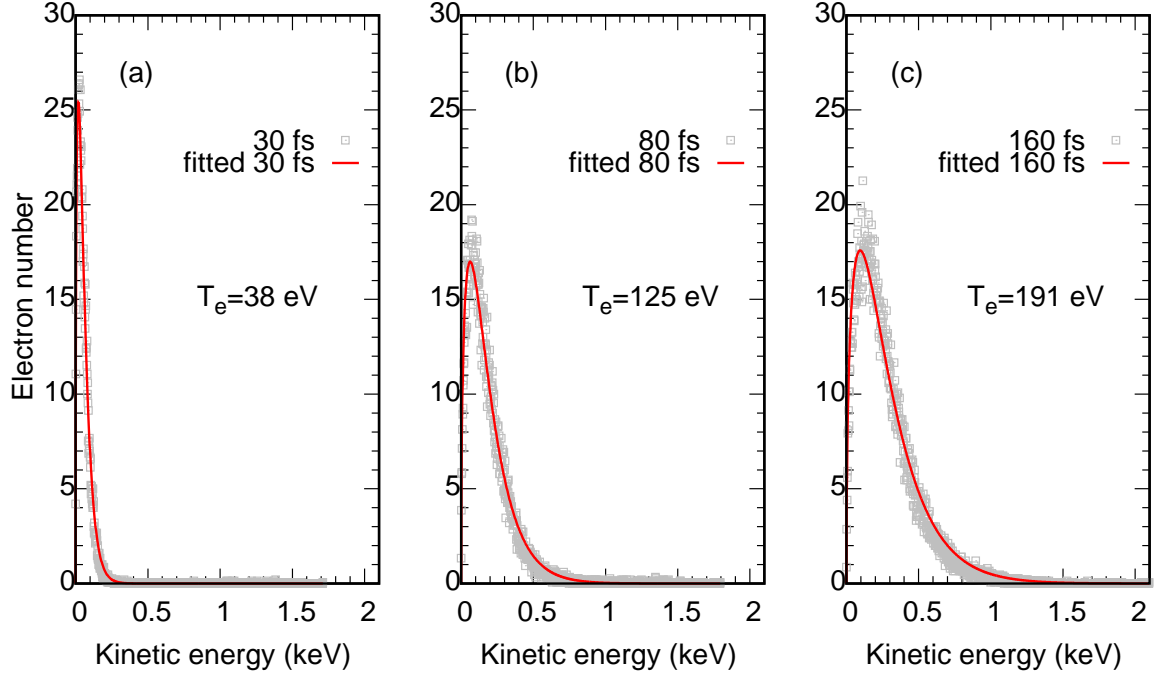}
	\caption{\label{fig:check-LTE} Maxwell--Boltzmann distribution fitting of electron kinetic energies for (a) $t = 30$~fs, (b) $t = 80$~fs, and (c) $t = 160$~fs.}
\end{figure}

In our simulation of the solid density aluminum target ($n_i$=2.7~g/cm$^3$=0.06026~\AA$^{-3}$), we employ XMDYN with the supercell approach~\cite{Abdullah2016,Abdullah2017,Abdullah2018}.
The number of atoms in the supercell should be sufficient to guarantee that stochastic x-ray interactions are properly described. 
We choose 500 atoms in the supercell with a lattice constant of 20.23~\AA, containing 5$\times$5$\times$5 fcc unit cells. 
We take 15 MC-MD realizations, starting from the same crystalline geometry where the atoms are all initially at rest.
The photon energy is fixed at 1850~eV.
The fluence is fixed at \fluence{1.0}{11} and we assume that atoms experience the same fluence throughout the supercell.
When an XFEL pulse is focused onto a target, the x-ray fluence value has a spatially nonuniform distribution in the focal spot, demanding volume integration for calculating physical observables~\cite{Toyota19}.
This spatial fluence distribution is not considered for simplicity.
The XFEL pulse shape is chosen as a Gaussian function with 80~fs FWHM and the peak is centered at 80~fs, which is plotted as the green dashed line in Fig.~\ref{fig:MD-CSD 1850}(a). 
The simulation is performed up to 160~fs. 
The time evolution of the average charge, $\bar{Q}(t)=\sum_{Q} Q p_Q(t)$, and the individual charge populations, $p_Q(t)$, from XMDYN simulation are shown in Figs.~\ref{fig:MD-CSD 1850}(a) and (b), respectively. 
As we define the atomic charge via $K$- and $L$-shell occupations, the initial charge state is +3. 

In order to check how far the electron plasma is from thermal equilibrium, we apply Maxwell--Boltzmann least-square fitting to the kinetic-energy distribution of the plasma electrons.
Note that if the mean interparticle distance for electrons ($d = 1/\sqrt[3]{n_e}$, where $n_e = \bar{Q} n_i$) is much larger than the thermal de Broglie wavelength~\cite{Pathria} ($\lambda_e = h / \sqrt{ 2 \pi m_e k T_e }$, where $m_e$ is the electron mass and $k$ is the Boltzmann constant), the electron plasma can be treated as a classical plasma and the kinetic-energy distribution follows the Maxwell--Boltzmann distribution upon thermalization.
If $d$ is smaller than $\lambda_e$, quantum degeneracy effects become important.
In our plasma condition at the end of the pulse ($\bar{Q} \sim +7$ and expected $T_e \sim 200$~eV), we have $d = 1.33$~\AA\ and $\lambda_e = 0.49$~\AA, i.e., $d > \lambda_e$ but they are comparable to each other.
Figure~\ref{fig:check-LTE} shows kinetic-energy distributions of plasma electrons 
(a) at the early stage [$t=30$~fs], 
(b) at the peak of the pulse [$t=80$~fs], and
(c) at the end of the pulse [$t=160$~fs]. 
One can see that they follow the Maxwell--Boltzmann distribution, indicating that the plasma electrons may be thermalized.
This could be explained by the fact that a relatively long pulse (80~fs) is applied, which provides enough time for thermalization at least for free electrons, during the pulse.
The discrepancies between the simulated electron spectrum and the Maxwell--Boltzmann distribution could be attributed to strong correlation between ions and electrons expected from the dense plasma.
At the same time, quantum electrons bound to individual atoms are far from equilibrium during the pulse, as clearly shown in the charge-state population dynamics in Fig.~\ref{fig:MD-CSD 1850}.
Therefore, it may not be straightforward to justify thermalization of classical and quantum electrons and to define a unified electron temperature $T_e$.
Only effective temperatures could be defined in the NLTE regime~\cite{Bauche2015}.

\begin{figure}[b]
	\includegraphics[width=\figurewidth]{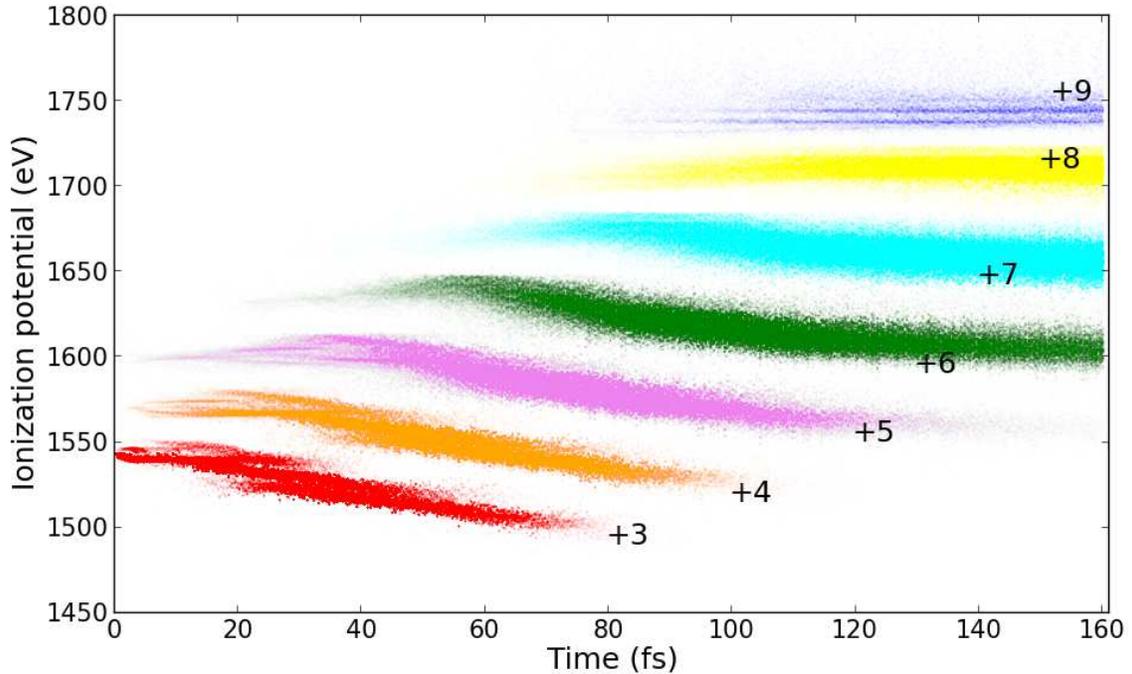}
	\caption{\label{fig:IP_evol_1850}The time dependent $\text{IP}^\text{pla}$ values from XMDYN simulation trajectories for a solid aluminum target irradiated by an intense x-ray pulse with a photon energy of 1850~eV and a fluence of \fluence{1.0}{11}. The values are grouped into individual charge states with different colors.}
\end{figure}

The IPs in the NLTE plasma environment, namely $\text{IP}^\text{pla}$, can be readily obtained by applying our approach to the real-time simulation results. 
The $\text{IP}^\text{pla}$ values for a specific atomic ion species form a distribution due to different electronic configurations and environmental fluctuations in different MC-MD simulation realizations. 
In Fig.~\ref{fig:IP_evol_1850}, we plot all IP data points as a function of time.
We group the $\text{IP}^\text{pla}$ distribution into individual charge states with different colors. 
Note that only the atomic ions with closed $K$ shells are considered, and IP values for single- or double-$K$-hole states are 100--200~eV higher than the closed $K$ shells.
As can be seen in Fig.~\ref{fig:IP_evol_1850}, initially the IP$^{\mathrm{pla}}$ values tend to increase with time. 
This is a consequence of the fact that the $M$-shell electrons are ionized rapidly in the early stage, such that their contribution to screening is reduced as they are being ionized. 
(The magnitude of this effect may be overestimated because of the way the $3s$ and $3p$ electrons are included in our atomic quantum-mechanical treatment.) 
This process is in competition with increasing screening by plasma electrons, which eventually leads to a decreasing IP until the number of plasma electrons is equilibrated.

\begin{figure}[b]
	\includegraphics[width=\figurewidth]{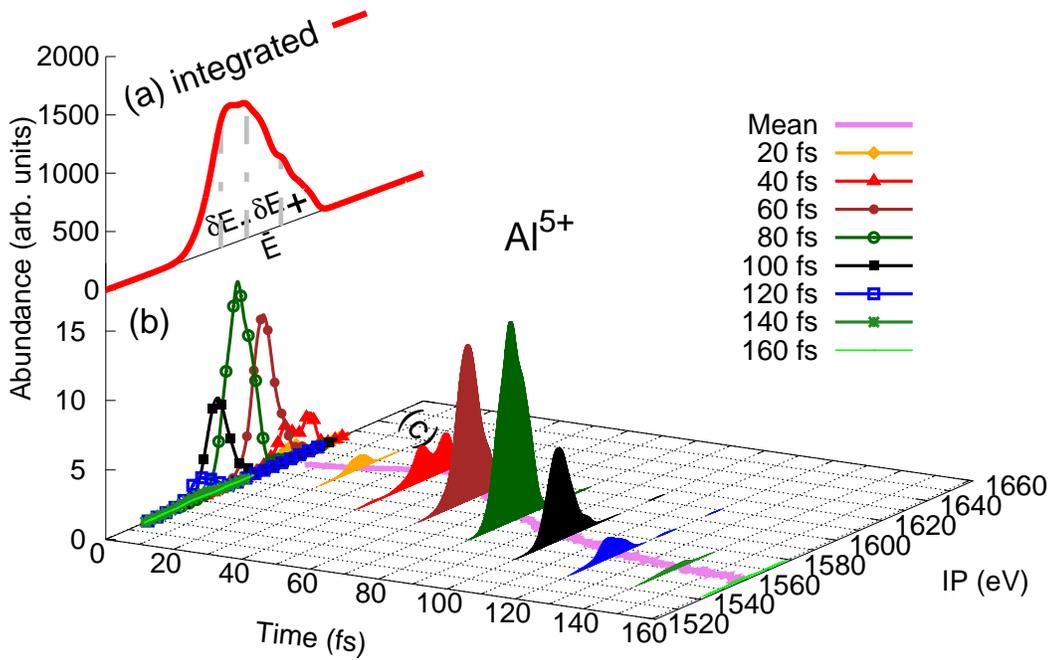}
	\caption{\label{fig:IP-Q5}The distribution of the Al$^{5+}$ IP in an x-ray-driven solid-density Al plasma. The distribution functions for eight times are shown as an example, the time-integrated distribution is shown as the thick red curve on the top left, with the corresponding mean value ($\bar{E}$) and asymmetric deviation ($\delta E_-$ and $\delta E_+$).}  
\end{figure} 

In our approach, the time evolution of the IP distributions can be obtained, signaling the evolution of environmental effects and the energy structures in an NLTE system. 
As an example the IP distributions for Al$^{5+}$ at selected times are shown as vertical colored shades in Fig.~\ref{fig:IP-Q5}. 
The energy distribution is obtained by manually broadening the discrete lines with a Gaussian of 4.7~eV FWHM. 
The mean IP values as a function of time are plotted on the bottom plane, when the charge population of Al$^{5+}$ is higher than 0.5\%.
The distribution corresponds mainly to the $1s^2 2s^2 2p^4$ electronic configuration. 
Again, the widely scattered IP values for the same electronic configuration originate from two factors. 
One is environmental fluctuations, and the other is the presence of $M$-shell electrons treated as quantum electrons. 
For instance, the configuration group represented by $1s^2 2s^2 2p^4 M^k\ (k \in [0,8])$ contributes to the same $K$- and $L$-shell configuration $1s^2 2s^2 2p^4$ under different environments.

Figure~\ref{fig:IP-Q5} also contains the time-integrated IP distribution (thick red curve on the top left). 
We characterize this asymmetric distribution though its mean value $\bar{E}$ and the asymmetric deviation $\delta E_{\pm}$ by identifying an energy interval with a probability equal to $68\%$, i.e., $P(\bar{E}<E\leq \bar{E}+\delta E_{+})=34\%$ and $P(\bar{E}-\delta E_{-}<E\leq \bar{E})=34\%$. 
Such an IP distribution has been reported in previous studies with classical MD~\cite{Calisti15a} and detailed configuration accounting~\cite{Iglesias14}. 
In both of them, LTE was assumed. The IP distribution reflects either sampling of the charged particles in the plasma environment~\cite{Calisti15a} or different electronic configurations taking into account $M$-shell electrons~\cite{Iglesias14}.
In contrast, our approach achieves both aspects: the ensemble of plasma environments via MC-MD simulations and detailed electronic configurations via atomic electronic-structure calculations.

\begin{figure}[b]
	\includegraphics[width=\figurewidth]{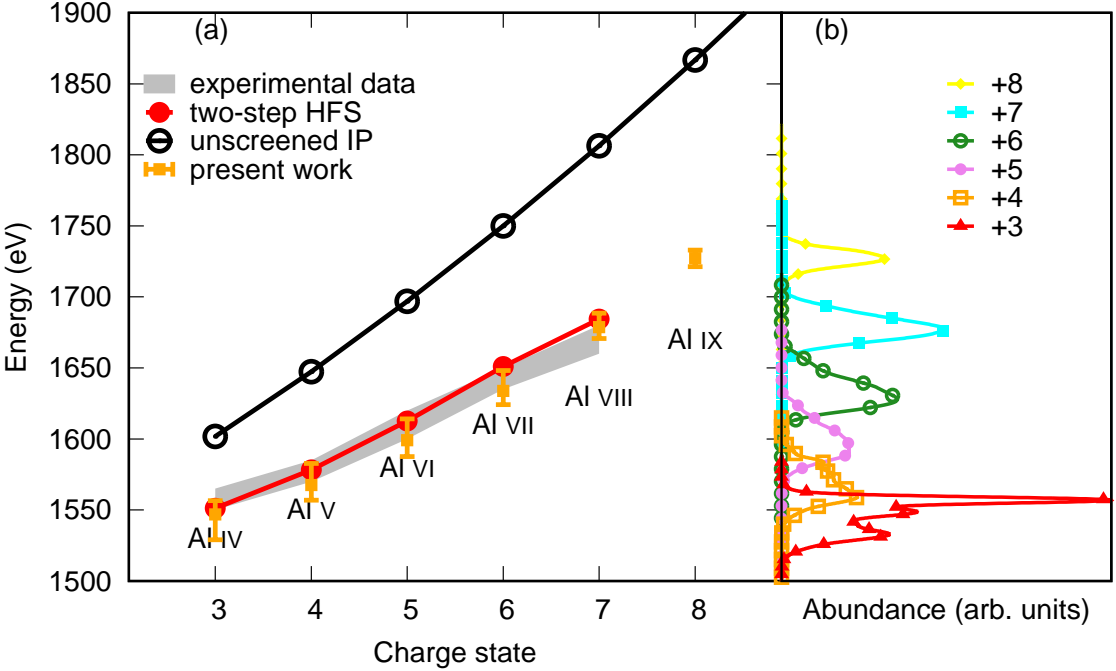}
	\caption{\label{fig:IP compare}(a) Time-integrated $K$-shell IP values (mean and asymmetric deviation) for different charge states in a solid-density Al plasma generated by an x-ray pulse with a photon energy of 1850~eV, a fluence of \fluence{1}{11}, and a pulse duration of 80~fs FWHM.
	The experimentally observed IP~\cite{Vinko2012,Ciricosta2012}, the theoretical calculation using the XATOM two-step model~\cite{xatom2014AA}, corresponding IP values for unscreened atoms are plotted for comparison.
	(b) Time-integrated $K$-shell IP distribution.}
\end{figure}

In Fig.~\ref{fig:IP compare}(a), the mean value $\bar{E}$ and the asymmetric deviation $\delta E_{\pm}$ of the time-integrated IP distributions for different charge states are compared with experimental data~\cite{Vinko2012, Ciricosta2012} and theoretical calculation with the XATOM two-step model~\cite{xatom2014AA}, which is based on the LTE condition. 
The isolated-atom IPs for the corresponding electronic configurations are also plotted for comparison. 
Note that the accuracy of the model is limited by the HFS description of the binding energy, with a typical relative uncertainty of $1\%$, so we apply a constant energy shift of $+16.5$~eV to all calculated IP values from now on, according to the difference between the mean $K$-shell IP value (1543.1~eV) of the Al$^{3+}$ ion in the cold plasma environment (at the beginning of the simulation) and the experimental $K$-shell IP value (1559.6 eV)~\cite{XrayBooklet}. 
The asymmetric deviation is obtained from the time-integrated IP distribution shown in Fig.~\ref{fig:IP compare}(b).
The time-integrated IP values calculated within the NLTE framework we have developed match the experimental data.
They are also in good agreement with theoretical LTE-based results.
It is currently unknown whether both LTE and NLTE calculations would be equally accurate in reproducing energetically more highly resolved XFEL data that reveal the detailed shapes of IP distributions in solid-density plasmas.   

\begin{figure}[b]
	\includegraphics[width=\figurewidth]{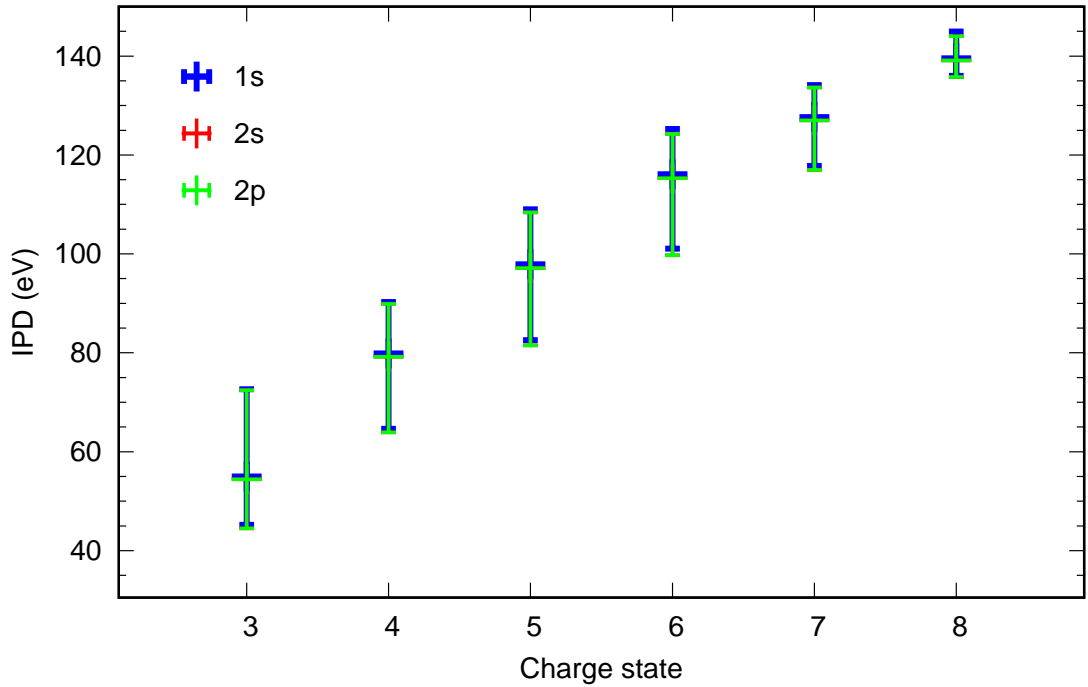}
	\caption{\label{fig:IPD compare}Mean and asymmetric deviation of time-integrated IPD values of $1s$, $2s$ and $2p$ orbitals for different charge states in a solid-density Al plasma generated by an x-ray pulse of 1850~eV, \fluence{1}{11}, and 80~fs FWHM.}
\end{figure}

We use Eq.~(\ref{eqn:ipd}) to calculate transient IPD values from the difference between $\text{IP}^\text{pla}$ and $\text{IP}^\text{iso}$ for the same bound-electronic configuration. 
The time-integrated IPD values from the time-resolved ones are calculated in the same procedure as used for the IP values. 
Figure~\ref{fig:IPD compare} shows the time-integrated IPD values for $1s$, $2s$, and $2p$ orbitals, calculated from the IPD distribution for each case.
The difference in IPD between $2s$ and $2p$ orbitals is at most 0.2~eV, the IPD values for $1s$ are at most 0.9~eV higher than those for $2s$.
This behavior is similar to the XATOM two-step model~\cite{xatom2014AA} and Debye-screened HFS model~\cite{Thiele2012}. Note that, strictly speaking, the IPD values calculated here are not universal properties. They apply to a specific case: a plasma generated by an x-ray pulse with 1850~eV, \fluence{1.0}{11}, and 80~fs FWHM.
However, we expect that the time-integrated IPD values are similar for other x-ray parameter sets, because the IPD for each charge state mainly depends on the plasma environment and time-integration could wash out some of the consequences of fluctuations.

\begin{figure}[b]
	\includegraphics[width=\figurewidth]{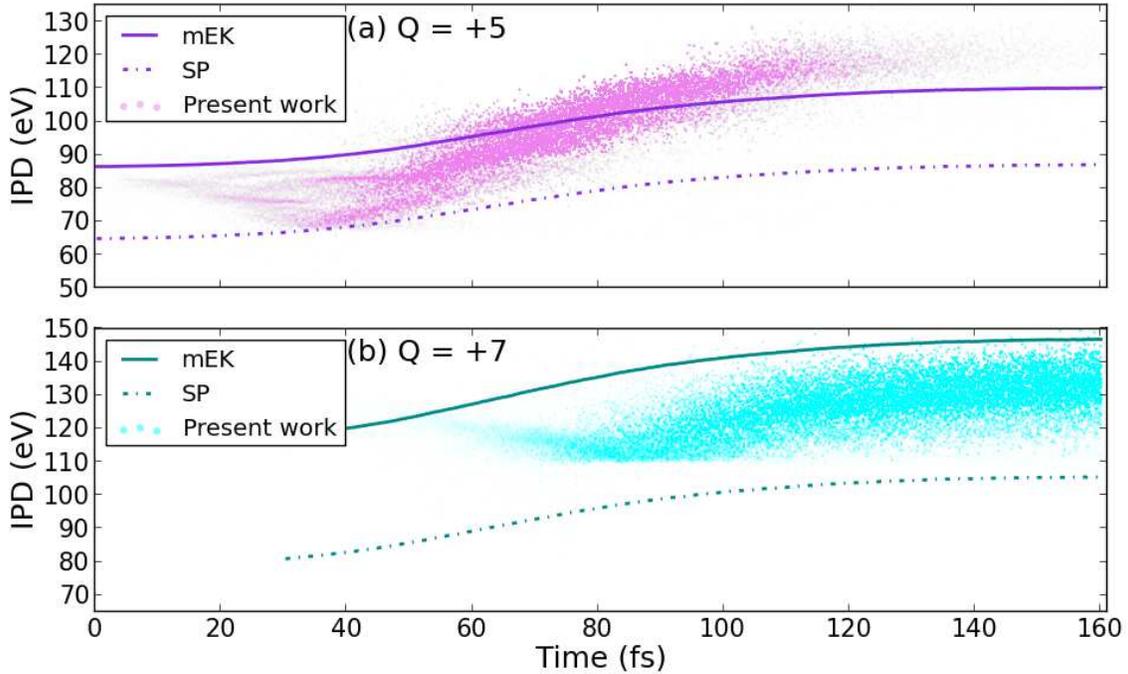} 
	\caption{\label{fig:ipd-xpot-EK}Transient IPD values calculated with our NLTE approach (scattered data points), in comparison with mEK (solid line) and SP (dashed line) for (a) Al$^{5+}$ and (b) Al$^{7+}$.}
\end{figure}

Using our NLTE calculations, we now investigate, in an internally consistent manner, to what degree standard LTE-based models can describe time-resolved IPD values. 
To this end, the average charge state $\bar{Q}$ and the plasma electron density $n_e$ are obtained as a function of time from our XMDYN simulations.
Using those inputs, we compute transient IPDs using the LTE-based modified Ecker--Kr\"oll (mEK)~\cite{E-K1963,Ciricosta2012} and the Stewart--Pyatt (SP)~\cite{S-P1966,Ciricosta2012} models,
\begin{subequations}
	\begin{align}
	\Delta E_\text{SP}(Q)  &=  \dfrac{3(Q+1)}{2r_\text{SP}} \label{eqn:SP},
	\\
	\Delta E_\text{mEK}(Q) &= \dfrac{C_\text{EK} {(Q+1)}}{r_\text{EK}} \label{eqn:EK},
	\end{align} 
\end{subequations}
where $r_\text{SP} = [3(Q+1)/(4\pi n_e)]^{1/3}$, $r_\text{EK} = [3/\{4\pi (n_e+n_i)\}]^{1/3}$, $n_i$ is the ion density, and the coefficient $C_\text{EK}$ is taken as 1 in the high-density regime. 
(These expressions are taken from Refs.~\cite{Ciricosta2012, Preston2013}.)
Figure~\ref{fig:ipd-xpot-EK} shows transient LTE IPD results obtained from mEK (solid line) and SP (dashed line) models, in comparison with our NLTE IPD values (scattered data points) for (a) $Q=+5$ and (b) $Q=+7$.
As can be seen in Fig.~\ref{fig:ipd-xpot-EK}, not only are the standard LTE-based approaches incapable of capturing the fluctuations giving rise to IPD distributions, the associated IPD values do not coincide with the mean value of the NLTE IPD distribution as a function of time.

\section{\label{sec:con}Conclusion}
In this work, we propose an NLTE approach to calculate IPD in dense plasmas.
It is a hybrid approach based on a quantum-mechanical electronic-structure calculation of atomic ions embedded in a plasma environment treated by combining Monte Carlo and classical molecular dynamics.
To do so, we develop a toolkit, XPOT, as an interface between the MC-MD simulation code XMDYN and the atomic structure code XATOM. 
In the current framework, XPOT takes the plasma environment from XMDYN and gives the calculated micro field to XATOM, in order to calculate atomic parameters affected by IPD. However, the modified binding energy values and atomic data are not yet plugged back into the XMDYN simulation.
An implementation of such IPD feedback into the dynamics simulation is in progress.

We apply this approach to describe IPD in a solid-density Al plasma generated by intense XFEL pulses.
Our NLTE approach allows us to track down the time evolution of transient IP and IPD values for individual atomic ions in a supercell.
In this way, we can examine time-resolved IP and IPD distributions and obtain time-integrated quantities for each charge state.
The mean values of time-integrated IPs are in good agreement with experimental IP data~\cite{Vinko2012,Ciricosta2012} and theoretical calculations based on the LTE assumption~\cite{xatom2014AA,Bekx2020}.
On the other hand, transient IPDs under nonequilibrium conditions show non-monotonic evolutions with time, which are not reproducible by standard LTE-based IPD models.
Our NLTE approach for IP and IPD calculation provides critical insight to understand ultrafast formation dynamics and fluctuation properties of dense plasmas induced by XFEL pulses, particularly for the early time scale ($\sim$100~fs), where electron thermalization is not fully guaranteed.
We expect that our computational approach provides detailed atomic data for NLTE kinetic simulation tools, avoiding usage of standard IPD models based on LTE.
It is worthwhile to note that the transient IP values reported here can be potentially measured by using single-color~\cite{Vinko20} and two-color~\cite{Lu18} pump-probe schemes at XFEL facilities.

\begin{acknowledgments}
R.J.\ would like to thank the Helmholtz-OCPC Postdoc Program for support. 
R.S.\ would like to thank Sherin Santra for her help with proofreading the manuscript.
\end{acknowledgments}

\normalem

\providecommand{\noopsort}[1]{}\providecommand{\singleletter}[1]{#1}%


\begin{thebibliography}{81}%
	\makeatletter
	\providecommand \@ifxundefined [1]{%
		\@ifx{#1\undefined}
	}%
	\providecommand \@ifnum [1]{%
		\ifnum #1\expandafter \@firstoftwo
		\else \expandafter \@secondoftwo
		\fi
	}%
	\providecommand \@ifx [1]{%
		\ifx #1\expandafter \@firstoftwo
		\else \expandafter \@secondoftwo
		\fi
	}%
	\providecommand \natexlab [1]{#1}%
	\providecommand \enquote  [1]{``#1''}%
	\providecommand \bibnamefont  [1]{#1}%
	\providecommand \bibfnamefont [1]{#1}%
	\providecommand \citenamefont [1]{#1}%
	\providecommand \href@noop [0]{\@secondoftwo}%
	\providecommand \href [0]{\begingroup \@sanitize@url \@href}%
	\providecommand \@href[1]{\@@startlink{#1}\@@href}%
	\providecommand \@@href[1]{\endgroup#1\@@endlink}%
	\providecommand \@sanitize@url [0]{\catcode `\\12\catcode `\$12\catcode
		`\&12\catcode `\#12\catcode `\^12\catcode `\_12\catcode `\%12\relax}%
	\providecommand \@@startlink[1]{}%
	\providecommand \@@endlink[0]{}%
	\providecommand \url  [0]{\begingroup\@sanitize@url \@url }%
	\providecommand \@url [1]{\endgroup\@href {#1}{\urlprefix }}%
	\providecommand \urlprefix  [0]{URL }%
	\providecommand \Eprint [0]{\href }%
	\providecommand \doibase [0]{https://doi.org/}%
	\providecommand \selectlanguage [0]{\@gobble}%
	\providecommand \bibinfo  [0]{\@secondoftwo}%
	\providecommand \bibfield  [0]{\@secondoftwo}%
	\providecommand \translation [1]{[#1]}%
	\providecommand \BibitemOpen [0]{}%
	\providecommand \bibitemStop [0]{}%
	\providecommand \bibitemNoStop [0]{.\EOS\space}%
	\providecommand \EOS [0]{\spacefactor3000\relax}%
	\providecommand \BibitemShut  [1]{\csname bibitem#1\endcsname}%
	\let\auto@bib@innerbib\@empty
	\bibitem [{\citenamefont {Taylor}(1994)}]{Taylor1994}%
	\BibitemOpen
	\bibfield  {author} {\bibinfo {author} {\bibfnamefont {R.~J.}\ \bibnamefont
			{Taylor}},\ }\href@noop {} {\emph {\bibinfo {title} {The Stars: Their
				Structure and Evolution}}},\ \bibinfo {edition} {2nd}\ ed.\ (\bibinfo
	{publisher} {Cambridge University Press},\ \bibinfo {address} {Cambridge,
		England},\ \bibinfo {year} {1994})\BibitemShut {NoStop}%
	\bibitem [{\citenamefont {Rogers}\ and\ \citenamefont
		{Iglesias}(1994)}]{Rogers1994}%
	\BibitemOpen
	\bibfield  {author} {\bibinfo {author} {\bibfnamefont {F.~J.}\ \bibnamefont
			{Rogers}}\ and\ \bibinfo {author} {\bibfnamefont {C.~A.}\ \bibnamefont
			{Iglesias}},\ }\bibfield  {title} {\bibinfo {title} {Astrophysical opacity},\
	}\href {https://doi.org/10.1088/1361-6633/aa7cca} {\bibfield  {journal}
		{\bibinfo  {journal} {Science}\ }\textbf {\bibinfo {volume} {263}},\ \bibinfo
		{pages} {50} (\bibinfo {year} {1994})}\BibitemShut {NoStop}%
	\bibitem [{\citenamefont {Chabrier}(2009)}]{Chabrier2009}%
	\BibitemOpen
	\bibfield  {author} {\bibinfo {author} {\bibfnamefont {G.}~\bibnamefont
			{Chabrier}},\ }\bibfield  {title} {\bibinfo {title} {Plasma physics and
			planetary astrophysics},\ }\href
	{https://doi.org/10.1088/0741-3335/51/12/124014} {\bibfield  {journal}
		{\bibinfo  {journal} {Plasma Phys. Control. Fusion.}\ }\textbf {\bibinfo
			{volume} {51}},\ \bibinfo {pages} {124014} (\bibinfo {year}
		{2009})}\BibitemShut {NoStop}%
	\bibitem [{\citenamefont {Helled}\ \emph {et~al.}(2010)\citenamefont {Helled},
		\citenamefont {Anderson}, \citenamefont {Podolak},\ and\ \citenamefont
		{Schubert}}]{Helled2010}%
	\BibitemOpen
	\bibfield  {author} {\bibinfo {author} {\bibfnamefont {R.}~\bibnamefont
			{Helled}}, \bibinfo {author} {\bibfnamefont {J.~D.}\ \bibnamefont
			{Anderson}}, \bibinfo {author} {\bibfnamefont {M.}~\bibnamefont {Podolak}},\
		and\ \bibinfo {author} {\bibfnamefont {G.}~\bibnamefont {Schubert}},\
	}\bibfield  {title} {\bibinfo {title} {Interior models of {U}ranus and
			{N}eptune},\ }\href {https://doi.org/10.1088/0004-637x/726/1/15} {\bibfield
		{journal} {\bibinfo  {journal} {Astrophys. J.}\ }\textbf {\bibinfo {volume}
			{726}},\ \bibinfo {pages} {15} (\bibinfo {year} {2010})}\BibitemShut
	{NoStop}%
	\bibitem [{\citenamefont {Guillot}(1999)}]{Guillot1999}%
	\BibitemOpen
	\bibfield  {author} {\bibinfo {author} {\bibfnamefont {T.}~\bibnamefont
			{Guillot}},\ }\bibfield  {title} {\bibinfo {title} {Interiors of giant
			planets inside and outside the solar system},\ }\href
	{https://doi.org/10.1126/science.286.5437.72} {\bibfield  {journal} {\bibinfo
			{journal} {Science}\ }\textbf {\bibinfo {volume} {286}},\ \bibinfo {pages}
		{72} (\bibinfo {year} {1999})}\BibitemShut {NoStop}%
	\bibitem [{\citenamefont {Militzer}\ \emph {et~al.}(2016)\citenamefont
		{Militzer}, \citenamefont {Soubiran}, \citenamefont {Wahl},\ and\
		\citenamefont {Hubbard}}]{Militzer2016}%
	\BibitemOpen
	\bibfield  {author} {\bibinfo {author} {\bibfnamefont {B.}~\bibnamefont
			{Militzer}}, \bibinfo {author} {\bibfnamefont {F.}~\bibnamefont {Soubiran}},
		\bibinfo {author} {\bibfnamefont {S.~M.}\ \bibnamefont {Wahl}},\ and\
		\bibinfo {author} {\bibfnamefont {W.}~\bibnamefont {Hubbard}},\ }\bibfield
	{title} {\bibinfo {title} {Understanding {J}upiter's interior},\ }\href
	{https://doi.org/10.1002/2016JE005080} {\bibfield  {journal} {\bibinfo
			{journal} {J. Geophys. Res. Planets}\ }\textbf {\bibinfo {volume} {121}},\
		\bibinfo {pages} {1552} (\bibinfo {year} {2016})}\BibitemShut {NoStop}%
	\bibitem [{\citenamefont {Lindl}(1995)}]{Lindl1995}%
	\BibitemOpen
	\bibfield  {author} {\bibinfo {author} {\bibfnamefont {J.}~\bibnamefont
			{Lindl}},\ }\bibfield  {title} {\bibinfo {title} {Development of the
			indirect‐drive approach to inertial confinement fusion and the target
			physics basis for ignition and gain},\ }\href
	{https://doi.org/10.1063/1.871025} {\bibfield  {journal} {\bibinfo  {journal}
			{Phys. Plasmas}\ }\textbf {\bibinfo {volume} {2}},\ \bibinfo {pages} {3933}
		(\bibinfo {year} {1995})}\BibitemShut {NoStop}%
	\bibitem [{\citenamefont {Glenzer}\ \emph {et~al.}(2010)\citenamefont
		{Glenzer}, \citenamefont {MacGowan}, \citenamefont {Michel}, \citenamefont
		{Meezan}, \citenamefont {Suter}, \citenamefont {Dixit}, \citenamefont
		{Kline}, \citenamefont {Kyrala}, \citenamefont {Bradley}, \citenamefont
		{Callahan}, \citenamefont {Dewald}, \citenamefont {Divol}, \citenamefont
		{Dzenitis}, \citenamefont {Edwards}, \citenamefont {Hamza}, \citenamefont
		{Haynam}, \citenamefont {Hinkel}, \citenamefont {Kalantar}, \citenamefont
		{Kilkenny}, \citenamefont {Landen}, \citenamefont {Lindl}, \citenamefont
		{LePape}, \citenamefont {Moody}, \citenamefont {Nikroo}, \citenamefont
		{Parham}, \citenamefont {Schneider}, \citenamefont {Town}, \citenamefont
		{Wegner}, \citenamefont {Widmann}, \citenamefont {Whitman}, \citenamefont
		{Young}, \citenamefont {Van~Wonterghem}, \citenamefont {Atherton},\ and\
		\citenamefont {Moses}}]{Glenzer2010}%
	\BibitemOpen
	\bibfield  {author} {\bibinfo {author} {\bibfnamefont {S.~H.}\ \bibnamefont
			{Glenzer}}, \bibinfo {author} {\bibfnamefont {B.~J.}\ \bibnamefont
			{MacGowan}}, \bibinfo {author} {\bibfnamefont {P.}~\bibnamefont {Michel}},
		\bibinfo {author} {\bibfnamefont {N.~B.}\ \bibnamefont {Meezan}}, \bibinfo
		{author} {\bibfnamefont {L.~J.}\ \bibnamefont {Suter}}, \bibinfo {author}
		{\bibfnamefont {S.~N.}\ \bibnamefont {Dixit}}, \bibinfo {author}
		{\bibfnamefont {J.~L.}\ \bibnamefont {Kline}}, \bibinfo {author}
		{\bibfnamefont {G.~A.}\ \bibnamefont {Kyrala}}, \bibinfo {author}
		{\bibfnamefont {D.~K.}\ \bibnamefont {Bradley}}, \bibinfo {author}
		{\bibfnamefont {D.~A.}\ \bibnamefont {Callahan}}, \bibinfo {author}
		{\bibfnamefont {E.~L.}\ \bibnamefont {Dewald}}, \bibinfo {author}
		{\bibfnamefont {L.}~\bibnamefont {Divol}}, \bibinfo {author} {\bibfnamefont
			{E.}~\bibnamefont {Dzenitis}}, \bibinfo {author} {\bibfnamefont {M.~J.}\
			\bibnamefont {Edwards}}, \bibinfo {author} {\bibfnamefont {A.~V.}\
			\bibnamefont {Hamza}}, \bibinfo {author} {\bibfnamefont {C.~A.}\ \bibnamefont
			{Haynam}}, \bibinfo {author} {\bibfnamefont {D.~E.}\ \bibnamefont {Hinkel}},
		\bibinfo {author} {\bibfnamefont {D.~H.}\ \bibnamefont {Kalantar}}, \bibinfo
		{author} {\bibfnamefont {J.~D.}\ \bibnamefont {Kilkenny}}, \bibinfo {author}
		{\bibfnamefont {O.~L.}\ \bibnamefont {Landen}}, \bibinfo {author}
		{\bibfnamefont {J.~D.}\ \bibnamefont {Lindl}}, \bibinfo {author}
		{\bibfnamefont {S.}~\bibnamefont {LePape}}, \bibinfo {author} {\bibfnamefont
			{J.~D.}\ \bibnamefont {Moody}}, \bibinfo {author} {\bibfnamefont
			{A.}~\bibnamefont {Nikroo}}, \bibinfo {author} {\bibfnamefont
			{T.}~\bibnamefont {Parham}}, \bibinfo {author} {\bibfnamefont {M.~B.}\
			\bibnamefont {Schneider}}, \bibinfo {author} {\bibfnamefont {R.~P.~J.}\
			\bibnamefont {Town}}, \bibinfo {author} {\bibfnamefont {P.}~\bibnamefont
			{Wegner}}, \bibinfo {author} {\bibfnamefont {K.}~\bibnamefont {Widmann}},
		\bibinfo {author} {\bibfnamefont {P.}~\bibnamefont {Whitman}}, \bibinfo
		{author} {\bibfnamefont {B.~K.~F.}\ \bibnamefont {Young}}, \bibinfo {author}
		{\bibfnamefont {B.}~\bibnamefont {Van~Wonterghem}}, \bibinfo {author}
		{\bibfnamefont {L.~J.}\ \bibnamefont {Atherton}},\ and\ \bibinfo {author}
		{\bibfnamefont {E.~I.}\ \bibnamefont {Moses}},\ }\bibfield  {title} {\bibinfo
		{title} {Symmetric inertial confinement fusion implosions at ultra-high laser
			energies},\ }\href {https://doi.org/10.1126/science.1185634} {\bibfield
		{journal} {\bibinfo  {journal} {Science}\ }\textbf {\bibinfo {volume}
			{327}},\ \bibinfo {pages} {1228} (\bibinfo {year} {2010})}\BibitemShut
	{NoStop}%
	\bibitem [{\citenamefont {Seddon}\ \emph {et~al.}(2017)\citenamefont {Seddon},
		\citenamefont {Clarke}, \citenamefont {Dunning}, \citenamefont
		{Masciovecchio}, \citenamefont {Milne}, \citenamefont {Parmigiani},
		\citenamefont {Rugg}, \citenamefont {Spence}, \citenamefont {Thompson},
		\citenamefont {Ueda}, \citenamefont {Vinko}, \citenamefont {Wark},\ and\
		\citenamefont {Wurth}}]{Seddon2017}%
	\BibitemOpen
	\bibfield  {author} {\bibinfo {author} {\bibfnamefont {E.~A.}\ \bibnamefont
			{Seddon}}, \bibinfo {author} {\bibfnamefont {J.~A.}\ \bibnamefont {Clarke}},
		\bibinfo {author} {\bibfnamefont {D.~J.}\ \bibnamefont {Dunning}}, \bibinfo
		{author} {\bibfnamefont {C.}~\bibnamefont {Masciovecchio}}, \bibinfo {author}
		{\bibfnamefont {C.~J.}\ \bibnamefont {Milne}}, \bibinfo {author}
		{\bibfnamefont {F.}~\bibnamefont {Parmigiani}}, \bibinfo {author}
		{\bibfnamefont {D.}~\bibnamefont {Rugg}}, \bibinfo {author} {\bibfnamefont
			{J.~C.~H.}\ \bibnamefont {Spence}}, \bibinfo {author} {\bibfnamefont {N.~R.}\
			\bibnamefont {Thompson}}, \bibinfo {author} {\bibfnamefont {K.}~\bibnamefont
			{Ueda}}, \bibinfo {author} {\bibfnamefont {S.~M.}\ \bibnamefont {Vinko}},
		\bibinfo {author} {\bibfnamefont {J.~S.}\ \bibnamefont {Wark}},\ and\
		\bibinfo {author} {\bibfnamefont {W.}~\bibnamefont {Wurth}},\ }\bibfield
	{title} {\bibinfo {title} {Short-wavelength free-electron laser sources and
			science: a review},\ }\href {https://doi.org/10.1088/1361-6633/aa7cca}
	{\bibfield  {journal} {\bibinfo  {journal} {Rep. Prog. Phys.}\ }\textbf
		{\bibinfo {volume} {80}},\ \bibinfo {pages} {115901} (\bibinfo {year}
		{2017})}\BibitemShut {NoStop}%
	\bibitem [{\citenamefont {Glenzer}\ \emph {et~al.}(2016)\citenamefont
		{Glenzer}, \citenamefont {Fletcher}, \citenamefont {Galtier}, \citenamefont
		{Nagler}, \citenamefont {Alonso-Mori}, \citenamefont {Barbrel}, \citenamefont
		{Brown}, \citenamefont {Chapman}, \citenamefont {Chen}, \citenamefont
		{Curry}, \citenamefont {Fiuza}, \citenamefont {Gamboa}, \citenamefont
		{Gauthier}, \citenamefont {Gericke}, \citenamefont {Gleason}, \citenamefont
		{Goede}, \citenamefont {Granados}, \citenamefont {Heimann}, \citenamefont
		{Kim}, \citenamefont {Kraus}, \citenamefont {MacDonald}, \citenamefont
		{Mackinnon}, \citenamefont {Mishra}, \citenamefont {Ravasio}, \citenamefont
		{Roedel}, \citenamefont {Sperling}, \citenamefont {Schumaker}, \citenamefont
		{Tsui}, \citenamefont {Vorberger}, \citenamefont {Zastrau}, \citenamefont
		{Fry}, \citenamefont {White}, \citenamefont {Hasting},\ and\ \citenamefont
		{Lee}}]{Glenzer2016}%
	\BibitemOpen
	\bibfield  {author} {\bibinfo {author} {\bibfnamefont {S.~H.}\ \bibnamefont
			{Glenzer}}, \bibinfo {author} {\bibfnamefont {L.~B.}\ \bibnamefont
			{Fletcher}}, \bibinfo {author} {\bibfnamefont {E.}~\bibnamefont {Galtier}},
		\bibinfo {author} {\bibfnamefont {B.}~\bibnamefont {Nagler}}, \bibinfo
		{author} {\bibfnamefont {R.}~\bibnamefont {Alonso-Mori}}, \bibinfo {author}
		{\bibfnamefont {B.}~\bibnamefont {Barbrel}}, \bibinfo {author} {\bibfnamefont
			{S.~B.}\ \bibnamefont {Brown}}, \bibinfo {author} {\bibfnamefont {D.~A.}\
			\bibnamefont {Chapman}}, \bibinfo {author} {\bibfnamefont {Z.}~\bibnamefont
			{Chen}}, \bibinfo {author} {\bibfnamefont {C.~B.}\ \bibnamefont {Curry}},
		\bibinfo {author} {\bibfnamefont {F.}~\bibnamefont {Fiuza}}, \bibinfo
		{author} {\bibfnamefont {E.}~\bibnamefont {Gamboa}}, \bibinfo {author}
		{\bibfnamefont {M.}~\bibnamefont {Gauthier}}, \bibinfo {author}
		{\bibfnamefont {D.~O.}\ \bibnamefont {Gericke}}, \bibinfo {author}
		{\bibfnamefont {A.}~\bibnamefont {Gleason}}, \bibinfo {author} {\bibfnamefont
			{S.}~\bibnamefont {Goede}}, \bibinfo {author} {\bibfnamefont
			{E.}~\bibnamefont {Granados}}, \bibinfo {author} {\bibfnamefont
			{P.}~\bibnamefont {Heimann}}, \bibinfo {author} {\bibfnamefont
			{J.}~\bibnamefont {Kim}}, \bibinfo {author} {\bibfnamefont {D.}~\bibnamefont
			{Kraus}}, \bibinfo {author} {\bibfnamefont {M.~J.}\ \bibnamefont
			{MacDonald}}, \bibinfo {author} {\bibfnamefont {A.~J.}\ \bibnamefont
			{Mackinnon}}, \bibinfo {author} {\bibfnamefont {R.}~\bibnamefont {Mishra}},
		\bibinfo {author} {\bibfnamefont {A.}~\bibnamefont {Ravasio}}, \bibinfo
		{author} {\bibfnamefont {C.}~\bibnamefont {Roedel}}, \bibinfo {author}
		{\bibfnamefont {P.}~\bibnamefont {Sperling}}, \bibinfo {author}
		{\bibfnamefont {W.}~\bibnamefont {Schumaker}}, \bibinfo {author}
		{\bibfnamefont {Y.~Y.}\ \bibnamefont {Tsui}}, \bibinfo {author}
		{\bibfnamefont {J.}~\bibnamefont {Vorberger}}, \bibinfo {author}
		{\bibfnamefont {U.}~\bibnamefont {Zastrau}}, \bibinfo {author} {\bibfnamefont
			{A.}~\bibnamefont {Fry}}, \bibinfo {author} {\bibfnamefont {W.~E.}\
			\bibnamefont {White}}, \bibinfo {author} {\bibfnamefont {J.~B.}\ \bibnamefont
			{Hasting}},\ and\ \bibinfo {author} {\bibfnamefont {H.~J.}\ \bibnamefont
			{Lee}},\ }\bibfield  {title} {\bibinfo {title} {Matter under extreme
			conditions experiments at the linac coherent light source},\ }\href
	{https://doi.org/10.1088/0953-4075/49/9/092001} {\bibfield  {journal}
		{\bibinfo  {journal} {J. Phys. B}\ }\textbf {\bibinfo {volume} {49}},\
		\bibinfo {pages} {092001} (\bibinfo {year} {2016})}\BibitemShut {NoStop}%
	\bibitem [{\citenamefont {Drake}(2006)}]{Drake2006}%
	\BibitemOpen
	\bibfield  {author} {\bibinfo {author} {\bibfnamefont {R.~P.}\ \bibnamefont
			{Drake}},\ }\href {https://doi.org/10.1007/3-540-29315-9} {\emph {\bibinfo
			{title} {High-Energy-Density Physics: Fundamentals, Inertial Fusion, and
				Experimental Astrophysics}}},\ \bibinfo {edition} {1st}\ ed.\ (\bibinfo
	{publisher} {Springer-Verlag Berlin Heidelberg},\ \bibinfo {address} {Berlin
		Heidelberg},\ \bibinfo {year} {2006})\ Chap.~\bibinfo {chapter} {1}, pp.\
	\bibinfo {pages} {5--8}\BibitemShut {NoStop}%
	\bibitem [{\citenamefont {{Vinko}}\ \emph {et~al.}(2012)\citenamefont
		{{Vinko}}, \citenamefont {Ciricosta}, \citenamefont {Cho}, \citenamefont
		{Engelhorn}, \citenamefont {Chung}, \citenamefont {Brown}, \citenamefont
		{Burian}, \citenamefont {Chalupsk{\'y}}, \citenamefont {Falcone},
		\citenamefont {Graves}, \citenamefont {H{\'a}jkov{\'a}}, \citenamefont
		{Higginbotham}, \citenamefont {Juha}, \citenamefont {Krzywinski},
		\citenamefont {Lee}, \citenamefont {Messerschmidt}, \citenamefont {Murphy},
		\citenamefont {Ping}, \citenamefont {Scherz}, \citenamefont {Schlotter},
		\citenamefont {Toleikis}, \citenamefont {Turner}, \citenamefont {Vysin},
		\citenamefont {Wang}, \citenamefont {Wu}, \citenamefont {Zastrau},
		\citenamefont {Zhu}, \citenamefont {Lee}, \citenamefont {Heimann},
		\citenamefont {Nagler},\ and\ \citenamefont {Wark}}]{Vinko2012}%
	\BibitemOpen
	\bibfield  {author} {\bibinfo {author} {\bibfnamefont {S.~M.}\ \bibnamefont
			{{Vinko}}}, \bibinfo {author} {\bibfnamefont {O.}~\bibnamefont {Ciricosta}},
		\bibinfo {author} {\bibfnamefont {B.~I.}\ \bibnamefont {Cho}}, \bibinfo
		{author} {\bibfnamefont {K.}~\bibnamefont {Engelhorn}}, \bibinfo {author}
		{\bibfnamefont {H.-K.}\ \bibnamefont {Chung}}, \bibinfo {author}
		{\bibfnamefont {C.~R.~D.}\ \bibnamefont {Brown}}, \bibinfo {author}
		{\bibfnamefont {T.}~\bibnamefont {Burian}}, \bibinfo {author} {\bibfnamefont
			{J.}~\bibnamefont {Chalupsk{\'y}}}, \bibinfo {author} {\bibfnamefont {R.~W.}\
			\bibnamefont {Falcone}}, \bibinfo {author} {\bibfnamefont {C.}~\bibnamefont
			{Graves}}, \bibinfo {author} {\bibfnamefont {V.}~\bibnamefont
			{H{\'a}jkov{\'a}}}, \bibinfo {author} {\bibfnamefont {A.}~\bibnamefont
			{Higginbotham}}, \bibinfo {author} {\bibfnamefont {L.}~\bibnamefont {Juha}},
		\bibinfo {author} {\bibfnamefont {J.}~\bibnamefont {Krzywinski}}, \bibinfo
		{author} {\bibfnamefont {H.~J.}\ \bibnamefont {Lee}}, \bibinfo {author}
		{\bibfnamefont {M.}~\bibnamefont {Messerschmidt}}, \bibinfo {author}
		{\bibfnamefont {C.~D.}\ \bibnamefont {Murphy}}, \bibinfo {author}
		{\bibfnamefont {Y.}~\bibnamefont {Ping}}, \bibinfo {author} {\bibfnamefont
			{A.}~\bibnamefont {Scherz}}, \bibinfo {author} {\bibfnamefont
			{W.}~\bibnamefont {Schlotter}}, \bibinfo {author} {\bibfnamefont
			{S.}~\bibnamefont {Toleikis}}, \bibinfo {author} {\bibfnamefont {J.~J.}\
			\bibnamefont {Turner}}, \bibinfo {author} {\bibfnamefont {L.}~\bibnamefont
			{Vysin}}, \bibinfo {author} {\bibfnamefont {T.}~\bibnamefont {Wang}},
		\bibinfo {author} {\bibfnamefont {B.}~\bibnamefont {Wu}}, \bibinfo {author}
		{\bibfnamefont {U.}~\bibnamefont {Zastrau}}, \bibinfo {author} {\bibfnamefont
			{D.}~\bibnamefont {Zhu}}, \bibinfo {author} {\bibfnamefont {R.~W.}\
			\bibnamefont {Lee}}, \bibinfo {author} {\bibfnamefont {P.~A.}\ \bibnamefont
			{Heimann}}, \bibinfo {author} {\bibfnamefont {B.}~\bibnamefont {Nagler}},\
		and\ \bibinfo {author} {\bibfnamefont {J.~S.}\ \bibnamefont {Wark}},\
	}\bibfield  {title} {\bibinfo {title} {Creation and diagnosis of a
			solid-density plasma with an x-ray free-electron laser},\ }\href
	{https://doi.org/10.1038/nature10746} {\bibfield  {journal} {\bibinfo
			{journal} {Nature}\ }\textbf {\bibinfo {volume} {482}},\ \bibinfo {pages}
		{59} (\bibinfo {year} {2012})}\BibitemShut {NoStop}%
	\bibitem [{\citenamefont {Bauche}\ \emph {et~al.}(2015)\citenamefont {Bauche},
		\citenamefont {Bauche-Arnoult},\ and\ \citenamefont {Peyrusse}}]{Bauche2015}%
	\BibitemOpen
	\bibfield  {author} {\bibinfo {author} {\bibfnamefont {J.}~\bibnamefont
			{Bauche}}, \bibinfo {author} {\bibfnamefont {C.}~\bibnamefont
			{Bauche-Arnoult}},\ and\ \bibinfo {author} {\bibfnamefont {O.}~\bibnamefont
			{Peyrusse}},\ }\href@noop {} {\emph {\bibinfo {title} {Atomic Properties in
				Hot Plasmas}}}\ (\bibinfo  {publisher} {Springer International Publishing},\
	\bibinfo {year} {2015})\BibitemShut {NoStop}%
	\bibitem [{\citenamefont {Agassi}(1984)}]{Agassi84}%
	\BibitemOpen
	\bibfield  {author} {\bibinfo {author} {\bibfnamefont {D.}~\bibnamefont
			{Agassi}},\ }\bibfield  {title} {\bibinfo {title} {Phenomenological model for
			pisosecond‐pulse laser annealing of semiconductors},\ }\href
	{https://doi.org/10.1063/1.333007} {\bibfield  {journal} {\bibinfo  {journal}
			{J. Appl. Phys.}\ }\textbf {\bibinfo {volume} {55}},\ \bibinfo {pages} {4376}
		(\bibinfo {year} {1984})}\BibitemShut {NoStop}%
	\bibitem [{\citenamefont {Silvestrelli}\ \emph {et~al.}(1996)\citenamefont
		{Silvestrelli}, \citenamefont {Alavi}, \citenamefont {Parrinello},\ and\
		\citenamefont {Frenkel}}]{Silvestrelli06}%
	\BibitemOpen
	\bibfield  {author} {\bibinfo {author} {\bibfnamefont {P.~L.}\ \bibnamefont
			{Silvestrelli}}, \bibinfo {author} {\bibfnamefont {A.}~\bibnamefont {Alavi}},
		\bibinfo {author} {\bibfnamefont {M.}~\bibnamefont {Parrinello}},\ and\
		\bibinfo {author} {\bibfnamefont {D.}~\bibnamefont {Frenkel}},\ }\bibfield
	{title} {\bibinfo {title} {\emph{Ab initio} molecular dynamics simulation of
			laser melting of silicon},\ }\href
	{https://doi.org/10.1103/PhysRevLett.77.3149} {\bibfield  {journal} {\bibinfo
			{journal} {Phys. Rev. Lett.}\ }\textbf {\bibinfo {volume} {77}},\ \bibinfo
		{pages} {3149} (\bibinfo {year} {1996})}\BibitemShut {NoStop}%
	\bibitem [{\citenamefont {Ostrikov}\ \emph {et~al.}(2016)\citenamefont
		{Ostrikov}, \citenamefont {Beg},\ and\ \citenamefont {Ng}}]{Ostrikov16}%
	\BibitemOpen
	\bibfield  {author} {\bibinfo {author} {\bibfnamefont {K.~K.}\ \bibnamefont
			{Ostrikov}}, \bibinfo {author} {\bibfnamefont {F.}~\bibnamefont {Beg}},\ and\
		\bibinfo {author} {\bibfnamefont {A.}~\bibnamefont {Ng}},\ }\bibfield
	{title} {\bibinfo {title} {Colloquium: Nanoplasmas generated by intense
			radiation},\ }\href {https://doi.org/10.1103/RevModPhys.88.011001} {\bibfield
		{journal} {\bibinfo  {journal} {Rev. Mod. Phys.}\ }\textbf {\bibinfo
			{volume} {88}},\ \bibinfo {pages} {011001} (\bibinfo {year}
		{2016})}\BibitemShut {NoStop}%
	\bibitem [{\citenamefont {Medvedev}\ \emph {et~al.}(2011)\citenamefont
		{Medvedev}, \citenamefont {Zastrau}, \citenamefont {F{\"o}rster},
		\citenamefont {Gericke},\ and\ \citenamefont {Rethfeld}}]{Medvedev11}%
	\BibitemOpen
	\bibfield  {author} {\bibinfo {author} {\bibfnamefont {N.}~\bibnamefont
			{Medvedev}}, \bibinfo {author} {\bibfnamefont {U.}~\bibnamefont {Zastrau}},
		\bibinfo {author} {\bibfnamefont {E.}~\bibnamefont {F{\"o}rster}}, \bibinfo
		{author} {\bibfnamefont {D.~O.}\ \bibnamefont {Gericke}},\ and\ \bibinfo
		{author} {\bibfnamefont {B.}~\bibnamefont {Rethfeld}},\ }\bibfield  {title}
	{\bibinfo {title} {Short-time electron dynamics in aluminum excited by
			femtosecond extreme ultraviolet radiation},\ }\href
	{https://doi.org/10.1103/PhysRevLett.107.165003} {\bibfield  {journal}
		{\bibinfo  {journal} {Phys. Rev. Lett.}\ }\textbf {\bibinfo {volume} {107}},\
		\bibinfo {pages} {165003} (\bibinfo {year} {2011})}\BibitemShut {NoStop}%
	\bibitem [{\citenamefont {Anisimov}\ \emph {et~al.}(1974)\citenamefont
		{Anisimov}, \citenamefont {Kapeliovich},\ and\ \citenamefont
		{Perel'man}}]{Anisimov74}%
	\BibitemOpen
	\bibfield  {author} {\bibinfo {author} {\bibfnamefont {S.~I.}\ \bibnamefont
			{Anisimov}}, \bibinfo {author} {\bibfnamefont {B.~L.}\ \bibnamefont
			{Kapeliovich}},\ and\ \bibinfo {author} {\bibfnamefont {T.~L.}\ \bibnamefont
			{Perel'man}},\ }\bibfield  {title} {\bibinfo {title} {Electron emission from
			metal surfaces exposed to ultrashort laser pulses},\ }\href
	{http://www.jetp.ac.ru/cgi-bin/e/index/e/39/2/p375?a=list} {\bibfield
		{journal} {\bibinfo  {journal} {Sov. Phys. JETP}\ }\textbf {\bibinfo {volume}
			{39}},\ \bibinfo {pages} {375} (\bibinfo {year} {1974})}\BibitemShut
	{NoStop}%
	\bibitem [{\citenamefont {Rethfeld}\ \emph {et~al.}(2017)\citenamefont
		{Rethfeld}, \citenamefont {Ivanov}, \citenamefont {Garcia},\ and\
		\citenamefont {Anisimov}}]{Rethfeld17}%
	\BibitemOpen
	\bibfield  {author} {\bibinfo {author} {\bibfnamefont {B.}~\bibnamefont
			{Rethfeld}}, \bibinfo {author} {\bibfnamefont {D.~S.}\ \bibnamefont
			{Ivanov}}, \bibinfo {author} {\bibfnamefont {M.~E.}\ \bibnamefont {Garcia}},\
		and\ \bibinfo {author} {\bibfnamefont {S.~I.}\ \bibnamefont {Anisimov}},\
	}\bibfield  {title} {\bibinfo {title} {Modelling ultrafast laser ablation},\
	}\href {https://doi.org/10.1088/1361-6463/50/19/193001} {\bibfield  {journal}
		{\bibinfo  {journal} {J. Phys. D: Appl. Phys.}\ }\textbf {\bibinfo {volume}
			{50}},\ \bibinfo {pages} {193001} (\bibinfo {year} {2017})}\BibitemShut
	{NoStop}%
	\bibitem [{\citenamefont {Chung}\ and\ \citenamefont {Lee}(2009)}]{Chung09}%
	\BibitemOpen
	\bibfield  {author} {\bibinfo {author} {\bibfnamefont {H.-K.}\ \bibnamefont
			{Chung}}\ and\ \bibinfo {author} {\bibfnamefont {R.~W.}\ \bibnamefont
			{Lee}},\ }\bibfield  {title} {\bibinfo {title} {Applications of {NLTE}
			population kinetics},\ }\href {https://doi.org/10.1016/j.hedp.2009.02.005}
	{\bibfield  {journal} {\bibinfo  {journal} {High Energy Density Phys.}\
		}\textbf {\bibinfo {volume} {5}},\ \bibinfo {pages} {1} (\bibinfo {year}
		{2009})}\BibitemShut {NoStop}%
	\bibitem [{\citenamefont {Hansen}\ \emph {et~al.}(2020)\citenamefont {Hansen},
		\citenamefont {Chung}, \citenamefont {Fontes}, \citenamefont {Ralchenko},
		\citenamefont {Scott},\ and\ \citenamefont {Stambulchik}}]{Hansen20}%
	\BibitemOpen
	\bibfield  {author} {\bibinfo {author} {\bibfnamefont {S.~B.}\ \bibnamefont
			{Hansen}}, \bibinfo {author} {\bibfnamefont {H.-K.}\ \bibnamefont {Chung}},
		\bibinfo {author} {\bibfnamefont {C.~J.}\ \bibnamefont {Fontes}}, \bibinfo
		{author} {\bibfnamefont {Y.}~\bibnamefont {Ralchenko}}, \bibinfo {author}
		{\bibfnamefont {H.~A.}\ \bibnamefont {Scott}},\ and\ \bibinfo {author}
		{\bibfnamefont {E.}~\bibnamefont {Stambulchik}},\ }\bibfield  {title}
	{\bibinfo {title} {Review of the 10th {Non-LTE} code comparison workshop},\
	}\href {https://doi.org/10.1016/j.hedp.2019.06.001} {\bibfield  {journal}
		{\bibinfo  {journal} {High Energy Density Phys.}\ }\textbf {\bibinfo {volume}
			{35}},\ \bibinfo {pages} {100693} (\bibinfo {year} {2020})}\BibitemShut
	{NoStop}%
	\bibitem [{\citenamefont {Scott}(2001)}]{Scott:2001aa}%
	\BibitemOpen
	\bibfield  {author} {\bibinfo {author} {\bibfnamefont {H.~A.}\ \bibnamefont
			{Scott}},\ }\bibfield  {title} {\bibinfo {title} {Cretin---a radiative
			transfer capability for laboratory plasmas},\ }\bibfield  {booktitle} {\emph
		{\bibinfo {booktitle} {Radiative Properties of Hot Dense Matter}},\ }\href
	{https://doi.org/https://doi.org/10.1016/S0022-4073(01)00109-1} {\bibfield
		{journal} {\bibinfo  {journal} {J. Quant. Spectrosc. \& Radiat. Transfer}\
		}\textbf {\bibinfo {volume} {71}},\ \bibinfo {pages} {689} (\bibinfo {year}
		{2001})}\BibitemShut {NoStop}%
	\bibitem [{\citenamefont {Ecker}\ and\ \citenamefont
		{Kr{\"o}ll}(1963)}]{E-K1963}%
	\BibitemOpen
	\bibfield  {author} {\bibinfo {author} {\bibfnamefont {G.}~\bibnamefont
			{Ecker}}\ and\ \bibinfo {author} {\bibfnamefont {W.}~\bibnamefont
			{Kr{\"o}ll}},\ }\bibfield  {title} {\bibinfo {title} {Lowering of the
			ionization energy for a plasma in thermodynamic equilibrium},\ }\href
	{https://doi.org/10.1063/1.1724509} {\bibfield  {journal} {\bibinfo
			{journal} {Phys. Fluids}\ }\textbf {\bibinfo {volume} {6}},\ \bibinfo {pages}
		{62} (\bibinfo {year} {1963})}\BibitemShut {NoStop}%
	\bibitem [{\citenamefont {Stewart}\ and\ \citenamefont
		{Pyatt}(1966)}]{S-P1966}%
	\BibitemOpen
	\bibfield  {author} {\bibinfo {author} {\bibfnamefont {J.~C.}\ \bibnamefont
			{Stewart}}\ and\ \bibinfo {author} {\bibfnamefont {K.~D.}\ \bibnamefont
			{Pyatt}, \bibfnamefont {Jr.}},\ }\bibfield  {title} {\bibinfo {title}
		{{Lowering of Ionization Potentials in Plasmas}},\ }\href
	{https://doi.org/10.1086/148714} {\bibfield  {journal} {\bibinfo  {journal}
			{\apj}\ }\textbf {\bibinfo {volume} {144}},\ \bibinfo {pages} {1203}
		(\bibinfo {year} {1966})}\BibitemShut {NoStop}%
	\bibitem [{\citenamefont {Ciricosta}\ \emph {et~al.}(2012)\citenamefont
		{Ciricosta}, \citenamefont {Vinko}, \citenamefont {Chung}, \citenamefont
		{Cho}, \citenamefont {Brown}, \citenamefont {Burian}, \citenamefont
		{Chalupsk\'y}, \citenamefont {Engelhorn}, \citenamefont {Falcone},
		\citenamefont {Graves}, \citenamefont {H\'ajkov\'a}, \citenamefont
		{Higginbotham}, \citenamefont {Juha}, \citenamefont {Krzywinski},
		\citenamefont {Lee}, \citenamefont {Messerschmidt}, \citenamefont {Murphy},
		\citenamefont {Ping}, \citenamefont {Rackstraw}, \citenamefont {Scherz},
		\citenamefont {Schlotter}, \citenamefont {Toleikis}, \citenamefont {Turner},
		\citenamefont {Vysin}, \citenamefont {Wang}, \citenamefont {Wu},
		\citenamefont {Zastrau}, \citenamefont {Zhu}, \citenamefont {Lee},
		\citenamefont {Heimann}, \citenamefont {Nagler},\ and\ \citenamefont
		{Wark}}]{Ciricosta2012}%
	\BibitemOpen
	\bibfield  {author} {\bibinfo {author} {\bibfnamefont {O.}~\bibnamefont
			{Ciricosta}}, \bibinfo {author} {\bibfnamefont {S.~M.}\ \bibnamefont
			{Vinko}}, \bibinfo {author} {\bibfnamefont {H.-K.}\ \bibnamefont {Chung}},
		\bibinfo {author} {\bibfnamefont {B.-I.}\ \bibnamefont {Cho}}, \bibinfo
		{author} {\bibfnamefont {C.~R.~D.}\ \bibnamefont {Brown}}, \bibinfo {author}
		{\bibfnamefont {T.}~\bibnamefont {Burian}}, \bibinfo {author} {\bibfnamefont
			{J.}~\bibnamefont {Chalupsk\'y}}, \bibinfo {author} {\bibfnamefont
			{K.}~\bibnamefont {Engelhorn}}, \bibinfo {author} {\bibfnamefont {R.~W.}\
			\bibnamefont {Falcone}}, \bibinfo {author} {\bibfnamefont {C.}~\bibnamefont
			{Graves}}, \bibinfo {author} {\bibfnamefont {V.}~\bibnamefont {H\'ajkov\'a}},
		\bibinfo {author} {\bibfnamefont {A.}~\bibnamefont {Higginbotham}}, \bibinfo
		{author} {\bibfnamefont {L.}~\bibnamefont {Juha}}, \bibinfo {author}
		{\bibfnamefont {J.}~\bibnamefont {Krzywinski}}, \bibinfo {author}
		{\bibfnamefont {H.~J.}\ \bibnamefont {Lee}}, \bibinfo {author} {\bibfnamefont
			{M.}~\bibnamefont {Messerschmidt}}, \bibinfo {author} {\bibfnamefont {C.~D.}\
			\bibnamefont {Murphy}}, \bibinfo {author} {\bibfnamefont {Y.}~\bibnamefont
			{Ping}}, \bibinfo {author} {\bibfnamefont {D.~S.}\ \bibnamefont {Rackstraw}},
		\bibinfo {author} {\bibfnamefont {A.}~\bibnamefont {Scherz}}, \bibinfo
		{author} {\bibfnamefont {W.}~\bibnamefont {Schlotter}}, \bibinfo {author}
		{\bibfnamefont {S.}~\bibnamefont {Toleikis}}, \bibinfo {author}
		{\bibfnamefont {J.~J.}\ \bibnamefont {Turner}}, \bibinfo {author}
		{\bibfnamefont {L.}~\bibnamefont {Vysin}}, \bibinfo {author} {\bibfnamefont
			{T.}~\bibnamefont {Wang}}, \bibinfo {author} {\bibfnamefont {B.}~\bibnamefont
			{Wu}}, \bibinfo {author} {\bibfnamefont {U.}~\bibnamefont {Zastrau}},
		\bibinfo {author} {\bibfnamefont {D.}~\bibnamefont {Zhu}}, \bibinfo {author}
		{\bibfnamefont {R.~W.}\ \bibnamefont {Lee}}, \bibinfo {author} {\bibfnamefont
			{P.}~\bibnamefont {Heimann}}, \bibinfo {author} {\bibfnamefont
			{B.}~\bibnamefont {Nagler}},\ and\ \bibinfo {author} {\bibfnamefont {J.~S.}\
			\bibnamefont {Wark}},\ }\bibfield  {title} {\bibinfo {title} {Direct
			measurements of the ionization potential depression in a dense plasma},\
	}\href {https://doi.org/10.1103/PhysRevLett.109.065002} {\bibfield  {journal}
		{\bibinfo  {journal} {Phys. Rev. Lett.}\ }\textbf {\bibinfo {volume} {109}},\
		\bibinfo {pages} {065002} (\bibinfo {year} {2012})}\BibitemShut {NoStop}%
	\bibitem [{\citenamefont {Hoarty}\ \emph
		{et~al.}(2013{\natexlab{a}})\citenamefont {Hoarty}, \citenamefont {Allan},
		\citenamefont {James}, \citenamefont {Brown}, \citenamefont {Hobbs},
		\citenamefont {Hill}, \citenamefont {Harris}, \citenamefont {Morton},
		\citenamefont {Brookes}, \citenamefont {Shepherd}, \citenamefont {Dunn},
		\citenamefont {Chen}, \citenamefont {Von~Marley}, \citenamefont
		{Beiersdorfer}, \citenamefont {Chung}, \citenamefont {Lee}, \citenamefont
		{Brown},\ and\ \citenamefont {Emig}}]{Hoarty13}%
	\BibitemOpen
	\bibfield  {author} {\bibinfo {author} {\bibfnamefont {D.~J.}\ \bibnamefont
			{Hoarty}}, \bibinfo {author} {\bibfnamefont {P.}~\bibnamefont {Allan}},
		\bibinfo {author} {\bibfnamefont {S.~F.}\ \bibnamefont {James}}, \bibinfo
		{author} {\bibfnamefont {C.~R.~D.}\ \bibnamefont {Brown}}, \bibinfo {author}
		{\bibfnamefont {L.~M.~R.}\ \bibnamefont {Hobbs}}, \bibinfo {author}
		{\bibfnamefont {M.~P.}\ \bibnamefont {Hill}}, \bibinfo {author}
		{\bibfnamefont {J.~W.~O.}\ \bibnamefont {Harris}}, \bibinfo {author}
		{\bibfnamefont {J.}~\bibnamefont {Morton}}, \bibinfo {author} {\bibfnamefont
			{M.~G.}\ \bibnamefont {Brookes}}, \bibinfo {author} {\bibfnamefont
			{R.}~\bibnamefont {Shepherd}}, \bibinfo {author} {\bibfnamefont
			{J.}~\bibnamefont {Dunn}}, \bibinfo {author} {\bibfnamefont {H.}~\bibnamefont
			{Chen}}, \bibinfo {author} {\bibfnamefont {E.}~\bibnamefont {Von~Marley}},
		\bibinfo {author} {\bibfnamefont {P.}~\bibnamefont {Beiersdorfer}}, \bibinfo
		{author} {\bibfnamefont {H.~K.}\ \bibnamefont {Chung}}, \bibinfo {author}
		{\bibfnamefont {R.~W.}\ \bibnamefont {Lee}}, \bibinfo {author} {\bibfnamefont
			{G.}~\bibnamefont {Brown}},\ and\ \bibinfo {author} {\bibfnamefont
			{J.}~\bibnamefont {Emig}},\ }\bibfield  {title} {\bibinfo {title}
		{Observations of the effect of ionization-potential depression in hot dense
			plasma},\ }\href {https://doi.org/10.1103/PhysRevLett.110.265003} {\bibfield
		{journal} {\bibinfo  {journal} {Phys. Rev. Lett.}\ }\textbf {\bibinfo
			{volume} {110}},\ \bibinfo {pages} {265003} (\bibinfo {year}
		{2013}{\natexlab{a}})}\BibitemShut {NoStop}%
	\bibitem [{\citenamefont {Hoarty}\ \emph
		{et~al.}(2013{\natexlab{b}})\citenamefont {Hoarty}, \citenamefont {Allan},
		\citenamefont {James}, \citenamefont {Brown}, \citenamefont {Hobbs},
		\citenamefont {Hill}, \citenamefont {Harris}, \citenamefont {Morton},
		\citenamefont {Brookes}, \citenamefont {Shepherd}, \citenamefont {Dunn},
		\citenamefont {Chen}, \citenamefont {Marley]}, \citenamefont {Beiersdorfer},
		\citenamefont {Chung}, \citenamefont {Lee}, \citenamefont {Brown},\ and\
		\citenamefont {Emig}}]{Hoarty2013}%
	\BibitemOpen
	\bibfield  {author} {\bibinfo {author} {\bibfnamefont {D.}~\bibnamefont
			{Hoarty}}, \bibinfo {author} {\bibfnamefont {P.}~\bibnamefont {Allan}},
		\bibinfo {author} {\bibfnamefont {S.}~\bibnamefont {James}}, \bibinfo
		{author} {\bibfnamefont {C.}~\bibnamefont {Brown}}, \bibinfo {author}
		{\bibfnamefont {L.}~\bibnamefont {Hobbs}}, \bibinfo {author} {\bibfnamefont
			{M.}~\bibnamefont {Hill}}, \bibinfo {author} {\bibfnamefont {J.}~\bibnamefont
			{Harris}}, \bibinfo {author} {\bibfnamefont {J.}~\bibnamefont {Morton}},
		\bibinfo {author} {\bibfnamefont {M.}~\bibnamefont {Brookes}}, \bibinfo
		{author} {\bibfnamefont {R.}~\bibnamefont {Shepherd}}, \bibinfo {author}
		{\bibfnamefont {J.}~\bibnamefont {Dunn}}, \bibinfo {author} {\bibfnamefont
			{H.}~\bibnamefont {Chen}}, \bibinfo {author} {\bibfnamefont {E.~V.}\
			\bibnamefont {Marley]}}, \bibinfo {author} {\bibfnamefont {P.}~\bibnamefont
			{Beiersdorfer}}, \bibinfo {author} {\bibfnamefont {H.}~\bibnamefont {Chung}},
		\bibinfo {author} {\bibfnamefont {R.}~\bibnamefont {Lee}}, \bibinfo {author}
		{\bibfnamefont {G.}~\bibnamefont {Brown}},\ and\ \bibinfo {author}
		{\bibfnamefont {J.}~\bibnamefont {Emig}},\ }\bibfield  {title} {\bibinfo
		{title} {The first data from the orion laser; measurements of the spectrum of
			hot, dense aluminium},\ }\href
	{https://doi.org/https://doi.org/10.1016/j.hedp.2013.06.005} {\bibfield
		{journal} {\bibinfo  {journal} {High Energy Density Phys.}\ }\textbf
		{\bibinfo {volume} {9}},\ \bibinfo {pages} {661 } (\bibinfo {year}
		{2013}{\natexlab{b}})}\BibitemShut {NoStop}%
	\bibitem [{\citenamefont {Ziaja}\ \emph {et~al.}(2013)\citenamefont {Ziaja},
		\citenamefont {Jurek}, \citenamefont {Medvedev}, \citenamefont {Son},
		\citenamefont {Thiele},\ and\ \citenamefont {Toleikis}}]{Ziaja13}%
	\BibitemOpen
	\bibfield  {author} {\bibinfo {author} {\bibfnamefont {B.}~\bibnamefont
			{Ziaja}}, \bibinfo {author} {\bibfnamefont {Z.}~\bibnamefont {Jurek}},
		\bibinfo {author} {\bibfnamefont {N.}~\bibnamefont {Medvedev}}, \bibinfo
		{author} {\bibfnamefont {S.-K.}\ \bibnamefont {Son}}, \bibinfo {author}
		{\bibfnamefont {R.}~\bibnamefont {Thiele}},\ and\ \bibinfo {author}
		{\bibfnamefont {S.}~\bibnamefont {Toleikis}},\ }\bibfield  {title} {\bibinfo
		{title} {Photoelectron spectroscopy method to reveal ionization potential
			lowering in nanoplasmas},\ }\href
	{https://doi.org/10.1088/0953-4075/46/16/164009} {\bibfield  {journal}
		{\bibinfo  {journal} {J. Phys. B: At. Mol. Opt. Phys.}\ }\textbf {\bibinfo
			{volume} {46}},\ \bibinfo {pages} {164009} (\bibinfo {year}
		{2013})}\BibitemShut {NoStop}%
	\bibitem [{\citenamefont {Preston}\ \emph {et~al.}(2013)\citenamefont
		{Preston}, \citenamefont {Vinko}, \citenamefont {Ciricosta}, \citenamefont
		{Chung}, \citenamefont {Lee},\ and\ \citenamefont {Wark}}]{Preston2013}%
	\BibitemOpen
	\bibfield  {author} {\bibinfo {author} {\bibfnamefont {T.~R.}\ \bibnamefont
			{Preston}}, \bibinfo {author} {\bibfnamefont {S.~M.}\ \bibnamefont {Vinko}},
		\bibinfo {author} {\bibfnamefont {O.}~\bibnamefont {Ciricosta}}, \bibinfo
		{author} {\bibfnamefont {H.-K.}\ \bibnamefont {Chung}}, \bibinfo {author}
		{\bibfnamefont {R.~W.}\ \bibnamefont {Lee}},\ and\ \bibinfo {author}
		{\bibfnamefont {J.~S.}\ \bibnamefont {Wark}},\ }\bibfield  {title} {\bibinfo
		{title} {The effects of ionization potential depression on the spectra
			emitted by hot dense aluminium plasmas},\ }\href
	{https://doi.org/https://doi.org/10.1016/j.hedp.2012.12.014} {\bibfield
		{journal} {\bibinfo  {journal} {High Energy Density Phys.}\ }\textbf
		{\bibinfo {volume} {9}},\ \bibinfo {pages} {258 } (\bibinfo {year}
		{2013})}\BibitemShut {NoStop}%
	\bibitem [{\citenamefont {Fletcher}\ \emph {et~al.}(2014)\citenamefont
		{Fletcher}, \citenamefont {Kritcher}, \citenamefont {Pak}, \citenamefont
		{Ma}, \citenamefont {D{\"o}ppner}, \citenamefont {Fortmann}, \citenamefont
		{Divol}, \citenamefont {Jones}, \citenamefont {Landen}, \citenamefont
		{Scott}, \citenamefont {Vorberger}, \citenamefont {Chapman}, \citenamefont
		{Gericke}, \citenamefont {Mattern}, \citenamefont {Seidler}, \citenamefont
		{Gregori}, \citenamefont {Falcone},\ and\ \citenamefont
		{Glenzer}}]{Fletcher14}%
	\BibitemOpen
	\bibfield  {author} {\bibinfo {author} {\bibfnamefont {L.}~\bibnamefont
			{Fletcher}}, \bibinfo {author} {\bibfnamefont {A.}~\bibnamefont {Kritcher}},
		\bibinfo {author} {\bibfnamefont {A.}~\bibnamefont {Pak}}, \bibinfo {author}
		{\bibfnamefont {T.}~\bibnamefont {Ma}}, \bibinfo {author} {\bibfnamefont
			{T.}~\bibnamefont {D{\"o}ppner}}, \bibinfo {author} {\bibfnamefont
			{C.}~\bibnamefont {Fortmann}}, \bibinfo {author} {\bibfnamefont
			{L.}~\bibnamefont {Divol}}, \bibinfo {author} {\bibfnamefont
			{O.}~\bibnamefont {Jones}}, \bibinfo {author} {\bibfnamefont
			{O.}~\bibnamefont {Landen}}, \bibinfo {author} {\bibfnamefont
			{H.}~\bibnamefont {Scott}}, \bibinfo {author} {\bibfnamefont
			{J.}~\bibnamefont {Vorberger}}, \bibinfo {author} {\bibfnamefont
			{D.}~\bibnamefont {Chapman}}, \bibinfo {author} {\bibfnamefont
			{D.}~\bibnamefont {Gericke}}, \bibinfo {author} {\bibfnamefont
			{B.}~\bibnamefont {Mattern}}, \bibinfo {author} {\bibfnamefont
			{G.}~\bibnamefont {Seidler}}, \bibinfo {author} {\bibfnamefont
			{G.}~\bibnamefont {Gregori}}, \bibinfo {author} {\bibfnamefont
			{R.}~\bibnamefont {Falcone}},\ and\ \bibinfo {author} {\bibfnamefont
			{S.}~\bibnamefont {Glenzer}},\ }\bibfield  {title} {\bibinfo {title}
		{Observations of continuum depression in warm dense matter with x-ray
			{T}homson scattering},\ }\href
	{https://doi.org/10.1103/PhysRevLett.112.145004} {\bibfield  {journal}
		{\bibinfo  {journal} {Phys. Rev. Lett.}\ }\textbf {\bibinfo {volume} {112}},\
		\bibinfo {pages} {145004} (\bibinfo {year} {2014})}\BibitemShut {NoStop}%
	\bibitem [{\citenamefont {Ciricosta}\ \emph {et~al.}(2016)\citenamefont
		{Ciricosta}, \citenamefont {Vinko}, \citenamefont {Barbrel}, \citenamefont
		{Rackstraw}, \citenamefont {Preston}, \citenamefont {Burian}, \citenamefont
		{Chalupsk{\'y}}, \citenamefont {Cho}, \citenamefont {Chung}, \citenamefont
		{Dakovski}, \citenamefont {Engelhorn}, \citenamefont {H{\'a}jkov{\'a}},
		\citenamefont {Heimann}, \citenamefont {Holmes}, \citenamefont {Juha},
		\citenamefont {Krzywinski}, \citenamefont {Lee}, \citenamefont {Toleikis},
		\citenamefont {Turner}, \citenamefont {Zastrau},\ and\ \citenamefont
		{Wark}}]{Ciricosta16a}%
	\BibitemOpen
	\bibfield  {author} {\bibinfo {author} {\bibfnamefont {O.}~\bibnamefont
			{Ciricosta}}, \bibinfo {author} {\bibfnamefont {S.~M.}\ \bibnamefont
			{Vinko}}, \bibinfo {author} {\bibfnamefont {B.}~\bibnamefont {Barbrel}},
		\bibinfo {author} {\bibfnamefont {D.~S.}\ \bibnamefont {Rackstraw}}, \bibinfo
		{author} {\bibfnamefont {T.~R.}\ \bibnamefont {Preston}}, \bibinfo {author}
		{\bibfnamefont {T.}~\bibnamefont {Burian}}, \bibinfo {author} {\bibfnamefont
			{J.}~\bibnamefont {Chalupsk{\'y}}}, \bibinfo {author} {\bibfnamefont {B.~I.}\
			\bibnamefont {Cho}}, \bibinfo {author} {\bibfnamefont {H.~K.}\ \bibnamefont
			{Chung}}, \bibinfo {author} {\bibfnamefont {G.~L.}\ \bibnamefont {Dakovski}},
		\bibinfo {author} {\bibfnamefont {K.}~\bibnamefont {Engelhorn}}, \bibinfo
		{author} {\bibfnamefont {V.}~\bibnamefont {H{\'a}jkov{\'a}}}, \bibinfo
		{author} {\bibfnamefont {P.}~\bibnamefont {Heimann}}, \bibinfo {author}
		{\bibfnamefont {M.}~\bibnamefont {Holmes}}, \bibinfo {author} {\bibfnamefont
			{L.}~\bibnamefont {Juha}}, \bibinfo {author} {\bibfnamefont {J.}~\bibnamefont
			{Krzywinski}}, \bibinfo {author} {\bibfnamefont {R.~W.}\ \bibnamefont {Lee}},
		\bibinfo {author} {\bibfnamefont {S.}~\bibnamefont {Toleikis}}, \bibinfo
		{author} {\bibfnamefont {J.~J.}\ \bibnamefont {Turner}}, \bibinfo {author}
		{\bibfnamefont {U.}~\bibnamefont {Zastrau}},\ and\ \bibinfo {author}
		{\bibfnamefont {J.~S.}\ \bibnamefont {Wark}},\ }\bibfield  {title} {\bibinfo
		{title} {Measurements of continuum lowering in solid-density plasmas created
			from elements and compounds},\ }\href {https://doi.org/10.1038/ncomms11713}
	{\bibfield  {journal} {\bibinfo  {journal} {Nat. Commun.}\ }\textbf {\bibinfo
			{volume} {7}},\ \bibinfo {pages} {11713} (\bibinfo {year}
		{2016})}\BibitemShut {NoStop}%
	\bibitem [{\citenamefont {Kraus}\ \emph {et~al.}(2019)\citenamefont {Kraus},
		\citenamefont {Bachmann}, \citenamefont {Barbrel}, \citenamefont {Falcone},
		\citenamefont {Fletcher}, \citenamefont {Frydrych}, \citenamefont {Gamboa},
		\citenamefont {Gauthier}, \citenamefont {Gericke}, \citenamefont {Glenzer},
		\citenamefont {G{\"o}de}, \citenamefont {Granados}, \citenamefont {Hartley},
		\citenamefont {Helfrich}, \citenamefont {Lee}, \citenamefont {Nagler},
		\citenamefont {Ravasio}, \citenamefont {Schumaker}, \citenamefont
		{Vorberger},\ and\ \citenamefont {D{\"o}ppner}}]{Kraus19}%
	\BibitemOpen
	\bibfield  {author} {\bibinfo {author} {\bibfnamefont {D.}~\bibnamefont
			{Kraus}}, \bibinfo {author} {\bibfnamefont {B.}~\bibnamefont {Bachmann}},
		\bibinfo {author} {\bibfnamefont {B.}~\bibnamefont {Barbrel}}, \bibinfo
		{author} {\bibfnamefont {R.~W.}\ \bibnamefont {Falcone}}, \bibinfo {author}
		{\bibfnamefont {L.~B.}\ \bibnamefont {Fletcher}}, \bibinfo {author}
		{\bibfnamefont {S.}~\bibnamefont {Frydrych}}, \bibinfo {author}
		{\bibfnamefont {E.~J.}\ \bibnamefont {Gamboa}}, \bibinfo {author}
		{\bibfnamefont {M.}~\bibnamefont {Gauthier}}, \bibinfo {author}
		{\bibfnamefont {D.~O.}\ \bibnamefont {Gericke}}, \bibinfo {author}
		{\bibfnamefont {S.~H.}\ \bibnamefont {Glenzer}}, \bibinfo {author}
		{\bibfnamefont {S.}~\bibnamefont {G{\"o}de}}, \bibinfo {author}
		{\bibfnamefont {E.}~\bibnamefont {Granados}}, \bibinfo {author}
		{\bibfnamefont {N.~J.}\ \bibnamefont {Hartley}}, \bibinfo {author}
		{\bibfnamefont {J.}~\bibnamefont {Helfrich}}, \bibinfo {author}
		{\bibfnamefont {H.~J.}\ \bibnamefont {Lee}}, \bibinfo {author} {\bibfnamefont
			{B.}~\bibnamefont {Nagler}}, \bibinfo {author} {\bibfnamefont
			{A.}~\bibnamefont {Ravasio}}, \bibinfo {author} {\bibfnamefont
			{W.}~\bibnamefont {Schumaker}}, \bibinfo {author} {\bibfnamefont
			{J.}~\bibnamefont {Vorberger}},\ and\ \bibinfo {author} {\bibfnamefont
			{T.}~\bibnamefont {D{\"o}ppner}},\ }\bibfield  {title} {\bibinfo {title}
		{Characterizing the ionization potential depression in dense carbon plasmas
			with high-precision spectrally resolved x-ray scattering},\ }\href
	{https://doi.org/10.1088/1361-6587/aadd6c} {\bibfield  {journal} {\bibinfo
			{journal} {Plasma Phys. Control. Fusion}\ }\textbf {\bibinfo {volume} {61}},\
		\bibinfo {pages} {014015} (\bibinfo {year} {2019})}\BibitemShut {NoStop}%
	\bibitem [{\citenamefont {Kasim}\ \emph {et~al.}(2018)\citenamefont {Kasim},
		\citenamefont {Wark},\ and\ \citenamefont {Vinko}}]{Kasim:2018aa}%
	\BibitemOpen
	\bibfield  {author} {\bibinfo {author} {\bibfnamefont {M.~F.}\ \bibnamefont
			{Kasim}}, \bibinfo {author} {\bibfnamefont {J.~S.}\ \bibnamefont {Wark}},\
		and\ \bibinfo {author} {\bibfnamefont {S.~M.}\ \bibnamefont {Vinko}},\
	}\bibfield  {title} {\bibinfo {title} {Validating continuum lowering models
			via multi-wavelength measurements of integrated x-ray emission},\ }\href
	{https://doi.org/10.1038/s41598-018-24410-2} {\bibfield  {journal} {\bibinfo
			{journal} {Sci. Rep.}\ }\textbf {\bibinfo {volume} {8}},\ \bibinfo {pages}
		{6276} (\bibinfo {year} {2018})}\BibitemShut {NoStop}%
	\bibitem [{\citenamefont {Crowley}(2014)}]{Crowley2014}%
	\BibitemOpen
	\bibfield  {author} {\bibinfo {author} {\bibfnamefont {B.}~\bibnamefont
			{Crowley}},\ }\bibfield  {title} {\bibinfo {title} {Continuum lowering -- a
			new perspective},\ }\href
	{https://doi.org/https://doi.org/10.1016/j.hedp.2014.04.003} {\bibfield
		{journal} {\bibinfo  {journal} {High Energy Density Phys.}\ }\textbf
		{\bibinfo {volume} {13}},\ \bibinfo {pages} {84 } (\bibinfo {year}
		{2014})}\BibitemShut {NoStop}%
	\bibitem [{\citenamefont {Iglesias}(2014)}]{Iglesias14}%
	\BibitemOpen
	\bibfield  {author} {\bibinfo {author} {\bibfnamefont {C.~A.}\ \bibnamefont
			{Iglesias}},\ }\bibfield  {title} {\bibinfo {title} {A plea for a
			reexamination of ionization potential depression measurements},\ }\href
	{https://doi.org/10.1016/j.hedp.2014.04.002} {\bibfield  {journal} {\bibinfo
			{journal} {High Energy Density Phys.}\ }\textbf {\bibinfo {volume} {12}},\
		\bibinfo {pages} {5} (\bibinfo {year} {2014})}\BibitemShut {NoStop}%
	\bibitem [{\citenamefont {Calisti}\ \emph
		{et~al.}(2015{\natexlab{a}})\citenamefont {Calisti}, \citenamefont {Ferri},\
		and\ \citenamefont {Talin}}]{Calisti2015}%
	\BibitemOpen
	\bibfield  {author} {\bibinfo {author} {\bibfnamefont {A.}~\bibnamefont
			{Calisti}}, \bibinfo {author} {\bibfnamefont {S.}~\bibnamefont {Ferri}},\
		and\ \bibinfo {author} {\bibfnamefont {B.}~\bibnamefont {Talin}},\ }\bibfield
	{title} {\bibinfo {title} {Ionization potential depression for non
			equilibrated aluminum plasmas},\ }\href
	{https://doi.org/10.1088/0953-4075/48/22/224003} {\bibfield  {journal}
		{\bibinfo  {journal} {J. Phys. B}\ }\textbf {\bibinfo {volume} {48}},\
		\bibinfo {pages} {224003} (\bibinfo {year} {2015}{\natexlab{a}})}\BibitemShut
	{NoStop}%
	\bibitem [{\citenamefont {Calisti}\ \emph
		{et~al.}(2015{\natexlab{b}})\citenamefont {Calisti}, \citenamefont {Ferri},\
		and\ \citenamefont {Talin}}]{Calisti15a}%
	\BibitemOpen
	\bibfield  {author} {\bibinfo {author} {\bibfnamefont {A.}~\bibnamefont
			{Calisti}}, \bibinfo {author} {\bibfnamefont {S.}~\bibnamefont {Ferri}},\
		and\ \bibinfo {author} {\bibfnamefont {B.}~\bibnamefont {Talin}},\ }\bibfield
	{title} {\bibinfo {title} {Ionization potential depression in hot dense
			plasmas through a pure classical model},\ }\href
	{https://doi.org/10.1002/ctpp.201400087} {\bibfield  {journal} {\bibinfo
			{journal} {Contrib. Plasma Phys.}\ }\textbf {\bibinfo {volume} {55}},\
		\bibinfo {pages} {360} (\bibinfo {year} {2015}{\natexlab{b}})}\BibitemShut
	{NoStop}%
	\bibitem [{\citenamefont {Stransky}(2016)}]{Stransky2016}%
	\BibitemOpen
	\bibfield  {author} {\bibinfo {author} {\bibfnamefont {M.}~\bibnamefont
			{Stransky}},\ }\bibfield  {title} {\bibinfo {title} {Monte carlo simulations
			of ionization potential depression in dense plasmas},\ }\href
	{https://doi.org/10.1063/1.4940313} {\bibfield  {journal} {\bibinfo
			{journal} {Phys. Plasmas}\ }\textbf {\bibinfo {volume} {23}},\ \bibinfo
		{pages} {012708} (\bibinfo {year} {2016})}\BibitemShut {NoStop}%
	\bibitem [{\citenamefont {Lin}\ \emph {et~al.}(2017)\citenamefont {Lin},
		\citenamefont {R\"opke}, \citenamefont {Kraeft},\ and\ \citenamefont
		{Reinholz}}]{Lin2017}%
	\BibitemOpen
	\bibfield  {author} {\bibinfo {author} {\bibfnamefont {C.}~\bibnamefont
			{Lin}}, \bibinfo {author} {\bibfnamefont {G.}~\bibnamefont {R\"opke}},
		\bibinfo {author} {\bibfnamefont {W.-D.}\ \bibnamefont {Kraeft}},\ and\
		\bibinfo {author} {\bibfnamefont {H.}~\bibnamefont {Reinholz}},\ }\bibfield
	{title} {\bibinfo {title} {Ionization-potential depression and dynamical
			structure factor in dense plasmas},\ }\href
	{https://doi.org/10.1103/PhysRevE.96.013202} {\bibfield  {journal} {\bibinfo
			{journal} {Phys. Rev. E}\ }\textbf {\bibinfo {volume} {96}},\ \bibinfo
		{pages} {013202} (\bibinfo {year} {2017})}\BibitemShut {NoStop}%
	\bibitem [{\citenamefont {R\"opke}\ \emph {et~al.}(2019)\citenamefont
		{R\"opke}, \citenamefont {Blaschke}, \citenamefont {D\"oppner}, \citenamefont
		{Lin}, \citenamefont {Kraeft}, \citenamefont {Redmer},\ and\ \citenamefont
		{Reinholz}}]{Roepke2019}%
	\BibitemOpen
	\bibfield  {author} {\bibinfo {author} {\bibfnamefont {G.}~\bibnamefont
			{R\"opke}}, \bibinfo {author} {\bibfnamefont {D.}~\bibnamefont {Blaschke}},
		\bibinfo {author} {\bibfnamefont {T.}~\bibnamefont {D\"oppner}}, \bibinfo
		{author} {\bibfnamefont {C.}~\bibnamefont {Lin}}, \bibinfo {author}
		{\bibfnamefont {W.-D.}\ \bibnamefont {Kraeft}}, \bibinfo {author}
		{\bibfnamefont {R.}~\bibnamefont {Redmer}},\ and\ \bibinfo {author}
		{\bibfnamefont {H.}~\bibnamefont {Reinholz}},\ }\bibfield  {title} {\bibinfo
		{title} {Ionization potential depression and {P}auli blocking in degenerate
			plasmas at extreme densities},\ }\href
	{https://doi.org/10.1103/PhysRevE.99.033201} {\bibfield  {journal} {\bibinfo
			{journal} {Phys. Rev. E}\ }\textbf {\bibinfo {volume} {99}},\ \bibinfo
		{pages} {033201} (\bibinfo {year} {2019})}\BibitemShut {NoStop}%
	\bibitem [{\citenamefont {Rosmej}(2018)}]{Rosmej18}%
	\BibitemOpen
	\bibfield  {author} {\bibinfo {author} {\bibfnamefont {F.~B.}\ \bibnamefont
			{Rosmej}},\ }\bibfield  {title} {\bibinfo {title} {Ionization potential
			depression in an atomic-solid-plasma picture},\ }\href
	{https://doi.org/10.1088/1361-6455/aab80f} {\bibfield  {journal} {\bibinfo
			{journal} {J. Phys. B: At. Mol. Opt. Phys.}\ }\textbf {\bibinfo {volume}
			{51}},\ \bibinfo {pages} {09LT01} (\bibinfo {year} {2018})}\BibitemShut
	{NoStop}%
	\bibitem [{\citenamefont {Pain}(2019)}]{Pain2019}%
	\BibitemOpen
	\bibfield  {author} {\bibinfo {author} {\bibfnamefont {J.-C.}\ \bibnamefont
			{Pain}},\ }\bibfield  {title} {\bibinfo {title} {On the {Li}-{Rosmej}
			analytical formula for energy level shifts in dense plasmas},\ }\href
	{https://doi.org/10.1016/j.hedp.2019.03.003} {\bibfield  {journal} {\bibinfo
			{journal} {High Energy Density Phys.}\ }\textbf {\bibinfo {volume} {31}},\
		\bibinfo {pages} {99} (\bibinfo {year} {2019})}\BibitemShut {NoStop}%
	\bibitem [{\citenamefont {Li}\ \emph {et~al.}(2019)\citenamefont {Li},
		\citenamefont {Rosmej}, \citenamefont {Lisitsa},\ and\ \citenamefont
		{Astapenko}}]{Li:2019aa}%
	\BibitemOpen
	\bibfield  {author} {\bibinfo {author} {\bibfnamefont {X.}~\bibnamefont
			{Li}}, \bibinfo {author} {\bibfnamefont {F.~B.}\ \bibnamefont {Rosmej}},
		\bibinfo {author} {\bibfnamefont {V.~S.}\ \bibnamefont {Lisitsa}},\ and\
		\bibinfo {author} {\bibfnamefont {V.~A.}\ \bibnamefont {Astapenko}},\
	}\bibfield  {title} {\bibinfo {title} {An analytical plasma screening
			potential based on the self-consistent-field ion-sphere model},\ }\href
	{https://doi.org/10.1063/1.5055689} {\bibfield  {journal} {\bibinfo
			{journal} {Phys. Plasmas}\ }\textbf {\bibinfo {volume} {26}},\ \bibinfo
		{pages} {033301} (\bibinfo {year} {2019})}\BibitemShut {NoStop}%
	\bibitem [{\citenamefont {Son}\ \emph {et~al.}(2014)\citenamefont {Son},
		\citenamefont {Thiele}, \citenamefont {Jurek}, \citenamefont {Ziaja},\ and\
		\citenamefont {Santra}}]{xatom2014AA}%
	\BibitemOpen
	\bibfield  {author} {\bibinfo {author} {\bibfnamefont {S.-K.}\ \bibnamefont
			{Son}}, \bibinfo {author} {\bibfnamefont {R.}~\bibnamefont {Thiele}},
		\bibinfo {author} {\bibfnamefont {Z.}~\bibnamefont {Jurek}}, \bibinfo
		{author} {\bibfnamefont {B.}~\bibnamefont {Ziaja}},\ and\ \bibinfo {author}
		{\bibfnamefont {R.}~\bibnamefont {Santra}},\ }\bibfield  {title} {\bibinfo
		{title} {Quantum-mechanical calculation of ionization-potential lowering in
			dense plasmas},\ }\href {https://doi.org/10.1103/PhysRevX.4.031004}
	{\bibfield  {journal} {\bibinfo  {journal} {Phys. Rev. X}\ }\textbf {\bibinfo
			{volume} {4}},\ \bibinfo {pages} {031004} (\bibinfo {year}
		{2014})}\BibitemShut {NoStop}%
	\bibitem [{\citenamefont {Vinko}\ \emph {et~al.}(2014)\citenamefont {Vinko},
		\citenamefont {Ciricosta},\ and\ \citenamefont {Wark}}]{Vinko2014}%
	\BibitemOpen
	\bibfield  {author} {\bibinfo {author} {\bibfnamefont {S.~M.}\ \bibnamefont
			{Vinko}}, \bibinfo {author} {\bibfnamefont {O.}~\bibnamefont {Ciricosta}},\
		and\ \bibinfo {author} {\bibfnamefont {J.~S.}\ \bibnamefont {Wark}},\
	}\bibfield  {title} {\bibinfo {title} {Density functional theory calculations
			of continuum lowering in strongly coupled plasmas},\ }\href
	{https://doi.org/10.1038/ncomms4533} {\bibfield  {journal} {\bibinfo
			{journal} {Nat. Commun.}\ }\textbf {\bibinfo {volume} {5}},\ \bibinfo {pages}
		{3533} (\bibinfo {year} {2014})}\BibitemShut {NoStop}%
	\bibitem [{\citenamefont {Bekx}\ \emph {et~al.}(2020)\citenamefont {Bekx},
		\citenamefont {Son}, \citenamefont {Ziaja},\ and\ \citenamefont
		{Santra}}]{Bekx2020}%
	\BibitemOpen
	\bibfield  {author} {\bibinfo {author} {\bibfnamefont {J.~J.}\ \bibnamefont
			{Bekx}}, \bibinfo {author} {\bibfnamefont {S.-K.}\ \bibnamefont {Son}},
		\bibinfo {author} {\bibfnamefont {B.}~\bibnamefont {Ziaja}},\ and\ \bibinfo
		{author} {\bibfnamefont {R.}~\bibnamefont {Santra}},\ }\bibfield  {title}
	{\bibinfo {title} {Electronic-structure calculations for nonisothermal warm
			dense matter},\ }\href {https://doi.org/10.1103/PhysRevResearch.2.033061}
	{\bibfield  {journal} {\bibinfo  {journal} {Phys. Rev. Res.}\ }\textbf
		{\bibinfo {volume} {2}},\ \bibinfo {pages} {033061} (\bibinfo {year}
		{2020})}\BibitemShut {NoStop}%
	\bibitem [{\citenamefont {Zeng}\ \emph {et~al.}(2020)\citenamefont {Zeng},
		\citenamefont {Li}, \citenamefont {Gao},\ and\ \citenamefont
		{Yuan}}]{Zeng2020}%
	\BibitemOpen
	\bibfield  {author} {\bibinfo {author} {\bibfnamefont {J.}~\bibnamefont
			{Zeng}}, \bibinfo {author} {\bibfnamefont {Y.}~\bibnamefont {Li}}, \bibinfo
		{author} {\bibfnamefont {C.}~\bibnamefont {Gao}},\ and\ \bibinfo {author}
		{\bibfnamefont {J.}~\bibnamefont {Yuan}},\ }\bibfield  {title} {\bibinfo
		{title} {Screening potential and continuum lowering in a dense plasma under
			solar-interior conditions},\ }\href
	{https://doi.org/10.1051/0004-6361/201937235} {\bibfield  {journal} {\bibinfo
			{journal} {Astron. Astrophys.}\ }\textbf {\bibinfo {volume} {634}},\
		\bibinfo {pages} {A117} (\bibinfo {year} {2020})}\BibitemShut {NoStop}%
	\bibitem [{\citenamefont {Driver}\ \emph {et~al.}(2018)\citenamefont {Driver},
		\citenamefont {Soubiran},\ and\ \citenamefont {Militzer}}]{Driver2018}%
	\BibitemOpen
	\bibfield  {author} {\bibinfo {author} {\bibfnamefont {K.~P.}\ \bibnamefont
			{Driver}}, \bibinfo {author} {\bibfnamefont {F.}~\bibnamefont {Soubiran}},\
		and\ \bibinfo {author} {\bibfnamefont {B.}~\bibnamefont {Militzer}},\
	}\bibfield  {title} {\bibinfo {title} {Path integral monte carlo simulations
			of warm dense aluminum},\ }\href {https://doi.org/10.1103/PhysRevE.97.063207}
	{\bibfield  {journal} {\bibinfo  {journal} {Phys. Rev. E}\ }\textbf {\bibinfo
			{volume} {97}},\ \bibinfo {pages} {063207} (\bibinfo {year}
		{2018})}\BibitemShut {NoStop}%
	\bibitem [{\citenamefont {Hu}(2017)}]{Hu2017}%
	\BibitemOpen
	\bibfield  {author} {\bibinfo {author} {\bibfnamefont {S.~X.}\ \bibnamefont
			{Hu}},\ }\bibfield  {title} {\bibinfo {title} {Continuum lowering and
			{F}ermi-surface rising in strongly coupled and degenerate plasmas},\ }\href
	{https://doi.org/10.1103/PhysRevLett.119.065001} {\bibfield  {journal}
		{\bibinfo  {journal} {Phys. Rev. Lett.}\ }\textbf {\bibinfo {volume} {119}},\
		\bibinfo {pages} {065001} (\bibinfo {year} {2017})}\BibitemShut {NoStop}%
	\bibitem [{\citenamefont {Hau-Riege}(2013)}]{Hau-Riege13}%
	\BibitemOpen
	\bibfield  {author} {\bibinfo {author} {\bibfnamefont {S.~P.}\ \bibnamefont
			{Hau-Riege}},\ }\bibfield  {title} {\bibinfo {title} {Nonequilibrium electron
			dynamics in materials driven by high-intensity x-ray pulses},\ }\href
	{https://doi.org/10.1103/PhysRevE.87.053102} {\bibfield  {journal} {\bibinfo
			{journal} {Phys. Rev. E}\ }\textbf {\bibinfo {volume} {87}},\ \bibinfo
		{pages} {053102} (\bibinfo {year} {2013})}\BibitemShut {NoStop}%
	\bibitem [{\citenamefont {Makita}\ \emph {et~al.}(2019)\citenamefont {Makita},
		\citenamefont {Vartiainen}, \citenamefont {Mohacsi}, \citenamefont {Caleman},
		\citenamefont {Diaz}, \citenamefont {J{\"o}nsson}, \citenamefont
		{Jurani{\'c}}, \citenamefont {Medvedev}, \citenamefont {Meents},
		\citenamefont {Mozzanica}, \citenamefont {Opara}, \citenamefont {Padeste},
		\citenamefont {Panneels}, \citenamefont {Saxena}, \citenamefont {Sikorski},
		\citenamefont {Song}, \citenamefont {Vera}, \citenamefont {Willmott},
		\citenamefont {Beaud}, \citenamefont {Milne}, \citenamefont {Ziaja-Motyka},\
		and\ \citenamefont {David}}]{Makita19}%
	\BibitemOpen
	\bibfield  {author} {\bibinfo {author} {\bibfnamefont {M.}~\bibnamefont
			{Makita}}, \bibinfo {author} {\bibfnamefont {I.}~\bibnamefont {Vartiainen}},
		\bibinfo {author} {\bibfnamefont {I.}~\bibnamefont {Mohacsi}}, \bibinfo
		{author} {\bibfnamefont {C.}~\bibnamefont {Caleman}}, \bibinfo {author}
		{\bibfnamefont {A.}~\bibnamefont {Diaz}}, \bibinfo {author} {\bibfnamefont
			{H.~O.}\ \bibnamefont {J{\"o}nsson}}, \bibinfo {author} {\bibfnamefont
			{P.}~\bibnamefont {Jurani{\'c}}}, \bibinfo {author} {\bibfnamefont
			{N.}~\bibnamefont {Medvedev}}, \bibinfo {author} {\bibfnamefont
			{A.}~\bibnamefont {Meents}}, \bibinfo {author} {\bibfnamefont
			{A.}~\bibnamefont {Mozzanica}}, \bibinfo {author} {\bibfnamefont {N.~L.}\
			\bibnamefont {Opara}}, \bibinfo {author} {\bibfnamefont {C.}~\bibnamefont
			{Padeste}}, \bibinfo {author} {\bibfnamefont {V.}~\bibnamefont {Panneels}},
		\bibinfo {author} {\bibfnamefont {V.}~\bibnamefont {Saxena}}, \bibinfo
		{author} {\bibfnamefont {M.}~\bibnamefont {Sikorski}}, \bibinfo {author}
		{\bibfnamefont {S.}~\bibnamefont {Song}}, \bibinfo {author} {\bibfnamefont
			{L.}~\bibnamefont {Vera}}, \bibinfo {author} {\bibfnamefont {P.~R.}\
			\bibnamefont {Willmott}}, \bibinfo {author} {\bibfnamefont {P.}~\bibnamefont
			{Beaud}}, \bibinfo {author} {\bibfnamefont {C.~J.}\ \bibnamefont {Milne}},
		\bibinfo {author} {\bibfnamefont {B.}~\bibnamefont {Ziaja-Motyka}},\ and\
		\bibinfo {author} {\bibfnamefont {C.}~\bibnamefont {David}},\ }\bibfield
	{title} {\bibinfo {title} {Femtosecond phase-transition in hard x-ray excited
			bismuth},\ }\href {https://doi.org/10.1038/s41598-018-36216-3} {\bibfield
		{journal} {\bibinfo  {journal} {Sci. Rep.}\ }\textbf {\bibinfo {volume}
			{9}},\ \bibinfo {pages} {602} (\bibinfo {year} {2019})}\BibitemShut {NoStop}%
	\bibitem [{\citenamefont {Hartley}\ \emph {et~al.}(2019)\citenamefont
		{Hartley}, \citenamefont {Grenzer}, \citenamefont {Lu}, \citenamefont
		{Huang}, \citenamefont {Inubushi}, \citenamefont {Kamimura}, \citenamefont
		{Katagiri}, \citenamefont {Kodama}, \citenamefont {Kon}, \citenamefont
		{Lipp}, \citenamefont {Makita}, \citenamefont {Matsuoka}, \citenamefont
		{Medvedev}, \citenamefont {Nakajima}, \citenamefont {Ozaki}, \citenamefont
		{Pikuz}, \citenamefont {Rode}, \citenamefont {Rohatsch}, \citenamefont
		{Sagae}, \citenamefont {Schuster}, \citenamefont {Tono}, \citenamefont
		{Vorberger}, \citenamefont {Yabuuchi},\ and\ \citenamefont
		{Kraus}}]{Hartley19}%
	\BibitemOpen
	\bibfield  {author} {\bibinfo {author} {\bibfnamefont {N.~J.}\ \bibnamefont
			{Hartley}}, \bibinfo {author} {\bibfnamefont {J.}~\bibnamefont {Grenzer}},
		\bibinfo {author} {\bibfnamefont {W.}~\bibnamefont {Lu}}, \bibinfo {author}
		{\bibfnamefont {L.~G.}\ \bibnamefont {Huang}}, \bibinfo {author}
		{\bibfnamefont {Y.}~\bibnamefont {Inubushi}}, \bibinfo {author}
		{\bibfnamefont {N.}~\bibnamefont {Kamimura}}, \bibinfo {author}
		{\bibfnamefont {K.}~\bibnamefont {Katagiri}}, \bibinfo {author}
		{\bibfnamefont {R.}~\bibnamefont {Kodama}}, \bibinfo {author} {\bibfnamefont
			{A.}~\bibnamefont {Kon}}, \bibinfo {author} {\bibfnamefont {V.}~\bibnamefont
			{Lipp}}, \bibinfo {author} {\bibfnamefont {M.}~\bibnamefont {Makita}},
		\bibinfo {author} {\bibfnamefont {T.}~\bibnamefont {Matsuoka}}, \bibinfo
		{author} {\bibfnamefont {N.}~\bibnamefont {Medvedev}}, \bibinfo {author}
		{\bibfnamefont {S.}~\bibnamefont {Nakajima}}, \bibinfo {author}
		{\bibfnamefont {N.}~\bibnamefont {Ozaki}}, \bibinfo {author} {\bibfnamefont
			{T.}~\bibnamefont {Pikuz}}, \bibinfo {author} {\bibfnamefont {A.~V.}\
			\bibnamefont {Rode}}, \bibinfo {author} {\bibfnamefont {K.}~\bibnamefont
			{Rohatsch}}, \bibinfo {author} {\bibfnamefont {D.}~\bibnamefont {Sagae}},
		\bibinfo {author} {\bibfnamefont {A.~K.}\ \bibnamefont {Schuster}}, \bibinfo
		{author} {\bibfnamefont {K.}~\bibnamefont {Tono}}, \bibinfo {author}
		{\bibfnamefont {J.}~\bibnamefont {Vorberger}}, \bibinfo {author}
		{\bibfnamefont {T.}~\bibnamefont {Yabuuchi}},\ and\ \bibinfo {author}
		{\bibfnamefont {D.}~\bibnamefont {Kraus}},\ }\bibfield  {title} {\bibinfo
		{title} {Ultrafast anisotropic disordering in graphite driven by intense hard
			x-ray pulses},\ }\href {https://doi.org/10.1016/j.hedp.2019.05.002}
	{\bibfield  {journal} {\bibinfo  {journal} {High Energy Density Phys.}\
		}\textbf {\bibinfo {volume} {32}},\ \bibinfo {pages} {63} (\bibinfo {year}
		{2019})}\BibitemShut {NoStop}%
	\bibitem [{\citenamefont {Murphy}\ \emph {et~al.}(2014)\citenamefont {Murphy},
		\citenamefont {Osipov}, \citenamefont {Jurek}, \citenamefont {Fang},
		\citenamefont {Son}, \citenamefont {Mucke}, \citenamefont {Eland},
		\citenamefont {Zha~unerchyk}, \citenamefont {Feifel}, \citenamefont {Avaldi},
		\citenamefont {Bolognesi}, \citenamefont {Bostedt}, \citenamefont {Bozek},
		\citenamefont {Grilj}, \citenamefont {Frasinski}, \citenamefont {Glownia},
		\citenamefont {Ha}, \citenamefont {Hoffmann}, \citenamefont {Kukk},
		\citenamefont {McFarland}, \citenamefont {Miron}, \citenamefont {Sistrunk},
		\citenamefont {Squibb}, \citenamefont {Ueda}, \citenamefont {Santra},\ and\
		\citenamefont {Berrah}}]{Murphy2014}%
	\BibitemOpen
	\bibfield  {author} {\bibinfo {author} {\bibfnamefont {B.~F.}\ \bibnamefont
			{Murphy}}, \bibinfo {author} {\bibfnamefont {T.}~\bibnamefont {Osipov}},
		\bibinfo {author} {\bibfnamefont {Z.}~\bibnamefont {Jurek}}, \bibinfo
		{author} {\bibfnamefont {L.}~\bibnamefont {Fang}}, \bibinfo {author}
		{\bibfnamefont {S.-K.}\ \bibnamefont {Son}}, \bibinfo {author} {\bibfnamefont
			{M.}~\bibnamefont {Mucke}}, \bibinfo {author} {\bibfnamefont
			{J.}~\bibnamefont {Eland}}, \bibinfo {author} {\bibfnamefont
			{V.}~\bibnamefont {Zha~unerchyk}}, \bibinfo {author} {\bibfnamefont
			{R.}~\bibnamefont {Feifel}}, \bibinfo {author} {\bibfnamefont
			{L.}~\bibnamefont {Avaldi}}, \bibinfo {author} {\bibfnamefont
			{P.}~\bibnamefont {Bolognesi}}, \bibinfo {author} {\bibfnamefont
			{C.}~\bibnamefont {Bostedt}}, \bibinfo {author} {\bibfnamefont {J.~D.}\
			\bibnamefont {Bozek}}, \bibinfo {author} {\bibfnamefont {M.}~\bibnamefont
			{Grilj}, \bibfnamefont {J.~an d~Guehr}}, \bibinfo {author} {\bibfnamefont
			{L.~J.}\ \bibnamefont {Frasinski}}, \bibinfo {author} {\bibfnamefont
			{J.}~\bibnamefont {Glownia}}, \bibinfo {author} {\bibfnamefont {D.~T.}\
			\bibnamefont {Ha}}, \bibinfo {author} {\bibfnamefont {K.}~\bibnamefont
			{Hoffmann}}, \bibinfo {author} {\bibfnamefont {E.}~\bibnamefont {Kukk}},
		\bibinfo {author} {\bibfnamefont {B.~K.}\ \bibnamefont {McFarland}}, \bibinfo
		{author} {\bibfnamefont {C.}~\bibnamefont {Miron}}, \bibinfo {author}
		{\bibfnamefont {E.}~\bibnamefont {Sistrunk}}, \bibinfo {author}
		{\bibfnamefont {R.~J.}\ \bibnamefont {Squibb}}, \bibinfo {author}
		{\bibfnamefont {K.}~\bibnamefont {Ueda}}, \bibinfo {author} {\bibfnamefont
			{R.}~\bibnamefont {Santra}},\ and\ \bibinfo {author} {\bibfnamefont
			{N.}~\bibnamefont {Berrah}},\ }\bibfield  {title} {\bibinfo {title}
		{Femtosecond x-ray-induced explosion of {C}$_{60}$ at extreme intensity},\
	}\href {https://doi.org/10.1038/ncomms5281} {\bibfield  {journal} {\bibinfo
			{journal} {Nat. Commun.}\ }\textbf {\bibinfo {volume} {5}},\ \bibinfo {pages}
		{4281} (\bibinfo {year} {2014})}\BibitemShut {NoStop}%
	\bibitem [{\citenamefont {Jurek}\ \emph {et~al.}(2016)\citenamefont {Jurek},
		\citenamefont {Son}, \citenamefont {Ziaja},\ and\ \citenamefont
		{Santra}}]{Zoltan2016}%
	\BibitemOpen
	\bibfield  {author} {\bibinfo {author} {\bibfnamefont {Z.}~\bibnamefont
			{Jurek}}, \bibinfo {author} {\bibfnamefont {S.-K.}\ \bibnamefont {Son}},
		\bibinfo {author} {\bibfnamefont {B.}~\bibnamefont {Ziaja}},\ and\ \bibinfo
		{author} {\bibfnamefont {R.}~\bibnamefont {Santra}},\ }\bibfield  {title}
	{\bibinfo {title} {{{\it XMDYN} and {\it XATOM}: versatile simulation tools
				for quantitative modeling of X-ray free-electron laser induced dynamics of
				matter}},\ }\href {https://doi.org/10.1107/S1600576716006014} {\bibfield
		{journal} {\bibinfo  {journal} {J. Appl. Cryst.}\ }\textbf {\bibinfo {volume}
			{49}},\ \bibinfo {pages} {1048} (\bibinfo {year} {2016})}\BibitemShut
	{NoStop}%
	\bibitem [{\citenamefont {Son}\ \emph {et~al.}(2011)\citenamefont {Son},
		\citenamefont {Young},\ and\ \citenamefont {Santra}}]{xatom2011}%
	\BibitemOpen
	\bibfield  {author} {\bibinfo {author} {\bibfnamefont {S.-K.}\ \bibnamefont
			{Son}}, \bibinfo {author} {\bibfnamefont {L.}~\bibnamefont {Young}},\ and\
		\bibinfo {author} {\bibfnamefont {R.}~\bibnamefont {Santra}},\ }\bibfield
	{title} {\bibinfo {title} {Impact of hollow-atom formation on coherent x-ray
			scattering at high intensity},\ }\href
	{https://doi.org/10.1103/PhysRevA.83.033402} {\bibfield  {journal} {\bibinfo
			{journal} {Phys. Rev. A}\ }\textbf {\bibinfo {volume} {83}},\ \bibinfo
		{pages} {033402} (\bibinfo {year} {2011})}\BibitemShut {NoStop}%
	\bibitem [{\citenamefont {Baczewski}\ \emph {et~al.}(2016)\citenamefont
		{Baczewski}, \citenamefont {Shulenburger}, \citenamefont {Desjarlais},
		\citenamefont {Hansen},\ and\ \citenamefont {Magyar}}]{Baczewski16}%
	\BibitemOpen
	\bibfield  {author} {\bibinfo {author} {\bibfnamefont {A.}~\bibnamefont
			{Baczewski}}, \bibinfo {author} {\bibfnamefont {L.}~\bibnamefont
			{Shulenburger}}, \bibinfo {author} {\bibfnamefont {M.}~\bibnamefont
			{Desjarlais}}, \bibinfo {author} {\bibfnamefont {S.}~\bibnamefont {Hansen}},\
		and\ \bibinfo {author} {\bibfnamefont {R.}~\bibnamefont {Magyar}},\
	}\bibfield  {title} {\bibinfo {title} {X-ray {T}homson scattering in warm
			dense matter without the {C}hihara decomposition},\ }\href
	{https://doi.org/10.1103/PhysRevLett.116.115004} {\bibfield  {journal}
		{\bibinfo  {journal} {Phys. Rev. Lett.}\ }\textbf {\bibinfo {volume} {116}},\
		\bibinfo {pages} {115004} (\bibinfo {year} {2016})}\BibitemShut {NoStop}%
	\bibitem [{\citenamefont {Berrah}\ \emph {et~al.}(2019)\citenamefont {Berrah},
		\citenamefont {Sanchez-Gonzalez}, \citenamefont {Jurek}, \citenamefont
		{Obaid}, \citenamefont {Xiong}, \citenamefont {Squibb}, \citenamefont
		{Osipov}, \citenamefont {Lutman}, \citenamefont {Fang}, \citenamefont
		{Barillot}, \citenamefont {Bozek}, \citenamefont {Cryan}, \citenamefont
		{Wolf}, \citenamefont {Rolles}, \citenamefont {Coffee}, \citenamefont
		{Schnorr}, \citenamefont {Augustin}, \citenamefont {Fukuzawa}, \citenamefont
		{Motomura}, \citenamefont {Niebuhr}, \citenamefont {Frasinski}, \citenamefont
		{Feifel}, \citenamefont {Schulz}, \citenamefont {Toyota}, \citenamefont
		{Son}, \citenamefont {Ueda}, \citenamefont {Pfeifer}, \citenamefont
		{Marangos},\ and\ \citenamefont {Santra}}]{Berrah2019}%
	\BibitemOpen
	\bibfield  {author} {\bibinfo {author} {\bibfnamefont {N.}~\bibnamefont
			{Berrah}}, \bibinfo {author} {\bibfnamefont {A.}~\bibnamefont
			{Sanchez-Gonzalez}}, \bibinfo {author} {\bibfnamefont {Z.}~\bibnamefont
			{Jurek}}, \bibinfo {author} {\bibfnamefont {R.}~\bibnamefont {Obaid}},
		\bibinfo {author} {\bibfnamefont {H.}~\bibnamefont {Xiong}}, \bibinfo
		{author} {\bibfnamefont {R.~J.}\ \bibnamefont {Squibb}}, \bibinfo {author}
		{\bibfnamefont {T.}~\bibnamefont {Osipov}}, \bibinfo {author} {\bibfnamefont
			{A.}~\bibnamefont {Lutman}}, \bibinfo {author} {\bibfnamefont
			{L.}~\bibnamefont {Fang}}, \bibinfo {author} {\bibfnamefont {T.}~\bibnamefont
			{Barillot}}, \bibinfo {author} {\bibfnamefont {J.~D.}\ \bibnamefont {Bozek}},
		\bibinfo {author} {\bibfnamefont {J.}~\bibnamefont {Cryan}}, \bibinfo
		{author} {\bibfnamefont {T.~J.~A.}\ \bibnamefont {Wolf}}, \bibinfo {author}
		{\bibfnamefont {D.}~\bibnamefont {Rolles}}, \bibinfo {author} {\bibfnamefont
			{R.}~\bibnamefont {Coffee}}, \bibinfo {author} {\bibfnamefont
			{K.}~\bibnamefont {Schnorr}}, \bibinfo {author} {\bibfnamefont
			{S.}~\bibnamefont {Augustin}}, \bibinfo {author} {\bibfnamefont
			{H.}~\bibnamefont {Fukuzawa}}, \bibinfo {author} {\bibfnamefont
			{K.}~\bibnamefont {Motomura}}, \bibinfo {author} {\bibfnamefont
			{N.}~\bibnamefont {Niebuhr}}, \bibinfo {author} {\bibfnamefont {L.~J.}\
			\bibnamefont {Frasinski}}, \bibinfo {author} {\bibfnamefont {R.}~\bibnamefont
			{Feifel}}, \bibinfo {author} {\bibfnamefont {C.~P.}\ \bibnamefont {Schulz}},
		\bibinfo {author} {\bibfnamefont {K.}~\bibnamefont {Toyota}}, \bibinfo
		{author} {\bibfnamefont {S.-K.}\ \bibnamefont {Son}}, \bibinfo {author}
		{\bibfnamefont {K.}~\bibnamefont {Ueda}}, \bibinfo {author} {\bibfnamefont
			{T.}~\bibnamefont {Pfeifer}}, \bibinfo {author} {\bibfnamefont {J.~P.}\
			\bibnamefont {Marangos}},\ and\ \bibinfo {author} {\bibfnamefont
			{R.}~\bibnamefont {Santra}},\ }\bibfield  {title} {\bibinfo {title}
		{Femtosecond-resolved observation of the fragmentation of
			buckminsterfullerene following x-ray multiphoton ionization},\ }\href
	{https://doi.org/10.1038/s41567-019-0665-7} {\bibfield  {journal} {\bibinfo
			{journal} {Nat. Phys.}\ }\textbf {\bibinfo {volume} {15}},\ \bibinfo {pages}
		{1279} (\bibinfo {year} {2019})}\BibitemShut {NoStop}%
	\bibitem [{\citenamefont {Tachibana}\ \emph {et~al.}(2015)\citenamefont
		{Tachibana}, \citenamefont {Jurek}, \citenamefont {Fukuzawa}, \citenamefont
		{Motomura}, \citenamefont {Nagaya}, \citenamefont {Wada}, \citenamefont
		{Johnsson}, \citenamefont {Siano}, \citenamefont {Mondal}, \citenamefont
		{Ito}, \citenamefont {Kimura}, \citenamefont {Sakai}, \citenamefont
		{Matsunami}, \citenamefont {Hayashita}, \citenamefont {Kajikawa},
		\citenamefont {Liu}, \citenamefont {Robert}, \citenamefont {Miron},
		\citenamefont {Feifel}, \citenamefont {Marangos}, \citenamefont {Tono},
		\citenamefont {Inubushi}, \citenamefont {Yabashi}, \citenamefont {Son},
		\citenamefont {Ziaja}, \citenamefont {Yao}, \citenamefont {Santra},\ and\
		\citenamefont {Ueda}}]{Tachibana2015}%
	\BibitemOpen
	\bibfield  {author} {\bibinfo {author} {\bibfnamefont {T.}~\bibnamefont
			{Tachibana}}, \bibinfo {author} {\bibfnamefont {Z.}~\bibnamefont {Jurek}},
		\bibinfo {author} {\bibfnamefont {H.}~\bibnamefont {Fukuzawa}}, \bibinfo
		{author} {\bibfnamefont {K.}~\bibnamefont {Motomura}}, \bibinfo {author}
		{\bibfnamefont {K.}~\bibnamefont {Nagaya}}, \bibinfo {author} {\bibfnamefont
			{S.}~\bibnamefont {Wada}}, \bibinfo {author} {\bibfnamefont {P.}~\bibnamefont
			{Johnsson}}, \bibinfo {author} {\bibfnamefont {M.}~\bibnamefont {Siano}},
		\bibinfo {author} {\bibfnamefont {S.}~\bibnamefont {Mondal}}, \bibinfo
		{author} {\bibfnamefont {Y.}~\bibnamefont {Ito}}, \bibinfo {author}
		{\bibfnamefont {M.}~\bibnamefont {Kimura}}, \bibinfo {author} {\bibfnamefont
			{T.}~\bibnamefont {Sakai}}, \bibinfo {author} {\bibfnamefont
			{K.}~\bibnamefont {Matsunami}}, \bibinfo {author} {\bibfnamefont
			{H.}~\bibnamefont {Hayashita}}, \bibinfo {author} {\bibfnamefont
			{J.}~\bibnamefont {Kajikawa}}, \bibinfo {author} {\bibfnamefont {X.~J.}\
			\bibnamefont {Liu}}, \bibinfo {author} {\bibfnamefont {E.}~\bibnamefont
			{Robert}}, \bibinfo {author} {\bibfnamefont {C.}~\bibnamefont {Miron}},
		\bibinfo {author} {\bibfnamefont {R.}~\bibnamefont {Feifel}}, \bibinfo
		{author} {\bibfnamefont {J.~P.}\ \bibnamefont {Marangos}}, \bibinfo {author}
		{\bibfnamefont {K.}~\bibnamefont {Tono}}, \bibinfo {author} {\bibfnamefont
			{Y.}~\bibnamefont {Inubushi}}, \bibinfo {author} {\bibfnamefont
			{M.}~\bibnamefont {Yabashi}}, \bibinfo {author} {\bibfnamefont {S.~K.}\
			\bibnamefont {Son}}, \bibinfo {author} {\bibfnamefont {B.}~\bibnamefont
			{Ziaja}}, \bibinfo {author} {\bibfnamefont {M.}~\bibnamefont {Yao}}, \bibinfo
		{author} {\bibfnamefont {R.}~\bibnamefont {Santra}},\ and\ \bibinfo {author}
		{\bibfnamefont {K.}~\bibnamefont {Ueda}},\ }\bibfield  {title} {\bibinfo
		{title} {Nanoplasma formation by high intensity hard x-rays},\ }\href
	{https://doi.org/10.1038/srep10977} {\bibfield  {journal} {\bibinfo
			{journal} {Sci. Rep.}\ }\textbf {\bibinfo {volume} {5}},\ \bibinfo {pages}
		{10977} (\bibinfo {year} {2015})}\BibitemShut {NoStop}%
	\bibitem [{\citenamefont {Kumagai}\ \emph {et~al.}(2018)\citenamefont
		{Kumagai}, \citenamefont {Jurek}, \citenamefont {Xu}, \citenamefont
		{Fukuzawa}, \citenamefont {Motomura}, \citenamefont {Iablonskyi},
		\citenamefont {Nagaya}, \citenamefont {Wada}, \citenamefont {Mondal},
		\citenamefont {Tachibana}, \citenamefont {Ito}, \citenamefont {Sakai},
		\citenamefont {Matsunami}, \citenamefont {Nishiyama}, \citenamefont
		{Umemoto}, \citenamefont {Nicolas}, \citenamefont {Miron}, \citenamefont
		{Togashi}, \citenamefont {Ogawa}, \citenamefont {Owada}, \citenamefont
		{Tono}, \citenamefont {Yabashi}, \citenamefont {Son}, \citenamefont {Ziaja},
		\citenamefont {Santra},\ and\ \citenamefont {Ueda}}]{Kumagai2018}%
	\BibitemOpen
	\bibfield  {author} {\bibinfo {author} {\bibfnamefont {Y.}~\bibnamefont
			{Kumagai}}, \bibinfo {author} {\bibfnamefont {Z.}~\bibnamefont {Jurek}},
		\bibinfo {author} {\bibfnamefont {W.}~\bibnamefont {Xu}}, \bibinfo {author}
		{\bibfnamefont {H.}~\bibnamefont {Fukuzawa}}, \bibinfo {author}
		{\bibfnamefont {K.}~\bibnamefont {Motomura}}, \bibinfo {author}
		{\bibfnamefont {D.}~\bibnamefont {Iablonskyi}}, \bibinfo {author}
		{\bibfnamefont {K.}~\bibnamefont {Nagaya}}, \bibinfo {author} {\bibfnamefont
			{S.-i.}\ \bibnamefont {Wada}}, \bibinfo {author} {\bibfnamefont
			{S.}~\bibnamefont {Mondal}}, \bibinfo {author} {\bibfnamefont
			{T.}~\bibnamefont {Tachibana}}, \bibinfo {author} {\bibfnamefont
			{Y.}~\bibnamefont {Ito}}, \bibinfo {author} {\bibfnamefont {T.}~\bibnamefont
			{Sakai}}, \bibinfo {author} {\bibfnamefont {K.}~\bibnamefont {Matsunami}},
		\bibinfo {author} {\bibfnamefont {T.}~\bibnamefont {Nishiyama}}, \bibinfo
		{author} {\bibfnamefont {T.}~\bibnamefont {Umemoto}}, \bibinfo {author}
		{\bibfnamefont {C.}~\bibnamefont {Nicolas}}, \bibinfo {author} {\bibfnamefont
			{C.}~\bibnamefont {Miron}}, \bibinfo {author} {\bibfnamefont
			{T.}~\bibnamefont {Togashi}}, \bibinfo {author} {\bibfnamefont
			{K.}~\bibnamefont {Ogawa}}, \bibinfo {author} {\bibfnamefont
			{S.}~\bibnamefont {Owada}}, \bibinfo {author} {\bibfnamefont
			{K.}~\bibnamefont {Tono}}, \bibinfo {author} {\bibfnamefont {M.}~\bibnamefont
			{Yabashi}}, \bibinfo {author} {\bibfnamefont {S.-K.}\ \bibnamefont {Son}},
		\bibinfo {author} {\bibfnamefont {B.}~\bibnamefont {Ziaja}}, \bibinfo
		{author} {\bibfnamefont {R.}~\bibnamefont {Santra}},\ and\ \bibinfo {author}
		{\bibfnamefont {K.}~\bibnamefont {Ueda}},\ }\bibfield  {title} {\bibinfo
		{title} {Radiation-induced chemical dynamics in ar clusters exposed to strong
			x-ray pulses},\ }\href {https://doi.org/10.1103/PhysRevLett.120.223201}
	{\bibfield  {journal} {\bibinfo  {journal} {Phys. Rev. Lett.}\ }\textbf
		{\bibinfo {volume} {120}},\ \bibinfo {pages} {223201} (\bibinfo {year}
		{2018})}\BibitemShut {NoStop}%
	\bibitem [{\citenamefont {Kumagai}\ \emph {et~al.}(2020)\citenamefont
		{Kumagai}, \citenamefont {Jurek}, \citenamefont {Xu}, \citenamefont
		{Fukuzawa}, \citenamefont {Motomura}, \citenamefont {Iablonskyi},
		\citenamefont {Nagaya}, \citenamefont {Wada}, \citenamefont {Mondal},
		\citenamefont {Tachibana}, \citenamefont {Ito}, \citenamefont {Sakai},
		\citenamefont {Matsunami}, \citenamefont {Nishiyama}, \citenamefont
		{Umemoto}, \citenamefont {Nicolas}, \citenamefont {Miron}, \citenamefont
		{Togashi}, \citenamefont {Ogawa}, \citenamefont {Owada}, \citenamefont
		{Tono}, \citenamefont {Yabashi}, \citenamefont {Son}, \citenamefont {Ziaja},
		\citenamefont {Santra},\ and\ \citenamefont {Ueda}}]{Kumagai2020}%
	\BibitemOpen
	\bibfield  {author} {\bibinfo {author} {\bibfnamefont {Y.}~\bibnamefont
			{Kumagai}}, \bibinfo {author} {\bibfnamefont {Z.}~\bibnamefont {Jurek}},
		\bibinfo {author} {\bibfnamefont {W.}~\bibnamefont {Xu}}, \bibinfo {author}
		{\bibfnamefont {H.}~\bibnamefont {Fukuzawa}}, \bibinfo {author}
		{\bibfnamefont {K.}~\bibnamefont {Motomura}}, \bibinfo {author}
		{\bibfnamefont {D.}~\bibnamefont {Iablonskyi}}, \bibinfo {author}
		{\bibfnamefont {K.}~\bibnamefont {Nagaya}}, \bibinfo {author} {\bibfnamefont
			{S.-i.}\ \bibnamefont {Wada}}, \bibinfo {author} {\bibfnamefont
			{S.}~\bibnamefont {Mondal}}, \bibinfo {author} {\bibfnamefont
			{T.}~\bibnamefont {Tachibana}}, \bibinfo {author} {\bibfnamefont
			{Y.}~\bibnamefont {Ito}}, \bibinfo {author} {\bibfnamefont {T.}~\bibnamefont
			{Sakai}}, \bibinfo {author} {\bibfnamefont {K.}~\bibnamefont {Matsunami}},
		\bibinfo {author} {\bibfnamefont {T.}~\bibnamefont {Nishiyama}}, \bibinfo
		{author} {\bibfnamefont {T.}~\bibnamefont {Umemoto}}, \bibinfo {author}
		{\bibfnamefont {C.}~\bibnamefont {Nicolas}}, \bibinfo {author} {\bibfnamefont
			{C.}~\bibnamefont {Miron}}, \bibinfo {author} {\bibfnamefont
			{T.}~\bibnamefont {Togashi}}, \bibinfo {author} {\bibfnamefont
			{K.}~\bibnamefont {Ogawa}}, \bibinfo {author} {\bibfnamefont
			{S.}~\bibnamefont {Owada}}, \bibinfo {author} {\bibfnamefont
			{K.}~\bibnamefont {Tono}}, \bibinfo {author} {\bibfnamefont {M.}~\bibnamefont
			{Yabashi}}, \bibinfo {author} {\bibfnamefont {S.-K.}\ \bibnamefont {Son}},
		\bibinfo {author} {\bibfnamefont {B.}~\bibnamefont {Ziaja}}, \bibinfo
		{author} {\bibfnamefont {R.}~\bibnamefont {Santra}},\ and\ \bibinfo {author}
		{\bibfnamefont {K.}~\bibnamefont {Ueda}},\ }\bibfield  {title} {\bibinfo
		{title} {Real-time observation of disintegration processes within argon
			clusters ionized by a hard-x-ray pulse of moderate fluence},\ }\href
	{https://doi.org/10.1103/PhysRevA.101.023412} {\bibfield  {journal} {\bibinfo
			{journal} {Phys. Rev. A}\ }\textbf {\bibinfo {volume} {101}},\ \bibinfo
		{pages} {023412} (\bibinfo {year} {2020})}\BibitemShut {NoStop}%
	\bibitem [{\citenamefont {Abdullah}\ \emph {et~al.}(2016)\citenamefont
		{Abdullah}, \citenamefont {Jurek}, \citenamefont {Son},\ and\ \citenamefont
		{Santra}}]{Abdullah2016}%
	\BibitemOpen
	\bibfield  {author} {\bibinfo {author} {\bibfnamefont {M.~M.}\ \bibnamefont
			{Abdullah}}, \bibinfo {author} {\bibfnamefont {Z.}~\bibnamefont {Jurek}},
		\bibinfo {author} {\bibfnamefont {S.-K.}\ \bibnamefont {Son}},\ and\ \bibinfo
		{author} {\bibfnamefont {R.}~\bibnamefont {Santra}},\ }\bibfield  {title}
	{\bibinfo {title} {Calculation of x-ray scattering patterns from nanocrystals
			at high x-ray intensity},\ }\href {https://doi.org/10.1063/1.4958887}
	{\bibfield  {journal} {\bibinfo  {journal} {Struct. Dyn.}\ }\textbf {\bibinfo
			{volume} {3}},\ \bibinfo {pages} {054101} (\bibinfo {year}
		{2016})}\BibitemShut {NoStop}%
	\bibitem [{\citenamefont {Abdullah}\ \emph {et~al.}(2017)\citenamefont
		{Abdullah}, \citenamefont {Anurag}, \citenamefont {Jurek}, \citenamefont
		{Son},\ and\ \citenamefont {Santra}}]{Abdullah2017}%
	\BibitemOpen
	\bibfield  {author} {\bibinfo {author} {\bibfnamefont {M.~M.}\ \bibnamefont
			{Abdullah}}, \bibinfo {author} {\bibnamefont {Anurag}}, \bibinfo {author}
		{\bibfnamefont {Z.}~\bibnamefont {Jurek}}, \bibinfo {author} {\bibfnamefont
			{S.-K.}\ \bibnamefont {Son}},\ and\ \bibinfo {author} {\bibfnamefont
			{R.}~\bibnamefont {Santra}},\ }\bibfield  {title} {\bibinfo {title}
		{Molecular-dynamics approach for studying the nonequilibrium behavior of
			x-ray-heated solid-density matter},\ }\href
	{https://doi.org/10.1103/PhysRevE.96.023205} {\bibfield  {journal} {\bibinfo
			{journal} {Phys. Rev. E}\ }\textbf {\bibinfo {volume} {96}},\ \bibinfo
		{pages} {023205} (\bibinfo {year} {2017})}\BibitemShut {NoStop}%
	\bibitem [{\citenamefont {Abdullah}\ \emph {et~al.}(2018)\citenamefont
		{Abdullah}, \citenamefont {Son}, \citenamefont {Jurek},\ and\ \citenamefont
		{Santra}}]{Abdullah2018}%
	\BibitemOpen
	\bibfield  {author} {\bibinfo {author} {\bibfnamefont {M.~M.}\ \bibnamefont
			{Abdullah}}, \bibinfo {author} {\bibfnamefont {S.-K.}\ \bibnamefont {Son}},
		\bibinfo {author} {\bibfnamefont {Z.}~\bibnamefont {Jurek}},\ and\ \bibinfo
		{author} {\bibfnamefont {R.}~\bibnamefont {Santra}},\ }\bibfield  {title}
	{\bibinfo {title} {{Towards the theoretical limitations of X-ray
				nanocrystallography at high intensity: the validity of the
				effective-form-factor description}},\ }\href
	{https://doi.org/10.1107/S2052252518011442} {\bibfield  {journal} {\bibinfo
			{journal} {IUCrJ}\ }\textbf {\bibinfo {volume} {5}},\ \bibinfo {pages} {699}
		(\bibinfo {year} {2018})}\BibitemShut {NoStop}%
	\bibitem [{\citenamefont {Yoon}\ \emph {et~al.}(2016)\citenamefont {Yoon},
		\citenamefont {Yurkov}, \citenamefont {Schneidmiller}, \citenamefont
		{Samoylova}, \citenamefont {Buzmakov}, \citenamefont {Jurek}, \citenamefont
		{Ziaja}, \citenamefont {Santra}, \citenamefont {Loh}, \citenamefont
		{Tschentscher},\ and\ \citenamefont {Mancuso}}]{Yoon2016}%
	\BibitemOpen
	\bibfield  {author} {\bibinfo {author} {\bibfnamefont {C.~H.}\ \bibnamefont
			{Yoon}}, \bibinfo {author} {\bibfnamefont {M.~V.}\ \bibnamefont {Yurkov}},
		\bibinfo {author} {\bibfnamefont {E.~A.}\ \bibnamefont {Schneidmiller}},
		\bibinfo {author} {\bibfnamefont {L.}~\bibnamefont {Samoylova}}, \bibinfo
		{author} {\bibfnamefont {A.}~\bibnamefont {Buzmakov}}, \bibinfo {author}
		{\bibfnamefont {Z.}~\bibnamefont {Jurek}}, \bibinfo {author} {\bibfnamefont
			{B.}~\bibnamefont {Ziaja}}, \bibinfo {author} {\bibfnamefont
			{R.}~\bibnamefont {Santra}}, \bibinfo {author} {\bibfnamefont {N.~D.}\
			\bibnamefont {Loh}}, \bibinfo {author} {\bibfnamefont {T.}~\bibnamefont
			{Tschentscher}},\ and\ \bibinfo {author} {\bibfnamefont {A.~P.}\ \bibnamefont
			{Mancuso}},\ }\bibfield  {title} {\bibinfo {title} {A comprehensive
			simulation framework for imaging single particles and biomolecules at the
			european x-ray free-electron laser},\ }\href
	{https://doi.org/10.1038/srep24791} {\bibfield  {journal} {\bibinfo
			{journal} {Sci. Rep.}\ }\textbf {\bibinfo {volume} {6}},\ \bibinfo {pages}
		{24791} (\bibinfo {year} {2016})}\BibitemShut {NoStop}%
	\bibitem [{\citenamefont {Fortmann-Grote}\ \emph {et~al.}(2017)\citenamefont
		{Fortmann-Grote}, \citenamefont {Buzmakov}, \citenamefont {Jurek},
		\citenamefont {Loh}, \citenamefont {Samoylova}, \citenamefont {Santra},
		\citenamefont {Schneidmiller}, \citenamefont {Tschentscher}, \citenamefont
		{Yakubov}, \citenamefont {Yoon}, \citenamefont {Yurkov}, \citenamefont
		{Ziaja-Motyka},\ and\ \citenamefont {Mancuso}}]{Fortmann-Grote2017}%
	\BibitemOpen
	\bibfield  {author} {\bibinfo {author} {\bibfnamefont {C.}~\bibnamefont
			{Fortmann-Grote}}, \bibinfo {author} {\bibfnamefont {A.}~\bibnamefont
			{Buzmakov}}, \bibinfo {author} {\bibfnamefont {Z.}~\bibnamefont {Jurek}},
		\bibinfo {author} {\bibfnamefont {N.-T.~D.}\ \bibnamefont {Loh}}, \bibinfo
		{author} {\bibfnamefont {L.}~\bibnamefont {Samoylova}}, \bibinfo {author}
		{\bibfnamefont {R.}~\bibnamefont {Santra}}, \bibinfo {author} {\bibfnamefont
			{E.~A.}\ \bibnamefont {Schneidmiller}}, \bibinfo {author} {\bibfnamefont
			{T.}~\bibnamefont {Tschentscher}}, \bibinfo {author} {\bibfnamefont
			{S.}~\bibnamefont {Yakubov}}, \bibinfo {author} {\bibfnamefont {C.~H.}\
			\bibnamefont {Yoon}}, \bibinfo {author} {\bibfnamefont {M.~V.}\ \bibnamefont
			{Yurkov}}, \bibinfo {author} {\bibfnamefont {B.}~\bibnamefont
			{Ziaja-Motyka}},\ and\ \bibinfo {author} {\bibfnamefont {A.~P.}\ \bibnamefont
			{Mancuso}},\ }\bibfield  {title} {\bibinfo {title} {{Start-to-end simulation
				of single-particle imaging using ultra-short pulses at the European X-ray
				Free-Electron Laser}},\ }\href {https://doi.org/10.1107/S2052252517009496}
	{\bibfield  {journal} {\bibinfo  {journal} {IUCrJ}\ }\textbf {\bibinfo
			{volume} {4}},\ \bibinfo {pages} {560} (\bibinfo {year} {2017})}\BibitemShut
	{NoStop}%
	\bibitem [{\citenamefont {Slater}(1951)}]{Slater1951}%
	\BibitemOpen
	\bibfield  {author} {\bibinfo {author} {\bibfnamefont {J.~C.}\ \bibnamefont
			{Slater}},\ }\bibfield  {title} {\bibinfo {title} {A simplification of the
			hartree-fock method},\ }\href {https://doi.org/10.1103/PhysRev.81.385}
	{\bibfield  {journal} {\bibinfo  {journal} {Phys. Rev.}\ }\textbf {\bibinfo
			{volume} {81}},\ \bibinfo {pages} {385} (\bibinfo {year} {1951})}\BibitemShut
	{NoStop}%
	\bibitem [{\citenamefont {Latter}(1955)}]{Latter1955}%
	\BibitemOpen
	\bibfield  {author} {\bibinfo {author} {\bibfnamefont {R.}~\bibnamefont
			{Latter}},\ }\bibfield  {title} {\bibinfo {title} {Atomic energy levels for
			the {T}homas-{F}ermi and {T}homas-{F}ermi-{D}irac potential},\ }\href
	{https://doi.org/10.1103/PhysRev.99.510} {\bibfield  {journal} {\bibinfo
			{journal} {Phys. Rev.}\ }\textbf {\bibinfo {volume} {99}},\ \bibinfo {pages}
		{510} (\bibinfo {year} {1955})}\BibitemShut {NoStop}%
	\bibitem [{\citenamefont {Slater}(1937)}]{Slater1937}%
	\BibitemOpen
	\bibfield  {author} {\bibinfo {author} {\bibfnamefont {J.~C.}\ \bibnamefont
			{Slater}},\ }\bibfield  {title} {\bibinfo {title} {Wave functions in a
			periodic potential},\ }\href {https://doi.org/10.1103/PhysRev.51.846}
	{\bibfield  {journal} {\bibinfo  {journal} {Phys. Rev.}\ }\textbf {\bibinfo
			{volume} {51}},\ \bibinfo {pages} {846} (\bibinfo {year} {1937})}\BibitemShut
	{NoStop}%
	\bibitem [{\citenamefont {Korringa}(1947)}]{Korringa1947}%
	\BibitemOpen
	\bibfield  {author} {\bibinfo {author} {\bibfnamefont {J.}~\bibnamefont
			{Korringa}},\ }\bibfield  {title} {\bibinfo {title} {On the calculation of
			the energy of a bloch wave in a metal},\ }\href
	{https://doi.org/https://doi.org/10.1016/0031-8914(47)90013-X} {\bibfield
		{journal} {\bibinfo  {journal} {Physica}\ }\textbf {\bibinfo {volume} {13}},\
		\bibinfo {pages} {392 } (\bibinfo {year} {1947})}\BibitemShut {NoStop}%
	\bibitem [{\citenamefont {Kohn}\ and\ \citenamefont
		{Rostoker}(1954)}]{Kohn1954}%
	\BibitemOpen
	\bibfield  {author} {\bibinfo {author} {\bibfnamefont {W.}~\bibnamefont
			{Kohn}}\ and\ \bibinfo {author} {\bibfnamefont {N.}~\bibnamefont
			{Rostoker}},\ }\bibfield  {title} {\bibinfo {title} {Solution of the
			schr\"odinger equation in periodic lattices with an application to metallic
			lithium},\ }\href {https://doi.org/10.1103/PhysRev.94.1111} {\bibfield
		{journal} {\bibinfo  {journal} {Phys. Rev.}\ }\textbf {\bibinfo {volume}
			{94}},\ \bibinfo {pages} {1111} (\bibinfo {year} {1954})}\BibitemShut
	{NoStop}%
	\bibitem [{\citenamefont {Methfessel}\ \emph {et~al.}(1989)\citenamefont
		{Methfessel}, \citenamefont {Rodriguez},\ and\ \citenamefont
		{Andersen}}]{Anderson1989}%
	\BibitemOpen
	\bibfield  {author} {\bibinfo {author} {\bibfnamefont {M.}~\bibnamefont
			{Methfessel}}, \bibinfo {author} {\bibfnamefont {C.~O.}\ \bibnamefont
			{Rodriguez}},\ and\ \bibinfo {author} {\bibfnamefont {O.~K.}\ \bibnamefont
			{Andersen}},\ }\bibfield  {title} {\bibinfo {title} {Fast full-potential
			calculations with a converged basis of atom-centered linear muffin-tin
			orbitals: Structural and dynamic properties of silicon},\ }\href
	{https://doi.org/10.1103/PhysRevB.40.2009} {\bibfield  {journal} {\bibinfo
			{journal} {Phys. Rev. B}\ }\textbf {\bibinfo {volume} {40}},\ \bibinfo
		{pages} {2009} (\bibinfo {year} {1989})}\BibitemShut {NoStop}%
	\bibitem [{\citenamefont {Skriver}(1984)}]{Skriver1984}%
	\BibitemOpen
	\bibfield  {author} {\bibinfo {author} {\bibfnamefont {H.~L.}\ \bibnamefont
			{Skriver}},\ }\href {https://doi.org/10.1007/978-3-642-81844-8} {\emph
		{\bibinfo {title} {The LMTO Method:Muffin-Tin Orbitals and Electronic
				Structure}}},\ \bibinfo {edition} {1st}\ ed.\ (\bibinfo  {publisher}
	{Springer-Verlag Berlin Heidelberg},\ \bibinfo {address} {Berlin
		Heidelberg},\ \bibinfo {year} {1984})\BibitemShut {NoStop}%
	\bibitem [{\citenamefont {Cho}\ \emph {et~al.}(2012)\citenamefont {Cho},
		\citenamefont {Engelhorn}, \citenamefont {Vinko}, \citenamefont {Chung},
		\citenamefont {Ciricosta}, \citenamefont {Rackstraw}, \citenamefont
		{Falcone}, \citenamefont {Brown}, \citenamefont {Burian}, \citenamefont
		{Chalupsk\'y}, \citenamefont {Graves}, \citenamefont {H\'ajkov\'a},
		\citenamefont {Higginbotham}, \citenamefont {Juha}, \citenamefont
		{Krzywinski}, \citenamefont {Lee}, \citenamefont {Messersmidt}, \citenamefont
		{Murphy}, \citenamefont {Ping}, \citenamefont {Rohringer}, \citenamefont
		{Scherz}, \citenamefont {Schlotter}, \citenamefont {Toleikis}, \citenamefont
		{Turner}, \citenamefont {Vysin}, \citenamefont {Wang}, \citenamefont {Wu},
		\citenamefont {Zastrau}, \citenamefont {Zhu}, \citenamefont {Lee},
		\citenamefont {Nagler}, \citenamefont {Wark},\ and\ \citenamefont
		{Heimann}}]{Cho2012}%
	\BibitemOpen
	\bibfield  {author} {\bibinfo {author} {\bibfnamefont {B.~I.}\ \bibnamefont
			{Cho}}, \bibinfo {author} {\bibfnamefont {K.}~\bibnamefont {Engelhorn}},
		\bibinfo {author} {\bibfnamefont {S.~M.}\ \bibnamefont {Vinko}}, \bibinfo
		{author} {\bibfnamefont {H.-K.}\ \bibnamefont {Chung}}, \bibinfo {author}
		{\bibfnamefont {O.}~\bibnamefont {Ciricosta}}, \bibinfo {author}
		{\bibfnamefont {D.~S.}\ \bibnamefont {Rackstraw}}, \bibinfo {author}
		{\bibfnamefont {R.~W.}\ \bibnamefont {Falcone}}, \bibinfo {author}
		{\bibfnamefont {C.~R.~D.}\ \bibnamefont {Brown}}, \bibinfo {author}
		{\bibfnamefont {T.}~\bibnamefont {Burian}}, \bibinfo {author} {\bibfnamefont
			{J.}~\bibnamefont {Chalupsk\'y}}, \bibinfo {author} {\bibfnamefont
			{C.}~\bibnamefont {Graves}}, \bibinfo {author} {\bibfnamefont
			{V.}~\bibnamefont {H\'ajkov\'a}}, \bibinfo {author} {\bibfnamefont
			{A.}~\bibnamefont {Higginbotham}}, \bibinfo {author} {\bibfnamefont
			{L.}~\bibnamefont {Juha}}, \bibinfo {author} {\bibfnamefont {J.}~\bibnamefont
			{Krzywinski}}, \bibinfo {author} {\bibfnamefont {H.~J.}\ \bibnamefont {Lee}},
		\bibinfo {author} {\bibfnamefont {M.}~\bibnamefont {Messersmidt}}, \bibinfo
		{author} {\bibfnamefont {C.}~\bibnamefont {Murphy}}, \bibinfo {author}
		{\bibfnamefont {Y.}~\bibnamefont {Ping}}, \bibinfo {author} {\bibfnamefont
			{N.}~\bibnamefont {Rohringer}}, \bibinfo {author} {\bibfnamefont
			{A.}~\bibnamefont {Scherz}}, \bibinfo {author} {\bibfnamefont
			{W.}~\bibnamefont {Schlotter}}, \bibinfo {author} {\bibfnamefont
			{S.}~\bibnamefont {Toleikis}}, \bibinfo {author} {\bibfnamefont {J.~J.}\
			\bibnamefont {Turner}}, \bibinfo {author} {\bibfnamefont {L.}~\bibnamefont
			{Vysin}}, \bibinfo {author} {\bibfnamefont {T.}~\bibnamefont {Wang}},
		\bibinfo {author} {\bibfnamefont {B.}~\bibnamefont {Wu}}, \bibinfo {author}
		{\bibfnamefont {U.}~\bibnamefont {Zastrau}}, \bibinfo {author} {\bibfnamefont
			{D.}~\bibnamefont {Zhu}}, \bibinfo {author} {\bibfnamefont {R.~W.}\
			\bibnamefont {Lee}}, \bibinfo {author} {\bibfnamefont {B.}~\bibnamefont
			{Nagler}}, \bibinfo {author} {\bibfnamefont {J.~S.}\ \bibnamefont {Wark}},\
		and\ \bibinfo {author} {\bibfnamefont {P.~A.}\ \bibnamefont {Heimann}},\
	}\bibfield  {title} {\bibinfo {title} {Resonant $\ensuremath{K\alpha}$
			spectroscopy of solid-density aluminum plasmas},\ }\href
	{https://doi.org/10.1103/PhysRevLett.109.245003} {\bibfield  {journal}
		{\bibinfo  {journal} {Phys. Rev. Lett.}\ }\textbf {\bibinfo {volume} {109}},\
		\bibinfo {pages} {245003} (\bibinfo {year} {2012})}\BibitemShut {NoStop}%
	\bibitem [{\citenamefont {Su}\ and\ \citenamefont {Goddard}(2007)}]{Su:2007aa}%
	\BibitemOpen
	\bibfield  {author} {\bibinfo {author} {\bibfnamefont {J.~T.}\ \bibnamefont
			{Su}}\ and\ \bibinfo {author} {\bibfnamefont {W.~A.}\ \bibnamefont
			{Goddard}},\ }\bibfield  {title} {\bibinfo {title} {Excited electron dynamics
			modeling of warm dense matter},\ }\href
	{https://doi.org/10.1103/PhysRevLett.99.185003} {\bibfield  {journal}
		{\bibinfo  {journal} {Phys. Rev. Lett.}\ }\textbf {\bibinfo {volume} {99}},\
		\bibinfo {pages} {185003} (\bibinfo {year} {2007})}\BibitemShut {NoStop}%
	\bibitem [{\citenamefont {Kim}\ \emph {et~al.}(2011)\citenamefont {Kim},
		\citenamefont {Su},\ and\ \citenamefont {Goddard}}]{Kim:2011aa}%
	\BibitemOpen
	\bibfield  {author} {\bibinfo {author} {\bibfnamefont {H.}~\bibnamefont
			{Kim}}, \bibinfo {author} {\bibfnamefont {J.~T.}\ \bibnamefont {Su}},\ and\
		\bibinfo {author} {\bibfnamefont {W.~A.}\ \bibnamefont {Goddard}},\
	}\bibfield  {title} {\bibinfo {title} {High-temperature high-pressure phases
			of lithium from electron force field ({eFF}) quantum electron dynamics
			simulations},\ }\href {https://doi.org/10.1073/pnas.1110322108} {\bibfield
		{journal} {\bibinfo  {journal} {Proc. Natl. Acad. Sci. U. S. A.}\ }\textbf
		{\bibinfo {volume} {108}},\ \bibinfo {pages} {15101} (\bibinfo {year}
		{2011})}\BibitemShut {NoStop}%
	\bibitem [{\citenamefont {Toyota}\ \emph {et~al.}(2019)\citenamefont {Toyota},
		\citenamefont {Jurek}, \citenamefont {Son}, \citenamefont {Fukuzawa},
		\citenamefont {Ueda}, \citenamefont {Berrah}, \citenamefont {Rudek},
		\citenamefont {Rolles}, \citenamefont {Rudenko},\ and\ \citenamefont
		{Santra}}]{Toyota19}%
	\BibitemOpen
	\bibfield  {author} {\bibinfo {author} {\bibfnamefont {K.}~\bibnamefont
			{Toyota}}, \bibinfo {author} {\bibfnamefont {Z.}~\bibnamefont {Jurek}},
		\bibinfo {author} {\bibfnamefont {S.-K.}\ \bibnamefont {Son}}, \bibinfo
		{author} {\bibfnamefont {H.}~\bibnamefont {Fukuzawa}}, \bibinfo {author}
		{\bibfnamefont {K.}~\bibnamefont {Ueda}}, \bibinfo {author} {\bibfnamefont
			{N.}~\bibnamefont {Berrah}}, \bibinfo {author} {\bibfnamefont
			{B.}~\bibnamefont {Rudek}}, \bibinfo {author} {\bibfnamefont
			{D.}~\bibnamefont {Rolles}}, \bibinfo {author} {\bibfnamefont
			{A.}~\bibnamefont {Rudenko}},\ and\ \bibinfo {author} {\bibfnamefont
			{R.}~\bibnamefont {Santra}},\ }\bibfield  {title} {\bibinfo {title}
		{\emph{xcalib}: a focal spot calibrator for intense x-ray free-electron laser
			pulses based on the charge state distributions of light atoms},\ }\href
	{https://doi.org/10.1107/S1600577519003564} {\bibfield  {journal} {\bibinfo
			{journal} {J. Synchrotron Radiat.}\ }\textbf {\bibinfo {volume} {26}},\
		\bibinfo {pages} {1017} (\bibinfo {year} {2019})}\BibitemShut {NoStop}%
	\bibitem [{\citenamefont {Pathria}\ and\ \citenamefont
		{Beale}(2011)}]{Pathria}%
	\BibitemOpen
	\bibfield  {author} {\bibinfo {author} {\bibfnamefont {R.~K.}\ \bibnamefont
			{Pathria}}\ and\ \bibinfo {author} {\bibfnamefont {P.~D.}\ \bibnamefont
			{Beale}},\ }\href@noop {} {\emph {\bibinfo {title} {Statistical
				Mechanics}}},\ \bibinfo {edition} {3rd}\ ed.\ (\bibinfo  {publisher}
	{Academic Press},\ \bibinfo {address} {Cambridge, England},\ \bibinfo {year}
	{2011})\ pp.\ \bibinfo {pages} {135--136}\BibitemShut {NoStop}%
	\bibitem [{\citenamefont {Thompson}\ \emph {et~al.}(2001)\citenamefont
		{Thompson} \emph {et~al.}}]{XrayBooklet}%
	\BibitemOpen
	\bibfield  {author} {\bibinfo {author} {\bibfnamefont {A.~C.}\ \bibnamefont
			{Thompson}} \emph {et~al.},\ }\href {https://xdb.lbl.gov/} {\bibinfo {title}
		{X-ray data booklet}},\ \bibinfo {howpublished} {Center for X-Ray Optics and
		Advanced Light Source},\ \bibinfo {address} {Lawrence Berkeley National
		Laboratory, Berkeley, CA} (\bibinfo {year} {2001})\BibitemShut {NoStop}%
	\bibitem [{\citenamefont {Thiele}\ \emph {et~al.}(2012)\citenamefont {Thiele},
		\citenamefont {Son}, \citenamefont {Ziaja},\ and\ \citenamefont
		{Santra}}]{Thiele2012}%
	\BibitemOpen
	\bibfield  {author} {\bibinfo {author} {\bibfnamefont {R.}~\bibnamefont
			{Thiele}}, \bibinfo {author} {\bibfnamefont {S.-K.}\ \bibnamefont {Son}},
		\bibinfo {author} {\bibfnamefont {B.}~\bibnamefont {Ziaja}},\ and\ \bibinfo
		{author} {\bibfnamefont {R.}~\bibnamefont {Santra}},\ }\bibfield  {title}
	{\bibinfo {title} {Effect of screening by external charges on the atomic
			orbitals and photoinduced processes within the hartree-fock-slater atom},\
	}\href {https://doi.org/10.1103/PhysRevA.86.033411} {\bibfield  {journal}
		{\bibinfo  {journal} {Phys. Rev. A}\ }\textbf {\bibinfo {volume} {86}},\
		\bibinfo {pages} {033411} (\bibinfo {year} {2012})}\BibitemShut {NoStop}%
	\bibitem [{\citenamefont {Vinko}\ \emph {et~al.}(2020)\citenamefont {Vinko},
		\citenamefont {Vozda}, \citenamefont {Andreasson}, \citenamefont {Bajt},
		\citenamefont {Bielecki}, \citenamefont {Burian}, \citenamefont {Chalupsky},
		\citenamefont {Ciricosta}, \citenamefont {Desjarlais}, \citenamefont
		{Fleckenstein}, \citenamefont {Hajdu}, \citenamefont {Hajkova}, \citenamefont
		{Hollebon}, \citenamefont {Juha}, \citenamefont {Kasim}, \citenamefont
		{McBride}, \citenamefont {Muehlig}, \citenamefont {Preston}, \citenamefont
		{Rackstraw}, \citenamefont {Roling}, \citenamefont {Toleikis}, \citenamefont
		{Wark},\ and\ \citenamefont {Zacharias}}]{Vinko20}%
	\BibitemOpen
	\bibfield  {author} {\bibinfo {author} {\bibfnamefont {S.~M.}\ \bibnamefont
			{Vinko}}, \bibinfo {author} {\bibfnamefont {V.}~\bibnamefont {Vozda}},
		\bibinfo {author} {\bibfnamefont {J.}~\bibnamefont {Andreasson}}, \bibinfo
		{author} {\bibfnamefont {S.}~\bibnamefont {Bajt}}, \bibinfo {author}
		{\bibfnamefont {J.}~\bibnamefont {Bielecki}}, \bibinfo {author}
		{\bibfnamefont {T.}~\bibnamefont {Burian}}, \bibinfo {author} {\bibfnamefont
			{J.}~\bibnamefont {Chalupsky}}, \bibinfo {author} {\bibfnamefont
			{O.}~\bibnamefont {Ciricosta}}, \bibinfo {author} {\bibfnamefont {M.~P.}\
			\bibnamefont {Desjarlais}}, \bibinfo {author} {\bibfnamefont
			{H.}~\bibnamefont {Fleckenstein}}, \bibinfo {author} {\bibfnamefont
			{J.}~\bibnamefont {Hajdu}}, \bibinfo {author} {\bibfnamefont
			{V.}~\bibnamefont {Hajkova}}, \bibinfo {author} {\bibfnamefont
			{P.}~\bibnamefont {Hollebon}}, \bibinfo {author} {\bibfnamefont
			{L.}~\bibnamefont {Juha}}, \bibinfo {author} {\bibfnamefont {M.~F.}\
			\bibnamefont {Kasim}}, \bibinfo {author} {\bibfnamefont {E.~E.}\ \bibnamefont
			{McBride}}, \bibinfo {author} {\bibfnamefont {K.}~\bibnamefont {Muehlig}},
		\bibinfo {author} {\bibfnamefont {T.~R.}\ \bibnamefont {Preston}}, \bibinfo
		{author} {\bibfnamefont {D.~S.}\ \bibnamefont {Rackstraw}}, \bibinfo {author}
		{\bibfnamefont {S.}~\bibnamefont {Roling}}, \bibinfo {author} {\bibfnamefont
			{S.}~\bibnamefont {Toleikis}}, \bibinfo {author} {\bibfnamefont {J.~S.}\
			\bibnamefont {Wark}},\ and\ \bibinfo {author} {\bibfnamefont
			{H.}~\bibnamefont {Zacharias}},\ }\bibfield  {title} {\bibinfo {title}
		{Time-resolved xuv opacity measurements of warm dense aluminum},\ }\href
	{https://doi.org/10.1103/PhysRevLett.124.225002} {\bibfield  {journal}
		{\bibinfo  {journal} {Phys. Rev. Lett.}\ }\textbf {\bibinfo {volume} {124}},\
		\bibinfo {pages} {225002} (\bibinfo {year} {2020})}\BibitemShut {NoStop}%
	\bibitem [{\citenamefont {Lu}\ \emph {et~al.}(2018)\citenamefont {Lu},
		\citenamefont {Friedrich}, \citenamefont {Noll}, \citenamefont {Zhou},
		\citenamefont {Hallmann}, \citenamefont {Ansaldi}, \citenamefont {Roth},
		\citenamefont {Serkez}, \citenamefont {Geloni}, \citenamefont {Madsen},\ and\
		\citenamefont {Eisebitt}}]{Lu18}%
	\BibitemOpen
	\bibfield  {author} {\bibinfo {author} {\bibfnamefont {W.}~\bibnamefont
			{Lu}}, \bibinfo {author} {\bibfnamefont {B.}~\bibnamefont {Friedrich}},
		\bibinfo {author} {\bibfnamefont {T.}~\bibnamefont {Noll}}, \bibinfo {author}
		{\bibfnamefont {K.}~\bibnamefont {Zhou}}, \bibinfo {author} {\bibfnamefont
			{J.}~\bibnamefont {Hallmann}}, \bibinfo {author} {\bibfnamefont
			{G.}~\bibnamefont {Ansaldi}}, \bibinfo {author} {\bibfnamefont
			{T.}~\bibnamefont {Roth}}, \bibinfo {author} {\bibfnamefont {S.}~\bibnamefont
			{Serkez}}, \bibinfo {author} {\bibfnamefont {G.}~\bibnamefont {Geloni}},
		\bibinfo {author} {\bibfnamefont {A.}~\bibnamefont {Madsen}},\ and\ \bibinfo
		{author} {\bibfnamefont {S.}~\bibnamefont {Eisebitt}},\ }\bibfield  {title}
	{\bibinfo {title} {Development of a hard x-ray split-and-delay line and
			performance simulations for two-color pump-probe experiments at the {European
				XFEL}},\ }\href {https://doi.org/10.1063/1.5027071} {\bibfield  {journal}
		{\bibinfo  {journal} {Rev. Sci. Instrum.}\ }\textbf {\bibinfo {volume}
			{89}},\ \bibinfo {pages} {063121} (\bibinfo {year} {2018})}\BibitemShut
	{NoStop}%
\end{thebibliography}
\end{document}